\documentclass[pra,twocolumn,showpacs,amsmath,amssymb]{revtex4}
\usepackage{amsfonts}
\usepackage{graphicx}
\usepackage{amsmath}
\usepackage{times}
\usepackage{amssymb}
\usepackage{color}
\usepackage{bm}
\usepackage{bbm}

\graphicspath{{figures/}} 
\usepackage{amsfonts, amsmath, amsthm, amssymb} 
\usepackage{array}
\usepackage[colorlinks,bookmarks=false,citecolor=blue,linkcolor=red,urlcolor=blue]{hyperref}

\newcommand{\be}{\begin{equation}}
\newcommand{\bee}{\begin{equation*}}
\newcommand{\ee}{\end{equation}}
\newcommand{\eee}{\end{equation*}}
\newcommand{\bearre}{\begin{eqnarray*}}
\newcommand{\eearre}{\end{eqnarray*}}
\newcommand{\bearr}{\begin{eqnarray}}
\newcommand{\eearr}{\end{eqnarray}}
\newcommand{\D}{\mathrm d}

\begin{document}

\title{Area law and its vioaltion: A microscopic inspection into the structure of entanglement and fluctuations}

\author{Ir\'en\'ee Fr\'erot %
  \footnote{Electronic address: \texttt{irenee.frerot@ens-lyon.fr}}}
\affiliation{Laboratoire de Physique, CNRS UMR 5672,  \'Ecole Normale Sup\'erieure de Lyon, 
Universit\'e de Lyon, 46 All\'ee d'Italie, Lyon, F-69364, France}

\author{Tommaso Roscilde%
  \footnote{Electronic address: \texttt{tommaso.roscilde@ens-lyon.fr}}}
\affiliation{Laboratoire de Physique, CNRS UMR 5672,  \'Ecole Normale Sup\'erieure de Lyon, 
Universit\'e de Lyon, 46 All\'ee d'Italie, Lyon, F-69364, France,
Institut Universitaire de France, 103 boulevard Saint-Michel, 75005 Paris, France}
\date{\today}

\begin{abstract}
Quantum fluctuations of local quantities can be a direct signature of entanglement in an extended quantum many-body system. Hence they may serve as a theoretical (as well as an experimental) tool to detect the spatial properties of the entanglement entropy of a subsystem -- more specifically, its scaling with the size of the subsystem itself. In the ground state of quantum many-body systems, this scaling is typically linear in the boundary of the subsystem (area law), with at most multiplicative logarithmic corrections. Here we propose a microscopic insight into the spatial structure of entanglement and  particle-number fluctuations using the concept of \emph{contour}, recently introduced to decompose the bipartite entanglement entropy of lattice free fermions between two extended regions $A$ and $B$ into contributions from single sites in $A$ \cite{entanglement-contour}. We generalize the notion of contour to the entanglement of any quadratic (bosonic or fermionic) lattice Hamiltonian, as well as to particle-number fluctuations. The entanglement and fluctuations contours are found to generally decay when moving away from the boundary between $A$ and $B$. We show that in the case of free fermions the decay of the entanglement contour follows closely that of the fluctuation contour: this establishes a microscopic link between the scaling of entanglement and that of { particle-number} fluctuations, and it allows to predict the presence (or violation) of entanglement area laws solely based on the density-density correlation function. In the case of Bose-condensed interacting bosons, treated via the Bogoliubov and spin-wave approximations, such a link cannot be established -- fluctuation and entanglement contours are found to be radically different, as they lead to a logarithmically violated area law for { particle-number} fluctuations, and to a strict area law of entanglement. Analyzing in depth the role of the zero-energy Goldstone mode of spin-wave theory, and of the corresponding lowest-energy mode in the entanglement spectrum, we unveil a subtle interplay between the special contour and energy scaling of the latter, and universal additive logarithmic corrections to entanglement area law discussed extensively in the recent literature. \end{abstract}

\maketitle

\section{Introduction}

 \emph{Entanglement in many-body systems.} 
  Entanglement properties of quantum many-body states physics has emerged as one of the most active fields in the study of complex quantum systems \cite{many-body,RMP-area-laws}. 
  The classification of quantum many-body states according to their entanglement properties is a priori completely independent with respect to the conventional classification based on observables that have a classical counterpart, such as order parameters and correlation functions.
  Entanglement between the degrees of freedom of an extended quantum system has the consequence that any subsystem is described by a reduced density matrix possessing a finite entropy -- the so-called entanglement entropy.  A fundamental characterization of entanglement in an extended many-body system is hence represented by the scaling of the entanglement entropy of a subsystem $A$ with its size.
 Such a scaling is generally sub-extensive for ground states of local Hamiltonians \cite{RMP-area-laws}, revealing a fundamental difference between entanglement entropy and equilibrium thermal entropy. This difference is even more marked when realizing that the scaling of entanglement entropy may depend crucially on the nature of the ground state phase (critical vs. short-ranged, symmetric vs. symmetry-breaking, etc.). The classification of states based on the scaling of entanglement is generally weaker than the one based on correlation properties. Nonetheless, given that entanglement has no classical analog, it is of fundamental importance in the characterization of phases that have no classical counterpart either, such as quantum critical points or incompressible quantum liquids \cite{Wen2012}. 
 
 \emph{Entanglement vs. fluctuations}. In this perspective, being able to measure tangible consequences of entanglement in an experiment would represent a fundamental progress in our ability to prepare and characterize quantum many-body states. A useful point of view is represented by the idea of interpreting the ground state entanglement entropy as the entropy of ground state quantum fluctuations. Several recent works \cite{bipartite-fluct,song,Songetal2010,jordan-buttiker,calabrese-mintchev-vicari} have been devoted to establishing a relationship between entanglement entropy and the local fluctuations of globally conserved quantities, such as particle-number (magnetization) fluctuations in lattice quantum gases (quantum spin systems with an axial symmetry). This turns out to be indeed possible in the particular cases of free fermions \cite{bipartite-fluct} and of one-dimensional Luttinger liquids \cite{Songetal2010}, but the problem is completely open for models going beyond these examples. To face this ambitious task, it is fundamental to understand the correspondences or differences among the scaling behavior of entanglement entropy and of fluctuations properties (namely of the various moments of the probability distribution of fluctuations), and the microscopic origin of such behavior. 
 
 The scaling of fluctuations is clearly controlled by the decay of correlation functions. In the case of entanglement, on the contrary, the decay of correlations can only establish an upper bound on the scaling, as entanglement cannot be longer-ranged than correlations \footnote{We specifically refer here to the dominant scaling term of the entanglement entropy. Additive topological constants in the entanglement entropy cannot obviously be accounted for by ordinary correlations, but can still be associated to non-local correlations -- such as string correlations.}. For exponentially decaying correlations, this upper bound implies the area law of entanglement, namely entanglement is proportional to the boundary surface of the subsystem, as it is exponentially confined around the boundary -- this statement is now proved rigorously for a wide class of systems \cite{Hastings2007,BrandaoH2015}. On the other hand, for critical correlations the scaling of entanglement can obey an area law (as in the case of Bose-condensed bosons \cite{Hastingsetal2010, Kallinetal2011, HumeniukR2012, Luitzetal2015, Kulchytskyyetal2015}) or a logarithmically violated area law (as in the case of $d$-dimensional free fermions with a finite density of states at the Fermi energy \cite{gioev-klich,wolf}). Most importantly, the two above cited examples display the same scaling of particle-number fluctuations \cite{gioev-klich,bipartite-fluct,calabrese-mintchev-vicari,astrakharchik,klawunn,song}, showing that the relationship between entanglement and fluctuations is a fundamental, constitutive aspect of the phase of the system, as well as of the statistics obeyed by its constituents. A fundamental question that one can ask in this respect is then the following: can we relate the (leading term of the) scaling of entanglement to that of a particular moment of fluctuations, and hence to the decay of a particular correlation function?    
  
 \emph{Entanglement and fluctuation contours.} In order to answer this question, a very useful tool is offered by the concept of \emph{entanglement contour}, introduced recently by Chen and Vidal \cite{entanglement-contour} in the specific case of free fermion models. The entanglement contour provides a decomposition into single-site contributions for the entanglement entropy of a subsystem immersed in an extended lattice quantum system. The contour decays when moving away from the boundary between the subsystem $A$ and its complement $B$, exposing in this way the spatial structure of entanglement in the many-body system. Here we generalize the study of the entanglement contour to a broad class of quadratic Hamiltonians, including free fermion models (Fermi liquids, semimetals, band insulators, integer quantum Hall insulators), as well as to weakly interacting Bose gases treated within the Bogoliubov approximation, and to hardcore bosons/ quantum spin models treated within spin-wave theory. Moreover we naturally introduce the concept of \emph{fluctuation contour} for particle number fluctuations, related to the correlations of the particle number fluctuations between each site of the subsystem $A$ and its complement $B$. The fluctuation contour allows therefore for a first, direct comparison between the spatial structure of entanglement and that of { particle-number} fluctuations. 
 
 Our central results are two-fold: 1) In the case of quadratic fermion Hamiltonians, we find that entanglement and particle-number fluctuations have the same spatial structure, as the two contours are essentially proportional to each other in all models we considered; 2) In the case of quadratic bosonic models the two contours are completely disconnected. In the particular case of the weakly interacting Bose gas, the area law of entanglement is found to originate from a contour decaying exponentially at short distance, with a decay length related to the healing length of the condensate; and decaying algebraically at long distance, with a decay exponent larger than unity, and seemingly dependent on the interaction strength. For hardcore bosons only the algebraic decay survives -- in the absence of a finite healing length -- with an exponent close to two. This is in stark contrast with the fluctuation contour, which decays like the inverse of the distance to the boundary between $A$ and $B$ in all cases, leading to a logarithmic violation of the area law for particle-number fluctuations. 
  These results establish that the entanglement contour is essentially a measurable quantity for fermions thanks to its correspondence with the fluctuation contour; on the other hand, the question of an observable signature of entanglement entropy and contour for bosonic systems remains completely open.  
  
  
   The structure of the paper is as follows: Sec.~\ref{s.ent-fluctu} discusses the rigorous link between local particle-number fluctuations and entanglement;  Sec.~\ref{tour-d'horizon} reviews the main results concerning the scaling of entanglement entropy and fluctuations in the ground state of quantum many-body systems; Sec.~\ref{sec_contours} introduces the concept of entanglement and fluctuation contours for quadratic Hamiltonians; Sec.~\ref{section_fermions} and Sec.~\ref{section_bosons} discuss the results for the contour of fermions and bosons respectively; Sec.~\ref{sec-zero_mode} discusses the special role of the lowest mode in the entanglement spectrum for bosonic Hamiltonians; and conclusions are drawn in Sec.~\ref{s.conclusion}.


\section{Bipartite fluctuations imply entanglement}
\label{s.ent-fluctu}

 In the following we consider a generic, spatially extended quantum system in $d$ spatial dimensions, which we divide into two portions $A$ and $B$, and show that, in a globally particle-number conserving model, local particle number fluctuations \emph{imply} entanglement between $A$ and $B$. While this implication is intuitive (the exchange of particles between $A$ and $B$ being the physical mechanism that mediates entanglement), its simple rigorous proof is not present in the literature to our knowledge. 
  
 We assume that the Hamiltonian ${\cal H}_{AB}$ describing the whole system conserves the total number of particles $N=N_A+N_B$, and consider any one of its eigenstates $|\Psi\rangle$ (including the ground state). The Schmidt decomposition of $|\Psi\rangle$ with respect to a $AB$  bipartition reads :
	\bee
		\vert \Psi \rangle = \sum_\alpha \lambda_\alpha \vert \psi_\alpha \rangle_A  \otimes \vert \phi_\alpha \rangle_B
	\eee
	where the basis $(\vert \psi_\alpha \rangle_A)$ diagonalizes the reduced density matrix $\rho_A = {\rm Tr}_B(\rho_{AB})$, and $\rho_{AB} = |\Psi \rangle \langle \Psi |$.
Since $N$ is a good quantum number of ${\cal H}_{AB}$, any eigenstate of ${\cal H}_{AB}$ is a simultaneous eigenstate of $N$, whence $[\rho_{AB},N_A+N_B] = 0$. 
 This in turn implies that 
	\bee
	{\rm Tr}_B[\rho_{AB},N_A+N_B] =  [{\rm Tr}_B(\rho_{AB}),N_A] = 0~, 
	\eee
so that the states $\vert \psi_\alpha \rangle_A=|N_A^{(\alpha)}, \{k^{(\alpha)}\}\rangle_A$ can be chosen to be a basis of eigenstates of $N_A$ (where $\{k^{(\alpha)}\}$ are the other quantum numbers labeling the states)
	\bee
      \vert \Psi \rangle = \sum_\alpha \lambda_\alpha~ |N_A^{(\alpha)}, \{k^{(\alpha)}\}\rangle_A \otimes  |N-N_A^{(\alpha)}, \{k'^{(\alpha)}\}\rangle_B~.
	\eee
Hence for any eigenstate $|\Psi\rangle$ the existence of a finite uncertainty on the local particle number $N_A$ implies a Schmidt decomposition with at least two terms, and a finite entanglement entropy $S_A = - {\rm Tr} \rho_A \ln \rho_A = - \sum_{\alpha} \lambda_\alpha \ln \lambda_\alpha$. This establishes particle-number fluctuations as a fundamental signature of entanglement. A similar reasoning applies to any model possessing a globally conserved quantity $O$, and exhibiting fluctuations of the corresponding local quantity $O_A$. An important example is that of quantum spin systems possessing an axial symmetry: fluctuations of the local magnetization along the symmetry axis imply entanglement.  { In what follows we shall focus on lattice-gas models, and on the variance of the particle-number distribution $\delta^2 N_A = \langle N_A^2 \rangle - \langle N_A \rangle^2$, a quantity we shall refer to simply as particle-number fluctuations.} 

\section{Scaling behavior of entanglement and fluctuations : a survey}
\label{tour-d'horizon}

In the following we specialize our attention to the ground state of the system, and to the situation in which $A$ and $B$ are connected, so that they admit only one connected, $(d-1)$-dimensional boundary. We indicate with $l$ the characteristic linear dimension of $A$, and with $L$ the linear dimension of the total $AB$ system. We shall consider only the case of translationally invariant Hamiltonians, leaving the case of disordered models for future discussion. 
As already mentioned in the introduction, a distinctive feature of different ground state phases is the scaling with $l$ of the entanglement entropy and of the local fluctuations of globally conserved quantities. In the following we shall shortly review known results, as well as original results from this work, concerning the \emph{leading term} of such scaling for both free and interacting models. We shall cite uniquely the case of lattice quantum gases, and of quantum spin models with a uniaxial symmetry. 

\emph{Gapless free fermions with a Fermi surface, Luttinger liquids.} Free fermions with a finite density of states $\rho(\varepsilon_F)$ at the Fermi energy $\varepsilon_F$ are known to obey a logarithmically violated area law for both entropy \cite{calabrese-cardy,gioev-klich,wolf} and fluctuations \cite{bipartite-fluct,calabrese-mintchev-vicari}:
				\be S_A \propto l^{d-1} \ln l ~~~~~;~~~~~~ \delta^2N_A \propto l^{d-1} \ln l~.
				\ee
Furthermore, entropy and fluctuations are related in the very same way as in a degenerate free-fermion gas in the grand-canonical ensemble \cite{calabrese-mintchev-vicari}, namely: 
				\be
				S_A = \frac{\pi^2}{3} \delta^2N_A + O(1)~.
				\label{prop_S_N}
				\ee
In the case of $d=1$, a logarithmic scaling of both entanglement entropy and fluctuations is a distinctive feature of all gapless critical phases -- described by Luttinger liquid theory --  and can be traced back to their conformal invariance \cite{calabrese-cardy}. Nonetheless the proportionality coefficient is not universal, as $S_A \approx c/3 \ln l$ where $c$ is the central charge \cite{calabrese-cardy}, while $\langle \delta^2 N_A \rangle \approx (K/\pi^2) \ln l$, with $K$ corresponding to the Luttinger parameter \cite{Songetal2010}.

\emph{Gapped free fermions}. If the Fermi energy falls inside a gap, free fermions obey a strict area law for both entropy and fluctuations \cite{gioev-klich,wolf,RMP-area-laws,storms-singh}: 
				\be S_A \propto l^{d-1}~~~~~;~~~~~~ \delta^2N_A \propto l^{d-1} ~.
				\ee
				This encompasses both (trivial) band insulators as well as topological insulators, such as integer quantum Hall states \cite{S-HE-entier} and quantum spin-Hall states. However, unlike in the previous case, the proportionality factor between the leading scaling terms of entropy and fluctuations is not known in general.
			
		\emph{Non-interacting bosons}. Free bosons with a unique ground state, forming a perfect Bose condensate, represent a very special case: fluctuations obey a volume law, and entanglement entropy scales like $\ln l$ \cite{ding-yang} : 
				\be S_A \propto \ln l ~~~~~;~~~~~~ \delta^2N_A \propto l^d~.
				\ee
Indeed for a pure condensate, the particle-number fluctuations within region $A$ are purely binomial, with a distribution $P(N_A) = {N\choose N_A} x_A^{N_A}(1-x_A)^{N-N_A} $ where $x_A = (l/L)^d$ is the weight of the condensate wave-function in region $A$ -- the associated variance is $\delta^2N_A = N x_A (1-x_A) \approx N_A$ for $x_A \ll 1$. The entanglement entropy $S_A$ is simply the entropy of the binomial distribution $P(N_A)$.
			
\emph{Condensed interacting bosons}. Interacting bosonic models with a Bose-condensed ground state in $d>1$, such as the weakly interacting Bose gas or hardcore bosons with a non-integer filling, display an entanglement area law, but a logarithmically violated area law for fluctuations   
			\be 
			S_A \propto l^{d-1}~~~~~;~~~~~~ \delta^2N_A \propto l^{d-1} \ln l
			\ee
In the case of the weakly interacting Bose gas, the scaling of fluctuations is indeed predicted by Bogoliubov theory \cite{astrakharchik,klawunn}; as it will be shown in Sec.~\ref{section_bosons}, the same theory predicts also the area law of entanglement. A similar behavior is found in the $S=1/2$ Heisenberg antiferromagnet on the square lattice both via quantum Monte Carlo \cite{Hastingsetal2010,Kallinetal2011,HumeniukR2012,song} and with spin-wave theory \cite{song, Luitzetal2015}; extension of this behavior to generic gapless hardcore bosons is discussed in Sec.~\ref{section_bosons}.

\emph{Gapped bosonic models.} Models of interacting bosons with a gapped spectrum, such as the XXZ model with Ising-like anisotropy, or dimerized Heisenberg antiferromagnets, have exponentially decaying correlations functions, and therefore obey area laws for both entanglement and fluctuations \cite{Hastings2007,BrandaoH2015}, but general predictions do not exist for the proportionality factor relating the two. 

\bigskip

 The variety of behaviors exhibited by the above-cited models seems to escape a simple classification. While a distinction is easily done in the case of free fermions between critical vs. short-range-correlated systems, such a distinction does not hold for bosons. In particular, area laws of entanglement are observed both for short-range-correlated as well as for critical models, suggesting that their origin must be different. It is the purpose of the following sections to unveil the microscopic origin of entanglement area laws in different quadratic models via the use of contours \cite{entanglement-contour}, and to relate/contrast them to the behavior of fluctuations.

\section{Entanglement Hamiltonian and contours}
\label{sec_contours}

\subsection{Entanglement Hamiltonian from correlation functions}
	\label{H_ent}
		In this section, we describe the general procedure (inspired by \cite{blaizot}) to extract the entanglement properties from correlation functions of a general quadratic Hamiltonian, fermionic or bosonic. A simplified treatment when the correlations functions are real was evoked in Refs.~\cite{Peschel-Eisler, casini-huerta} and detailed in Ref.~\cite{botero-reznik} for bosons, but our approach is fully general and treats bosons and fermions on the same footing. Without loss of generality, we shall consider a $d$-dimensional hypercubic lattice with size $L^d$.  The Hamiltonian has the form
		\be
			{\cal H} = \sum_{ij} \left[ a_i^\dagger H_{ij} a_j + \frac{1}{2} \left( a_i^\dagger G_{ij} a_j^\dagger +  a_j G_{ij}^* a_i \right) \right ]~.
		\label{eq_H}
		\ee
where $H_{ij}$ and $G_{ij}$  are $L^d\times L^d$ matrices, and the field operators satisfy the commutation relations $a_i a_j^{\dagger} +{\epsilon} a_j^{\dagger} a_i = \delta_{ij}$, $a_i^{(\dagger)} a_j^{(\dagger)} +{\epsilon} a_j^{(\dagger)} a_i^{(\dagger)} = 0$, with $\epsilon = -1$ for bosons and $+1$ for fermions.  		
 
  Being semi-positive definite, the reduced density matrix $\rho_A$ can always be cast in the form $\rho_A = \exp(-{\cal H}_E)$ where ${\cal H}_E$ (the so-called entanglement Hamiltonian) is an Hermitian operator, and for quadratic models it is also a quadratic form. Indeed, if $\rho_{AB}=|\Psi\rangle \langle \Psi |$ with $|\Psi\rangle$ the ground state of a quadratic Hamiltonian, it is a Gaussian state and any correlation function factorizes according to the prescriptions of Wick's theorem. But as correlation functions concerning degrees of freedom in $A$ can also be calculated using $\rho_A$, applying Wick's theorem in reverse actually implies that $\rho_A$ is the exponential of a quadratic form. \footnote{Note that a pure condensate cannot be submitted to this treatment, but the Schmidt decomposition can be easily calculated anyway.}
		
 Without loss of generality, we can consider a subsystem $A$ with size $V_A = a l^d$, where $l$ is the characteristic linear dimension and the $a$ prefactor takes care of the aspect ratio of $A$. The entanglement Hamiltonian ${\cal H}_E$ can then be written in terms of two $V_A \times V_A$ matrices $\mathcal{A}$ and $\mathcal{B}$ as follows:
		\be
			{\cal H}_E = \sum_{ij} \left[ a_i^\dagger \mathcal{A}_{ij} a_j + \frac{1}{2} \left(  a_i^\dagger \mathcal{B}_{ij} a_j^\dagger + a_j \mathcal{B}_{ij}^* a_i \right )\right ]
		\label{eq_H_quadra}
		\ee
	where $\mathcal{A}$ is Hermitian. As shown in Ref.~\onlinecite{blaizot}, and as further discussed in Appendix \ref{section-wick}, there exists an invertible matrix $U$ such that: 
		\be
		\eta \begin{pmatrix} \mathcal{A} & \mathcal{B}\\  \mathcal{B}^*& \mathcal{A}^* \end{pmatrix}  = 
		U \begin{pmatrix} \omega & 0 \\ 0 & - \omega \end{pmatrix} U^{-1}
		\ee
 where $\omega = {\rm diag}(\omega_1,\cdots,\omega_{V_A})>0$, $\eta = {\rm diag}(1_{V_A},\epsilon 1_{V_A})$, and $1_{V_A}$  is the $V_A\times V_A$ identity matrix. In fact the $U$ matrix satisfies the identity $\eta U^{-1} \eta = U^T$, namely it is unitary in the case of fermions and 
 ``$\eta$-unitary" in the case of bosons. The $\omega_\alpha$ ($\alpha = 1, ..., V_A$) eigenvalues form the so-called (single-mode) entanglement spectrum, and the associated eigenvectors (encoded in the columns of $U$) will be hereafter denoted as \emph{entanglement eigenmodes}. The matrix $U$ writes as 

 \be
		U = \begin{pmatrix} \{ {\bm u}_{\alpha} \} & \{ {\bm v}^*_{\alpha} \}  \\   \{ {\bm v}_{\alpha} \} & \{ {\bm u}^*_{\alpha} \} \end{pmatrix}
		\ee
 where ${\bm u}_{\alpha} = \left ( u_{\alpha}(i_1), ..., u_{\alpha}(i_{V_A}) \right)$ and ${\bm v}_{\alpha} = \left ( v_{\alpha}(i_1), ..., v_{\alpha}(i_{V_A}) \right )$ are the mode eigenfunctions, satisfying the normalization condition 
 \be
 \sum_i \left ( |u_{\alpha}(i)|^2 + \epsilon |v_{\alpha}(i)|^2 \right ) = 1~.
 \ee
 
 Both the single-mode spectrum and the corresponding eigenmodes can be obtained via the diagonalization of the generalized correlation matrix \cite{Peschel-Eisler}, composed by the one-body correlation matrices $C_{ij} = \langle a_i^\dagger a_j \rangle$ and $F_{ij} = \langle a_i a_j \rangle$, and sharing the same eigenmodes with the entanglement Hamiltonian: 
		\be
		 \begin{pmatrix} \epsilon - C^* &  F\\ - F^*& C \end{pmatrix} = U \begin{pmatrix} {\rm diag}(\epsilon- n_\alpha) & 0 & \\ 
								0 &~ {\rm diag}(n_\alpha)& \end{pmatrix} U^{-1}
		\label{relation_C_H}
		\ee
		where $n_\alpha = 1/(\exp{(\omega_\alpha)}+\epsilon)$ are the mode occupations. The above relation implies that the knowledge of the spectrum of a quadratic hamiltonian ${\cal H}_E$, and the knowledge of the one-body correlation functions obtained from it, are equivalent on the basis of the Wick's theorem. Obviously, the procedure works for any reduced density matrix of Gaussian form, which includes the case of finite temperatures $T$ as well. 		
Entanglement (thermal) entropy at $T=0$ ($T>0$) is then simply expressed as the thermal entropy associated with the populations $n_\alpha$ of the (fermionic or bosonic) eigenmodes:
\begin{equation}
S = \sum_\alpha S_\alpha = \sum_\alpha \left[ -\epsilon (1 - \epsilon n_\alpha) \ln(1-\epsilon n_\alpha) - n_\alpha \ln n_\alpha~\right]~.
\label{mode_entanglement}
\end{equation}

	
\subsection{Entanglement contour from the local density of states of the entanglement Hamiltonian}

\subsubsection{Entanglement contour: generalities}

	The concept of \emph{entanglement contour} was proposed recently \cite{entanglement-contour} as a way to decompose the entanglement entropy of region $A$ into contributions coming from the individual degrees of freedom contained in that region. The entanglement contour $\mathcal{C}_s(i)$ is a lattice function defined on $A$ such that
	\be S_A = \sum_{i\in A} \mathcal{C}_s(i)~, 
	\label{sum_rule_cont_s}
	\ee
so that $\mathcal{C}_s(i)\ge0$ represents the contribution of site $x$ to entanglement entropy between $A$ and the remainder $B$. This quantity offers a deep insight into the origin of different entanglement scalings, and the possibility of clarifying how widely different models may show the same (dominant) entanglement scaling. Note that the contour accounts for the entanglement of site $i$ with $B$ in all its possible forms, namely as bipartite entanglement between $i$ and degrees of freedom of $B$, as well as multipartite entanglement involving $i$, further degrees of freedom of $A$ and degrees of freedom of $B$ -- yet the contour does not account for the entanglement involving exclusively degrees of freedom in $A$.

 In principle many functions $\mathcal{C}_s(i)$ can satisfy the sum rule in Eq.~\eqref{sum_rule_cont_s}, and one may conclude that the contour is not well defined. 
Ref.~\cite{entanglement-contour} states a list of plausible conditions which a well-defined contour should satisfy -- and which, aside from positivity, basically require that the contour has the same { spatial symmetries as those of the $A$ region}. While these conditions narrow down the possible functions $\mathcal{C}_s(i)$, they cannot identify one unique definition of $\mathcal{C}_s(i)$. Hence a specific definition requires a physically motivated, constructive approach, which can be taken in the case of the quadratic models of our interest. The modal decomposition of entanglement of Eq.~\eqref{mode_entanglement} suggests that, introducing a (mode-dependent) normalized weight function $w_\alpha(i)$ satisfying $\sum_{i\in A} w_\alpha(i) = 1$, one has
\begin{equation}
S = \sum_\alpha S_\alpha  = \sum_i \sum_\alpha w_\alpha(i) S_\alpha
\end{equation}
which leads to the contour 
\begin{equation}
\mathcal{C}_s(i) = \sum_\alpha w_\alpha(i) S_\alpha~.
\end{equation}  
The choice of $w_\alpha(i)$ satisfying the criteria of Ref.~\cite{entanglement-contour} is a priori arbitrary mathematically, but we argue that, from a physical point of view, only one choice is meaningful. 

\subsubsection{From the local density of states to the local entropy}

 It is physically compelling to require that $\mathcal{C}_s(i)$ has the same form as the local thermodynamic entropy at site $i$ in a system $A$ (without any complement $B$) coupled to a thermal reservoir -- as entanglement entropy and thermal entropy must be continuously connected when $T$ is varied from zero to a finite value. To obtain a general expression (valid both for bosons and fermions) for the local entropy of quadratic Hamiltonians such as ${\cal H}_E$, it is helpful to use Green's function theory, which allows to define a local density of states. 
 For both bosons and fermions, one can introduce the $T=0$ retarded Green's function \cite{Fetter-Walecka}:
 \begin{equation}
 G^R(i,j; t) = - i~\theta(t) ~ \langle \Psi_0 | [a_i(t),a_j^{\dagger}(0)]_{\epsilon} | \Psi_0 \rangle 
 \end{equation}
 where $[...]_{\epsilon}$ is the commutator for bosons and anticommutator for fermions and $|\Psi_0\rangle$ is the many-body ground state. From the temporal Fourier transform of the Green's function one can naturally define the \emph{local spectral weight}
 \begin{equation}
 W(i,\omega) = \frac{1}{\pi} {\rm Im} \left [ G^R(i,i;\omega) \right ]
 \end{equation}
 which takes the explicit, general form 
 \begin{eqnarray}
 W(i,\omega) &=& \sum_n  \Big ( |\langle \Psi_n | a_i^{\dagger} |\Psi_0 \rangle|^2~ \delta (\omega-\omega_{n0}) \nonumber \\
 && ~+ ~\epsilon |\langle \Psi_n | a_i |\Psi_0 \rangle|^2 ~\delta (\omega+\omega_{n0}) \Big )
 \label{e.Wi}
 \end{eqnarray}
 where $|\Psi_n\rangle$ are the many-body Hamiltonian eigenstates with energies $E_n$, and $\omega_{n0} = E_n - E_0$. From the local spectral weight one can then define a  \emph{(one-body) local density of states} (LDOS) at positive frequency $\omega>0$, given by  
 \begin{equation}
 \rho_i(\omega) =  W(i,\omega) + W(i,-\omega) 
 \label{e.rhoi}
 \end{equation}
 with the meaning of a density of states for excitations ($\omega > 0$) induced by adding/subtracting a single particle from the ground state at site $i$. 
 
 We now restrict our attention to a system which admits a description in terms of free (quasi-)particles,  and which is diagonalized by $({\bm u}_{\alpha}, {\bm v}_{\alpha})$ modes, as it is the case of all the models of interest to this work. Under this circumstance the excited states $|\Psi_n\rangle$ created by adding/subtracting a particle from the ground state are readily obtained from the quasi-particle eigenmodes, and the LDOS takes the form
 \begin{equation}
 \rho_i(\omega) = \sum_{\alpha} \left( |u_{\alpha}(i)|^2 +\epsilon |v_{\alpha}(i)|^2 \right) \delta(\omega - \omega_{\alpha})~.
 \end{equation}
 This expression is well known for fermions -- as it provides the response of the system in scanning tunneling microscopy experiments \cite{TersoffH1985, Kreiseletal2014} -- and it readily generalizes to the case of bosons via the above treatment.  
 The total density of states for quasi-particle eigenmodes is then obtained by integrating the LDOS over the whole system, $\rho(\omega) = \sum_i \rho_i(\omega)$. 
 
 Given the general relationship between the entropy and the density of states in a quadratic (bosonic or fermionic) model:
\begin{eqnarray}
S = - \int d\omega ~\rho(\omega) && \big\{\epsilon (1 - \epsilon n(\omega)) \ln\left[1-\epsilon n(\omega)\right ] \nonumber \\ 
&& +n(\omega) \ln n(\omega) ~\big \} 
\end{eqnarray}
it follows immediately that the local entropy is obtained by replacing $\rho$ with $\rho_i$ in the previous equation. 

We then require the entanglement contour to be equal to the local thermodynamic entropy of a fictitious system described by the quadratic entanglement Hamiltionan ${\cal H}_E$ at finite $T$. Hence it takes the form 
\begin{equation}
{\cal C}_s(i) = \sum_{\alpha} \left( |u_{\alpha}(i)|^2 +\epsilon |v_{\alpha}(i)|^2 \right) S_{\alpha}~,
\label{definition_Cs}
\end{equation}
namely $w_{\alpha}(i) = |u_{\alpha}(i)|^2 +\epsilon |v_{\alpha}(i)|^2 $.  The above expression generalizes the definition of entanglement contour of Ref.~\cite{entanglement-contour} (restricted to fermions) to the case of generic quadratic Hamiltonians for fermions and bosons alike. In all the examples considered below we find that the contour is positive definite -- although this is not obvious in the case of bosons, since $\epsilon =-1$.



\subsection{Fluctuation contour}		
		
	As one of our main goals is to compare the spatial structure of entanglement with that of fluctuations, we introduce the \emph{contour of particle-number fluctuations} ${\cal C}_n(i)$ (hereafter called simply the \emph{fluctuation contour}) via the obvious decomposition of the particle-number fluctuations in $A$ into local contributions : 
	\be \mathcal{C}_n(i) = \langle \delta n_i \delta N_A \rangle \ee
	such that $\delta^2N_A = \sum_{i \in A} \mathcal{C}_n(i)$. $\mathcal{C}_n$ has the same spatial symmetries as those of the $A$ region, and can have positive or negative sign (like any correlation function). At $T=0$ (or at finite $T$ in the canonical ensemble) $\delta N_A = - \delta N_B$, so that $ \mathcal{C}_n(i) = - \langle \delta n_i \delta N_B \rangle $, showing that  $\mathcal{C}_n(i)$ represents the correlation function between the density fluctuations at site $i$ and those in the whole complement $B$. 
As we shall later see, in the case of fermions the comparison between the contours of entanglement and fluctuations further corroborates the expression of the entanglement contour; while for bosons it fundamentally justifies the difference between the scalings of entanglement and fluctuations.   

		
\begin{figure}
			\centering
			\includegraphics[width = \linewidth]{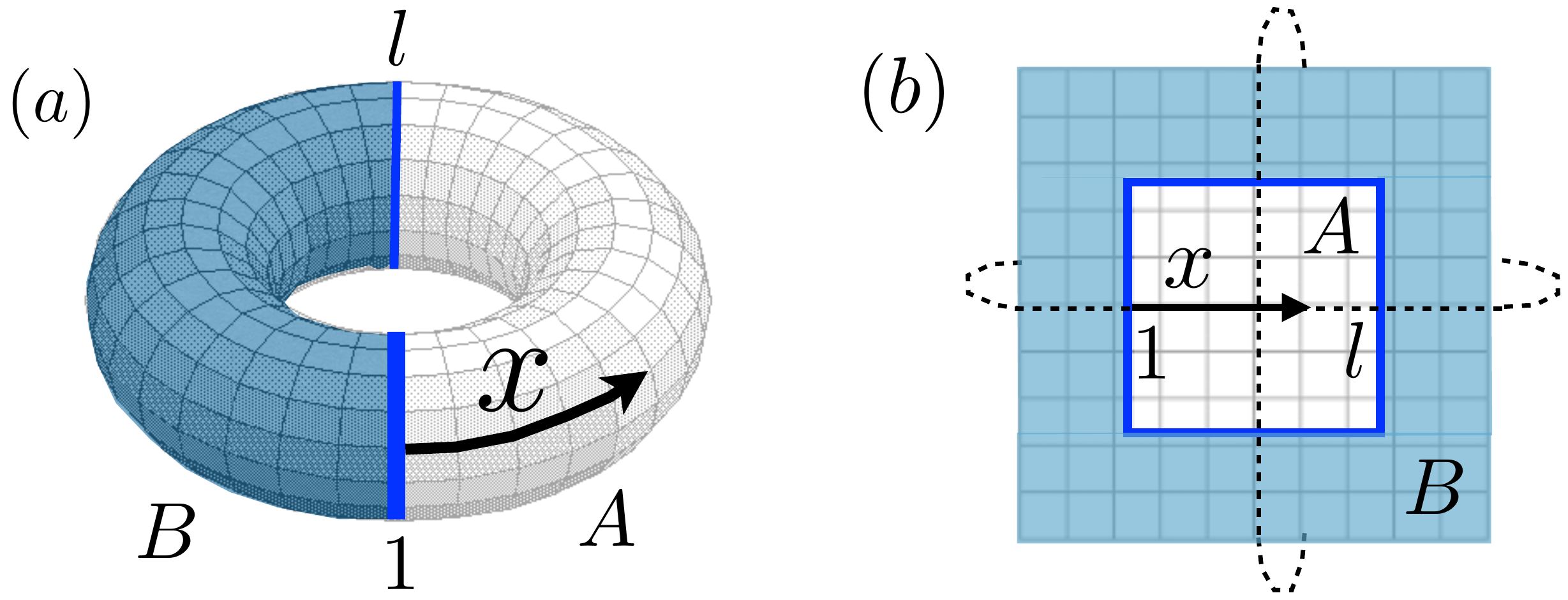}
			\caption{Geometries of $A$ subsystems used in the calculations. (a) $A$ is the $l^d$ region obtained by cutting in half a (hyper)torus of size $2l \times l^{d-1}$; (b) $A$ is a (hyper)cube of size $l^d$ carved from a (hyper)torus of much larger size $L^d$.}
			\label{cont_geom}
			\end{figure}		
		
\section{Entanglement and fluctuation contours for fermions}
\label{section_fermions}

  We begin our discussion by considering the case of quadratic fermionic Hamiltonians, whose entanglement contour has already been extensively studied in Ref.~\cite{entanglement-contour}. Our goal in this context is the comparison between the entanglement contour and the fluctuation contour. As we will see, this provides a microscopic real-space insight into the tight relationship between entanglement and fluctuations for fermions.  In this perspective we will restrict our attention to particle-number conserving Hamiltonians (namely we will consider ${\cal B}_{ij}=0$ in Eq.~\eqref{eq_H_quadra}), mainly because local particle-number fluctuations take unphysical contributions in the absence of total particle-number conservation. \footnote{In the case of bosons, we shall indeed consider  ${\cal B}_{ij} \neq 0$ as required by Bogoliubov theory, inducing spurious extensive particle-number fluctuations. Yet such fluctuations can be systematically removed in a controlled manner from the density-density correlation function.}
  
 For free fermions with particle-number conservation, the entanglement eigenmodes reduce to ordinary single-particle lattice wavefunctions ${\bm u}_{\alpha}$ (namely ${\bm v}_{\alpha} = 0$), and both entanglement and fluctuation contours have simple expressions in terms of such eigenmodes: 
		\bearr
		\mathcal{C}_s(i)& =& \sum_{\alpha} \vert u_\alpha(i)\vert^2 S_\alpha\nonumber \\
		\mathcal{C}_n(i) &= &\sum_{\alpha} \vert u_\alpha(i)\vert^2 n_\alpha(1-n_\alpha)~.
		\label{formule_cont_s_n_fermions}
		\eearr

In the following we inspect the relationship between entanglement and fluctuation contours for a wide class of translationally invariant, particle-number conserving models, realizing gapless metals and semimetals, as well as gapped trivial band insulators and topological Chern insulators. 
                 
		\subsection{Gapped and gapless free fermions}
			To investigate topologically trivial metals and insulators, we consider the model Hamiltonian defined on a $d$-dimensional hypercubic lattice:
			\be
			H = -t \sum_{\langle ij \rangle}  (c_i^\dagger c_j + {\rm h. c.}) + \sum_i (-1)^i \Delta c_i^\dagger c_i - \mu \sum_i c_i^\dagger c_i
			\label{H_FL}
			\ee
 whose spectrum, for a filling $0 < n < 1$, is gapless when $n \neq 1/2$ at any value of $\Delta$; and gapped for $n=1/2$ when $\Delta \neq 0$, with a gap given by $2 \Delta$ 
  As already mentioned in Sec.~\ref{tour-d'horizon}, entanglement and fluctuations obey an area law in the presence of a gap (implying exponentially decaying correlations); while, if $\rho(\varepsilon_F)\neq 0$,  the area law for both quantities is violated by a logarithmic multiplicative factor, and their leading scaling behavior is identical up to a proportionality factor, as in Eq.~\eqref{prop_S_N}. These scaling behaviors are easily understood at the level of the contours. Without loss of generality we can imagine that $A+B$ is a hypertorus of size $2l\times l^{d-1}$, and that the $A$ subsystem is half of the total system, having size $l^d$ (see Fig.~\ref{cont_geom}(a)). As the contours are constant along the boundary of $A$, they are uniquely characterized by their evolution as one moves orthogonally to the boundary along the $x$ direction. Indicating with ${\bm r} = (x,y \dots)$ the $d$ coordinates of site $i$, we shall then denote ${\cal C}_{n,s}(x) =: {\cal C}_{n,s}(x,l/2,\dots,l/2)$ the contours calculated along a trajectory orthogonal to the boundary; ${\cal C}_{n,s}(x)$ are periodic functions of period $l$. 
With this geometry, the scaling of entanglement entropy and particle-number fluctuations as a function of $l$ are then immediately obtained as
\begin{equation}
S_A = l^{d-1} \sum_{x=1}^{l} \mathcal{C}_s(x)	~~~~~~~~ \langle \delta^2 N_A \rangle = l^{d-1} \sum_{x=1}^{l} \mathcal{C}_n(x).		
\label{integrals}	
\end{equation}
An area law will be therefore respected or violated depending on the convergent/divergent nature of the integrals of the $\mathcal{C}_{s,n}(x)$ functions. 
A similar conclusion holds also for different geometries of the subsystem $A$, namely in the case in which $A$ is a $l^d$ (hyper)cube carved out of a (hyper)torus, as shown in Fig.~\ref{cont_geom}(b); in this case one can characterize the behavior of the contours from their variation as one moves along one of the axes of symmetry of the cube (as implied by the definition of ${\cal C}_{n,s}(x)$). 
 
 \begin{figure}
			\centering
			\includegraphics[width = \linewidth]{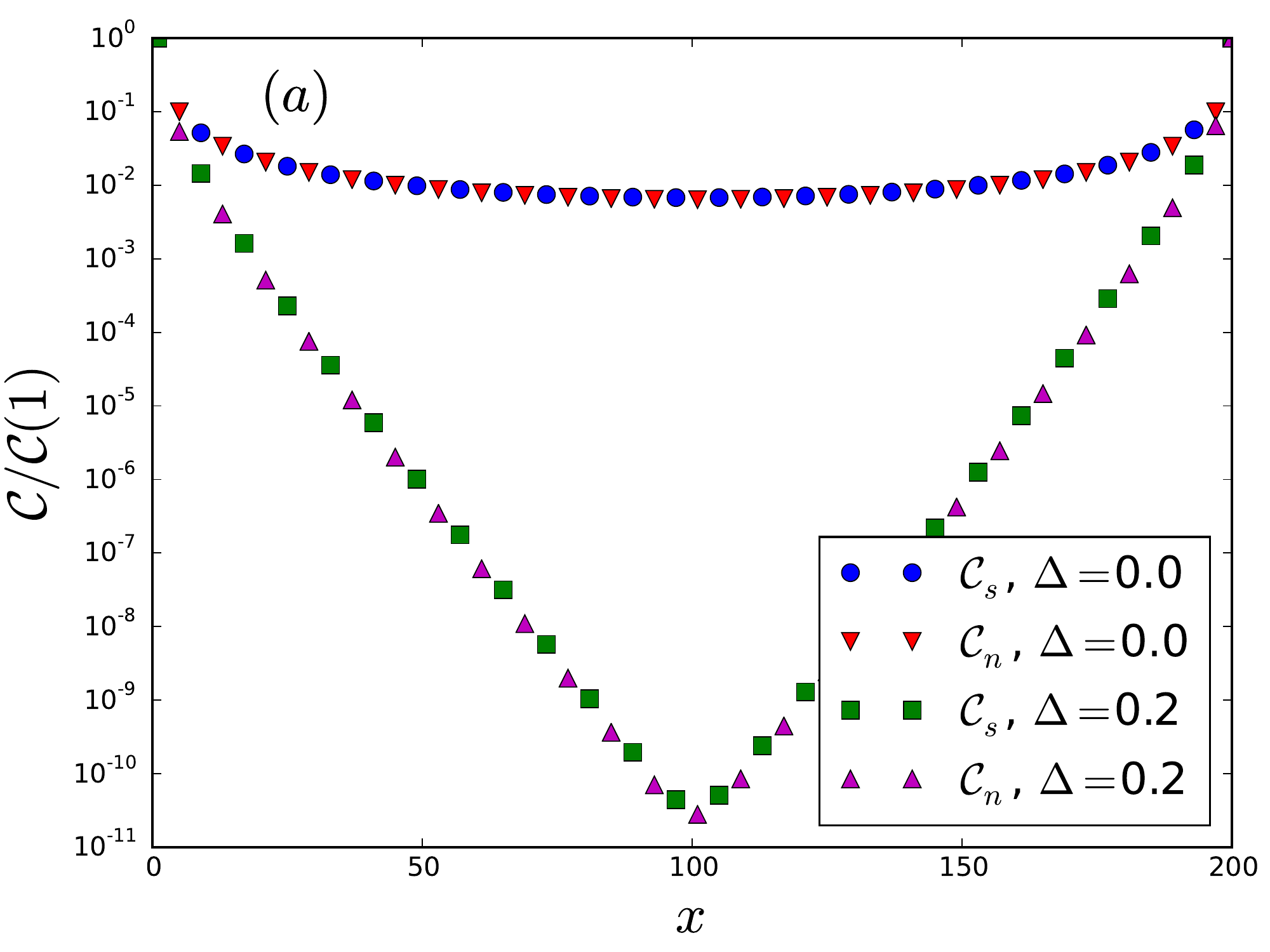}
			\includegraphics[width = \linewidth]{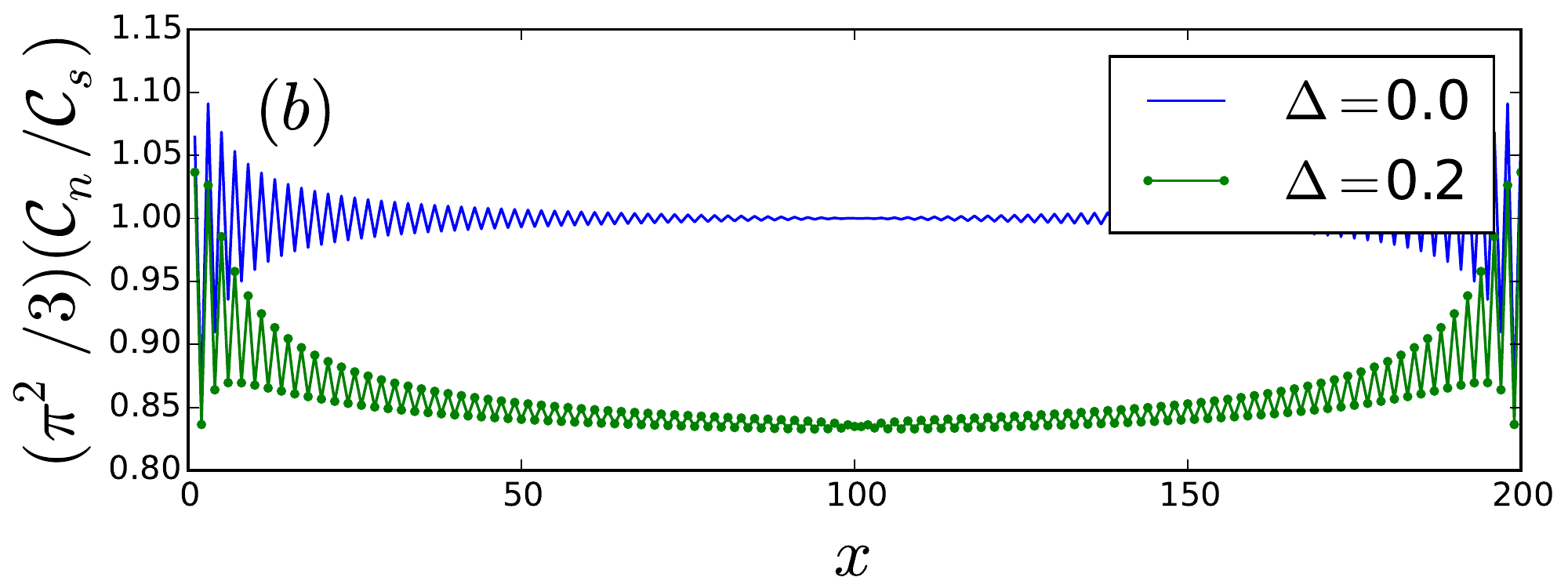}
			\caption{$(a)$ Fluctuation contour and entanglement contour for gapped and gapless free fermions described by Hamiltonian (\ref{H_FL}). $A$ is a line of $200$ sites in a circle of $400$ sites ($d$=1), and $\mu=0$ (half-filling). Contours are normalized by their value at the boundary $\mathcal{C}(1)$. $(b)$ ratio $(\pi^2/3) \times \mathcal{C}_n / \mathcal{C}_s$.}
			\label{cont_s_n_FL}
			\end{figure}
 
 Our findings for free fermions are summarized in Fig.~\ref{cont_s_n_FL}, where we observe that, sufficiently far from the boundary
\bearr
			\mathcal{C}_n, \mathcal{C}_s \sim 1/x
		\label{1_x}	
		\eearr
for gapless fermions, whereas		
		\bearr
			\mathcal{C}_n,  \mathcal{C}_s  \sim e^{-x/\xi}
		\label{exp_x}
			\eearr
for gapped fermions. Actually the most appropriate functional form is that of a symmetric function around $x=l/2$, namely $x^{-1} + (l+1-x)^{-1}$ and $e^{-x/\xi} + e^{-(l+1-x)/\xi}$, respectively. 
 In the case of gapped fermions, the correlation length is proportional to the inverse of the gap, $\xi\approx c/\Delta$.
 
In both cases, the decay of the fluctuation contour is readily obtained from the known decay of density-density correlations; indeed one has that 
\be
			\mathcal{C}_n(i) = -\langle \delta n_i \delta N_B \rangle = - \sum_{j \in B} \langle \delta n_i \delta n_{j} \rangle~.
			\label{intuitive_decay_cont_n}
			\ee
Starting from the toroidal geometry of Fig.~\ref{cont_geom}(a), we may further simplify the calculation by assuming that the system is in fact a semi-infinite cylinder (whose radius tends to infinity), where the $A$ subsystem corresponds to the coordinates
 $1 \leq x \leq l$ along the cylinder, and $B$ corresponds to $-\infty \leq x \leq 0$. Hence
\be
			\mathcal{C}_n(i) = - \left ( \sum_{x_j = -\infty}^{0} \sum_{y_j = -\infty}^{\infty} ... \right ) ~~ \langle \delta n_i \delta n_{j} \rangle
			\label{intuitive_decay_cont_n_2}
			\ee			

The asymptotic behavior of the density-density correlation function for free fermions with a finite Fermi surface is well known \cite{Wenbook}, and it decreases as 
$\langle \delta n_i \delta n_{j} \rangle \sim - |{\bm r}_i - {\bm r}_j|^{-(d+1)}$ (the case $d=1$ is discussed explicitly in the Appendix \ref{dens-dens-fermions}). The $d$ integrations implied by Eq.~\eqref{intuitive_decay_cont_n_2} eliminate the dependence on all coordinates of site $i$ but $x_i$, and leave out a $1/x_i$ dependence for ${\cal C}_n(i)$. On the other hand, for exponentially decaying correlations the integration over the $B$ region clearly leads to an exponentially decaying contour.

 As a consequence, putting together Eqs.~\eqref{integrals}, \eqref{1_x} and \eqref{exp_x} one readily obtains a logarithmic violation of the area law in the gapless case, and an area law in the gapped one. In the case of entanglement entropy and contour, our observations simply confirm the results of Ref.~\cite{entanglement-contour}. Our original finding consists in the fact that, in the case of gapless fermions, the fluctuation contour appears to be \emph{proportional} to the entanglement contour, with the same proportionality factor $\pi^2/3$ relating the dominant scaling behavior of the entanglement entropy and particle-number fluctuations, as in Eq.~\eqref{prop_S_N}. A closer inspection in the contours shows that exact proportionality holds only when contours are calculated deep in the bulk of $A$, while closer to the boundary the ratio $\mathcal{C}_s/\mathcal{C}_n$ shows Friedel-like oscillations at wavevector $2 k_F = 2\pi n$.  \footnote{Friedel oscillations at $2k_F$ are expected in the entanglement contour as its integral (the block entanglement entropy) possesses such oscillations, as proven rigorously for gapless fermions in $d=1$ in Ref.~\cite{FagottiC2011}; similar oscillations are also expected in the fluctuation contour as integral of the (Friedel oscillating) density-density correlations (see Appendix \ref{dens-dens-fermions}). The oscillations of the two contours have an amplitude ratio differing from that of the dominant non-oscillating part, and therefore it shows up in Fig.~\ref{cont_s_n_FL}.}
 
This result corroborates the choice of the expression of the entanglement contour for free fermions from a physical point of view. Indeed the entanglement contour  (expressing the contribution of a site in $A$ to the entanglement between $A$ and $B$) bears the same relationship to the fluctuation contour (expressing the contribution of the site to the correlations in particle-number fluctuations between $A$ and $B$ ) as that relating the global entanglement entropy of $A$ with the global particle-number fluctuations in $A$. 
Turning this argument on its head, one may conclude that the entanglement contour being proportional to the fluctuation contour, the logarithmic violation of the area law for gapless fermions can be viewed as a consequence of the peculiar power-law scaling of density-density correlations, namely one can \emph{infer entanglement scaling from the behavior of correlation functions}. A similar reasoning also applies to the gapped case. Indeed Fig.~\ref{cont_s_n_FL} shows that an approximate proportionality holds also in the case of gapped fermions, although the proportionality factor appears to be non-universal and to depend on the gap. As already mentioned in Sec.~\ref{tour-d'horizon}, the $\pi^2/3$ prefactor, relating ${\cal C}_s$ and ${\cal C}_n$ in the gapless case, is the same as the prefactor relating $S$ and $\delta^2 N$ in a gas of free gapless fermions at low temperature: this coincidence may suggest the existence of local thermodynamic relations linking the contours, analog to the relations valid for the thermodynamics of the bulk system. This idea resonates with recent works \cite{Wong2013,Swingle2013}, where \emph{e.g.} the reduced density matrix of conformally invariant quantum field theories has been shown to admit the form of the exponential of the microscopic energy density, modulated in strength by an ``entanglement temperature" varying spatially like $1/x$. For $d=1$, in which the conformal invariance applies to free fermions, the spatial dependence of the contours we observe is recovered by positing that such contours depend on the entanglement temperature in the same way as the thermal entropy and fluctuations depend on temperature. Further elaborations on the link between contours and the notion of local entanglement thermodynamics go beyond the scope of the present paper, and will be the subject of future work.

\begin{figure}
			\centering
			{\includegraphics[width=0.5\linewidth]{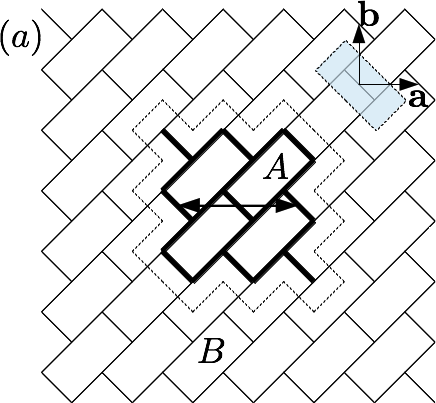}}
			{\includegraphics[width=\linewidth]{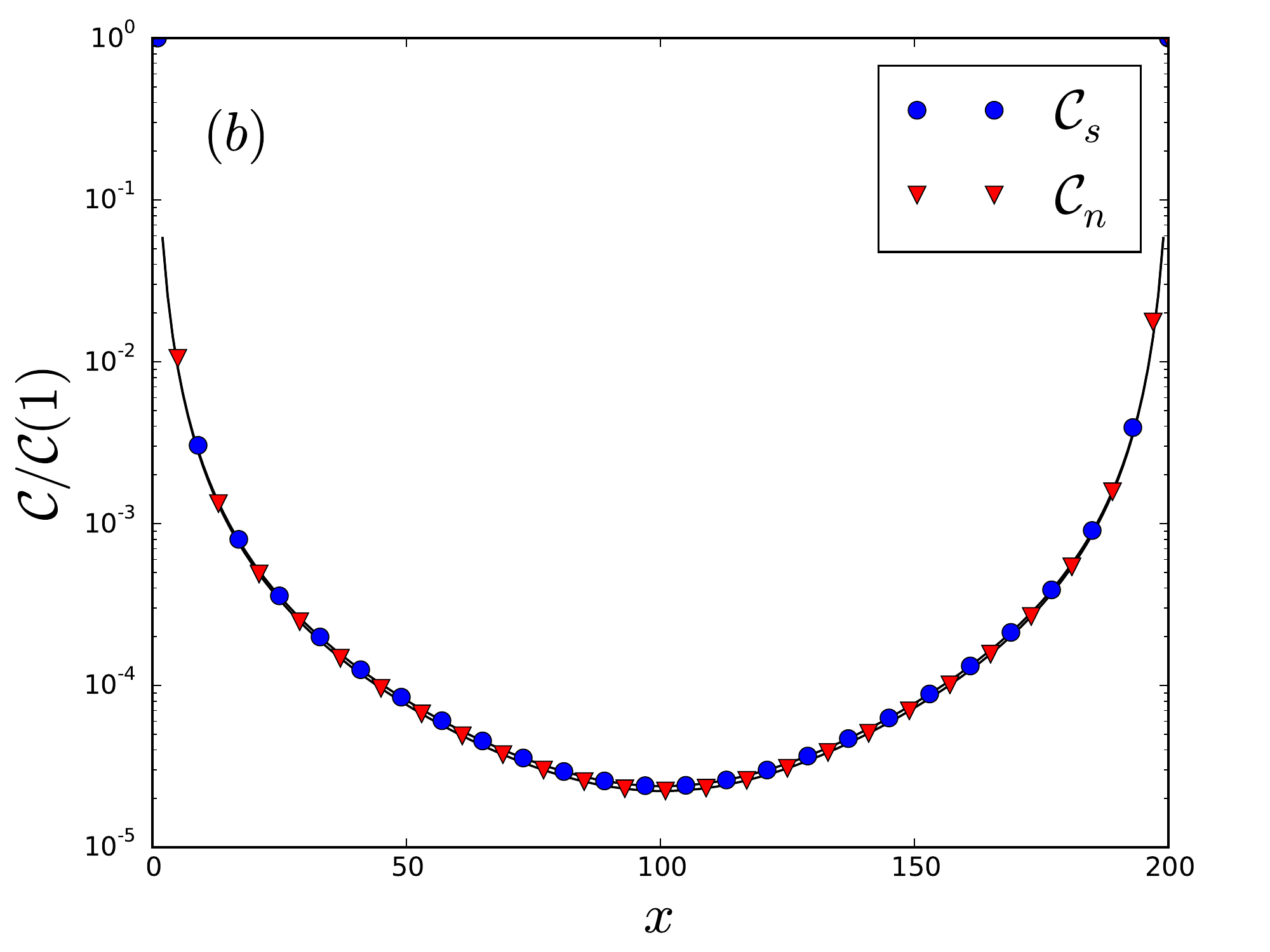}}
			{\includegraphics[width=\linewidth]{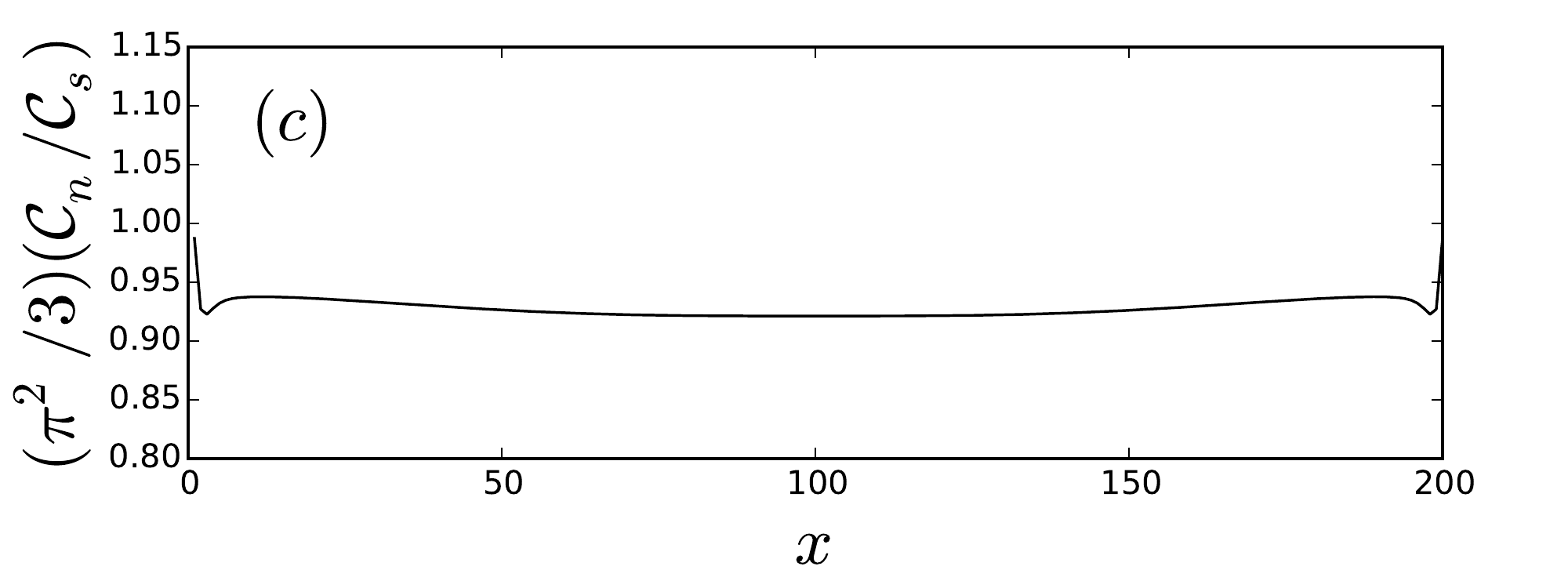}}
			\caption{$(a)$ Brick-wall lattice. The lattice is rotated by $\pi/4$ to align the directions of the toroidal simulation box with the natural axes of the square Brillouin zone of the system. The unit cell, containing two sites, is represented, as well as a typical region $A$ of $3\times 3$ unit cells. The double black arrow indicates the trajectory along which the contour are calculated. $(b)$ Fluctuation contour and entanglement contour for the brick-wall Hamiltonian at half-filling. $A$ is a cylinder of $200 \times 200$ unit cells in a torus of $200\times 400$ unit cells. Solid lines show fits of the form $\mathcal{C}_n = a[1/x^{\alpha_n} + 1/(l+1-x)^{\alpha_n}] + b$ and $\mathcal{C}_s = a'[1/x^{\alpha_s} + 1/(l+1-x)^{\alpha_s}] + b'$, with $\alpha_n \approx \alpha_s \approx 2$. $(c)$ Ratio $(\pi^2/3) \times \mathcal{C}_n / \mathcal{C}_s$ as a function of $x$.}
			\label{cont_s_n_mdb}
			\end{figure}

		\subsection{Semi-metals}
		\label{semi-metal}
		
	In the previous section we have proposed a quantitative link between the scaling of the entanglement entropy and the behavior of the density correlation function in free fermionic models via the use of contours.  In particular we have discussed well-known models of fermions having a finite density of states at the Fermi energy, or displaying a gap in the spectrum. We shall now test it in a less conventional situation, namely the case of semimetals having a gapless spectrum but a vanishing density of states. For this scope, we consider the minimal model of graphene, namely the tight-binding model on a honeycomb lattice at half filling - which, for the convenience of the calculation, is deformed into a brick-wall lattice (Fig.~\ref{cont_s_n_mdb}(a)). 
		
	The fundamental result of Ref.~\cite{gioev-klich} and \cite{wolf} relates the prefactor of a logarithmically violated area law of entanglement and particle-number fluctuations to an integral over the Fermi surface, which vanishes if such a surface has dimensions $(d-2)$ or lower (as in semimetals). Hence in a semimetal the area law is expected to dominate the scaling of entanglement and fluctuations, as it does in insulators -- in the case of entanglement, the area-law scaling has indeed been numerically verified in Ref.~\cite{You2014}. Nonetheless the model is gapless and critical, and it exhibits power-law correlations, as in the case of metals. Therefore for an area law to be respected, the decay of correlations and of contours must be sufficiently fast. Fig.~\ref{cont_s_n_mdb}(b) shows the entanglement and fluctuation contours for the fermionic brick-wall lattice at half filling, remarkably showing that also in the case of semimetals the two contours appear to be exactly proportional in the asymptotic limit of large distance from the boundaries, and that the proportionality factor is again close to (but systematically lower than) $\pi^2/3$. Doping the system away from half filling, we fall back to the case of a properly defined metal as described in the previous section, for which one expects the $\pi^2/3$ factor to be exactly verified; hence half filling appears to be a singular case in this respect.     

The decay of contours has an algebraic form, which we fit numerically finding $1/x^{\alpha}$ with $\alpha\approx 2$; in the case of the fluctuation contour this is expected from a density-density correlation function decaying as $1/x^4$, which we verified numerically to be the case.  
		
		\begin{figure}
			\centering
			\includegraphics[width = \linewidth]{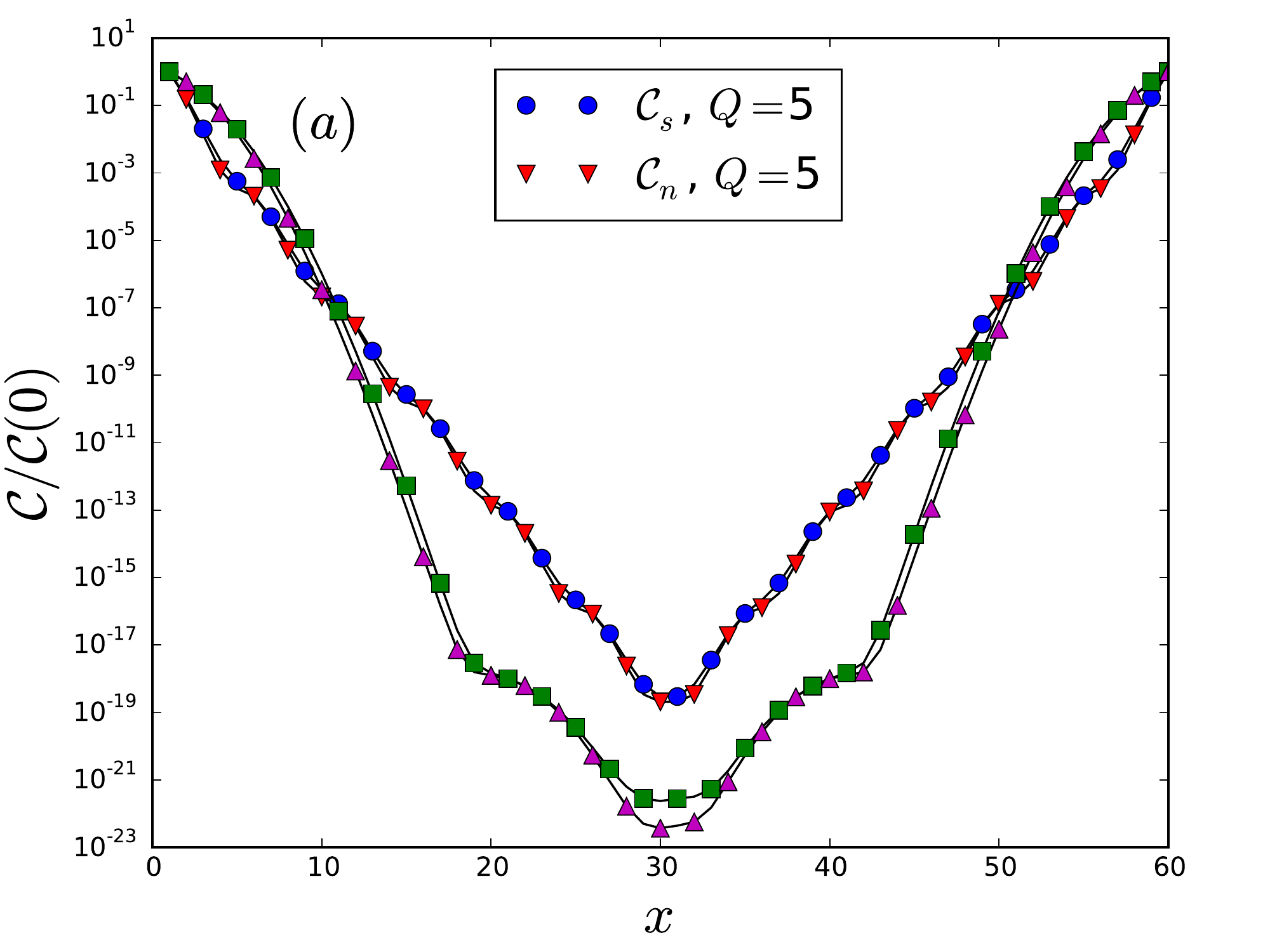}
			\includegraphics[width = \linewidth]{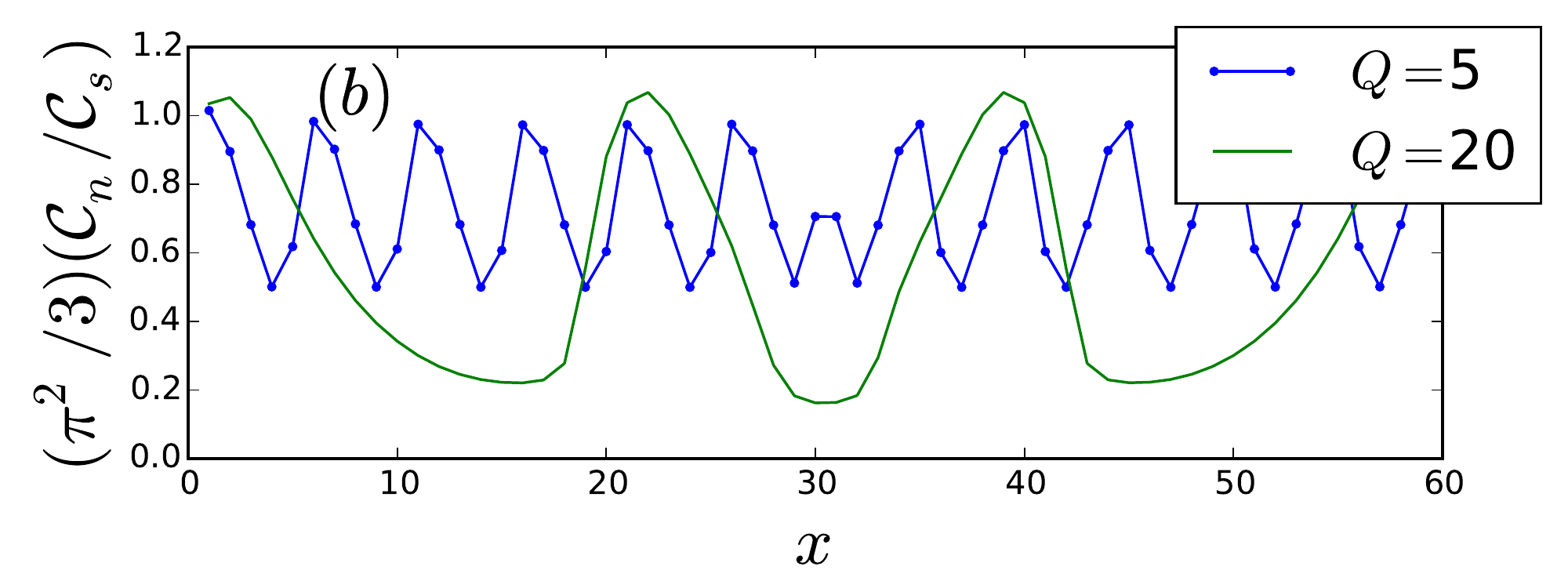}
			\caption{$(a)$ Fluctuation and entanglement contours for a quantum Hall state at $\nu=1$. $A$ is a cylinder of $60\time 60$ sites in a torus of $60\times 120$ sites. Solid lines are guides for the eye. $(b)$ Ratio $(\pi^2/3) \times \mathcal{C}_n / \mathcal{C}_s$ as a function of $x$.}
			\label{cont_s_n_HE}
			\end{figure}	
		
		\subsection{Chern insulators}	
		
		\label{chern-insulator}
		
		To conclude this section on fermions we consider the simplest model of a topologically non-trivial insulator, namely an integer quantum Hall state on a square lattice at filling (=number of flux quanta per particle) given by $\nu=1$. The magnetic field has a flux $\phi$ per unit cell which is an integer fraction of a flux quantum, namely $\phi=\phi_0/Q$ with $\phi_0 =h/e$ the flux quantum and $Q$ an integer. Within the Landau gauge, the magnetic field is encoded in the complex hoppings in the $y$ direction, with a $x$-dependent phase shift : 
		\be
		H = -t \sum_{j} (c_{j+\hat{x}}^\dagger c_{j} + e^{2i\pi x/Q} c_{j+\hat{y}}^\dagger c_{j} + {\rm h.c.}) -  \mu \sum_{j} c_{j}^\dagger c_{j}~.
		\label{H_HE}
		\ee
		The original cosine band of the square lattice is split into $Q$ sub-bands, and we choose a chemical potential $\mu$ so as to fill the lowest subband, possessing a nonzero Chern number. The gapped spectrum guarantees the area laws to be strict for both entanglement and fluctuations -- contrary to what is found in the case of the fractional quantum Hall effect \cite{levin-wen,kitaev-preskill}, no universal subleading corrections are expected in the integer quantum Hall case (as verified in Ref.~\cite{S-HE-entier}), being a non-interacting topological phase. Fig.~\ref{cont_s_n_HE} shows that the area laws originate from exponentially decaying contours for both entanglement and fluctuations, with superposed fluctuations induced by the gauge potential. Indeed the model in Eq.~\eqref{H_HE} has a $Q\times 1$ unit cell due to the gauge choice; the contours, being gauge-invariant, display a $Q\times Q$ unit cell which restores the symmetry between the $x$ and $y$ axis. We observe again that the contours are nearly proportional to each other, modulo the $Q$-periodic fluctuations, and in particular that the proportionality factor is very close to $\pi^2/3$ close to the boundary. 
			
	\subsection{Discussion}
			
	We have observed that the fundamental proportionality exhibited by entanglement and particle-number fluctuations of free fermions for widely different phases (metallic, semi-metallic, trivial band insulating and topological band insulating) originates at the microscopic level from a close relationship between the entanglement and fluctuation contours, exhibiting essentially the same spatial dependence. For gapped phases both contours decay exponentially, and they do so with the same decay rate, given by the correlation length of density-density correlations. 
	
	For gapless phases the contours decays algebraically, and they appear to be strictly proportional in the asymptotic limit of large distance from the boundaries of the $A$ region:
 \be
 \mathcal{C}_s(x \to l/2) = \gamma ~\mathcal{C}_n(x \to l/2)  ~~~~ (l \gg1)
\label{e.prop}
 \ee
 where $\gamma=\pi^2/3$ in the metallic phase, and generically $\alpha \lesssim \pi^2/3$ for the other phases. Actually, we remark that the entanglement eigenmodes $\alpha$ which significantly contribute to the contours Eq.~\eqref{formule_cont_s_n_fermions} have an amplitude $|u_{\alpha}(l/2,l/2)|$ which is very weakly dependent on $\alpha$. For $x= l/2$, Eq.~\eqref{formule_cont_s_n_fermions} can thus be rewritten as: 
 \bearr
 {\cal C}_s(l/2) &\approx& |u(l/2,l/2)|^2 ~\sum_\alpha S_\alpha = |u(l/2,l/2)|^2 ~S_A~~~~~~~~~~~~\\
 {\cal C}_n(l/2) &\approx& |u(l/2,l/2)|^2 ~\sum_\alpha n_\alpha(1-n_\alpha) \nonumber \\
 &=& |u(l/2,l/2)|^2 ~\delta^2N_A
 \eearr
 Hence, the fact that ${\cal C}_s(l/2) = (\pi^2/3) {\cal C}_n(l/2)$ is implied by the proportionality $S_A = (\pi^2/3) \delta^2N_A$ -- this remains true also in the vicinity of $x=l/2$. However, the proportionality holds even away form $x=l/2$, namely even in regions the eigenmodes profile strongly depends on $\alpha$. 
 The proportionality in Eq.~\eqref{e.prop} implies that the leading scaling behavior of the entanglement entropy and fluctuations, given by the integral of the respective contours, is necessarily the same when such integrals are dominated by the tail -- this is indeed the case when such integrals diverge in the limit $l\to\infty$. Hence our result provides a real-space insight into the origin of the relation Eq.~\eqref{prop_S_N}. 
 
 On the other hand, the short-range behavior of the contours accounts for a strict area law and for further subleading scaling terms, which are expected to differ between entanglement and fluctuations as the proportionality does not hold among contours at short distance. Nonetheless our observations  show that contours, even though not strictly proportional, mimic each other very closely even at short distance. 
 
	The above conclusions may appear unsurprising in the case of gapped phases with a finite correlation length - as the decay of contours is expected to be exponential with the same decay rate. One could argue that it is also to be expected for metals exhibiting a logarithmically violated area law, as the logarithm naturally stems from the integral of a $1/x$ decay of both contours. Nonetheless in the latter case it is remarkable that all fluctuations at wavevectors which are multiples of $k_F$, characterizing correlation functions, nearly disappear in a similar manner in both contours, showing that the correspondence among contours goes well beyond the simple power-law decay.  Moreover in the case of semimetals the area law exhibited by entanglement and fluctuations could in principle stem from different power-law decays with a convergent integral, which is found not to be the case. These fundamental correspondences show that measuring density-density correlations of free fermions allows to obtain essentially all features of the entanglement contour.  As we shall see in the next section, this is definitely not the case for bosons. 		
			
	\section{Decrease of contours for bosons}
	\label{section_bosons}
	
	In this section we investigate the entanglement and fluctuation contours for lattice bosons featuring Bose-Einstein condensation in the widely different regimes of weak interactions, and of infinite (hardcore) interactions. In both cases appropriate theoretical descriptions reduce the Hamiltonian to a quadratic form, lending itself to an exact analysis of the entanglement Hamiltonian and in particular of its eigenmodes, as described in Sec.~\ref{sec_contours}. To the best of our knowledge, an analysis of the spatial structure of entanglement eigenmodes has not been performed before in the case of quadratic bosonic Hamiltonians (apart from the case $d=1$ ~\cite{botero-reznik}).   
Willing to describe Bose-condensed systems, we focus on \emph{two-dimensional} lattices (unless otherwise specified). Two spatial dimensions have the advantage of featuring proper Bose condensation in the ground state on the one hand; and, on the other hand, they allow to keep the numerical effort of diagonalization of the correlation matrix to a minimum.  	

The common starting point of all the following sections is the minimal model of lattice bosons 	with contact interactions (namely the Bose-Hubbard model) 
\begin{equation}
{\cal H} = - J \sum_{\langle ij \rangle} \left( b_i^{\dagger} b_j + {\rm h.c.} \right ) + \frac{g}{2} \sum_i b_i^{\dagger} b_i^{\dagger} b_i b_i~.
\label{BH}
\end{equation}
This interacting model will be then reduced to a quadratic form of the kind Eq.\eqref{eq_H_quadra} via Bogoliubov theory in the case of weakly interacting bosons, and via spin-wave theory in the case of hardcore bosons. Before discussing the models directly, we shall provide some general details of the diagonalization procedure of the quadratic Hamiltonian. 
	
		\subsection{Quadratic Hamiltonian and Bogoliubov diagonalization}
All the models we shall consider later on are defined on translationally invariant lattices, and hence are most conveniently expressed in Fourier space. Their quadratic Hamiltonian, having the form Eq.~\eqref{eq_H_quadra} in real space, reduces to the following form in $k$-space: 
		\be
		{\cal H} = \frac{1}{2} \sum_{{\bm k}} \begin{pmatrix} b_{{\bm k}}^\dagger &  b_{-{\bm k}} \end{pmatrix} 
				\begin{pmatrix} A_{{\bm k}} & B_{{\bm k}} \\ B_{{\bm k}} & A_{{\bm k}} \end{pmatrix} 
				\begin{pmatrix} b_{{\bm k}} \\  b_{-{\bm k}}^\dagger \end{pmatrix}~
		\label{H_quadra}
		\ee		
		where $A_{\bm k}$ and $B_{\bm k}$ are real coefficients.
		The above Hamiltonian can be diagonalized by a canonical Bogoliubov transformation  
		\be
		b_{\bm k} = u_{\bm k} \beta_{\bm k} - v_{\bm k} \beta_{-{\bm k}}^\dagger
		\ee
		where $\beta_{\bm k}, \beta^{\dagger}_{\bm k}$ are bosonic operators destroying/creating Bogoliubov quasiparticles.
		 
Requiring the above transformation to diagonalize ${\cal H}$ and to satisfy bosonic commutation relations for $\beta_{\bm k}$, $\beta_{\bm k}^\dagger$, leads to the following expressions for the $u_{\bm k}$ and $v_{\bm k}$ coefficients: 
		\bearr
		 u_{\bm k}  &=&\displaystyle{ \frac{1}{\sqrt{2}}\left( \frac{A_{\bm k}}{\sqrt{A_{\bm k}^2 - B_{\bm k}^2}} +1 \right)^{1/2}}\\
		v_{\bm k}  &=& \frac{A_{\bm k}}{\vert A_{\bm k} \vert} \frac{1}{\sqrt{2}} \left( \frac{A_{\bm k}}{\sqrt{A_{\bm k}^2 - B_{\bm k}^2}} -1\right)^{1/2}
		 \label{expressions_u_v}
		 \eearr

	This reduces ${\cal H}$ to the form ${\cal H} = \sum_{\bm k} E_{\bm k} \beta_{\bm k}^\dagger \beta_{\bm k} + {\rm const.}$, where  
		 \be
		  E_{\bm k} = \sqrt{A_{\bm k}^2-B_{\bm k}^2}~.
		 \ee
In order to calculate entanglement properties, we only need to know the regular and anomalous (one-body) correlation functions $\langle b_{{\bm r}}^\dagger b_{{\bm r}'} \rangle$ and $\langle b_{{\bm r}} b_{{\bm r}'} \rangle$. In terms of $A_{\bm k}$ and $B_{\bm k}$, they may be expressed as : 
		\be 
		\langle b_{{\bm r}}^\dagger b_{{\bm r}'} \rangle = -\frac{1}{2} \delta_{{\bm r},{\bm r}'} + f({\bm r} - {\bm r}') 
		~~~~~~~~~ \langle b_{{\bm r}} b_{{\bm r}'} \rangle = g({\bm r} - {\bm r}') 
		\ee
		where 
		\bearr
		f({\bm r}) & = & \frac{1}{2V} \sum_{{\bm k}} e^{i{\bm k} \cdot {\bm r}} \frac{A_{{\bm k}}}{\sqrt{A_{{\bm k}}^2-B_{{\bm k}}^2}} \nonumber \\
		g({\bm r}) & = & \frac{1}{2V} \sum_{{\bm k}} e^{i{\bm k} \cdot {\bm r}} \frac{-B_{{\bm k}}}{\sqrt{A_{{\bm k}}^2-B_{{\bm k}}^2}} 
		\label{def_fg}
		\eearr
		 
	To calculate fluctuation properties we need the two-body correlation function, whose calculation is slightly more elaborate. Its precise expression  will be discussed model by model.
		 
		\subsection{Weakly interacting Bosons}
	\subsubsection{Bogoliubov theory}	
	In the case of weak interactions, namely $g n \ll1$, a quantitatively accurate approach of the model in Eq.~\eqref{BH} is represented by the Bogoliubov approximation, which, in its simplest formulation, amounts to regarding the ground state of the system as described by a coherent state in the ${\bm k}=(0,0)$ mode, plus weak quantum fluctuations around it. This allows to replace the field operator $b_i$ by $\sqrt{n_0} + \delta b_i$, where $n_0\approx n$ is the condensate density and $\delta b_i$ incorporates the weak quantum fluctuations around the coherent-state limit. Expanding the interacting Hamiltonian up to quadratic order in the fluctuations, one obtains the well-known Bogoliubov Hamiltonian \cite{PitaevskiiStringari}:
\be
		\label{H-bogo}
		{\cal H}_{\rm Bogo} = E_0 + \sum_{{\bm k} \neq 0} \left[ (\epsilon_{{\bm k}} + g n) b_{{\bm k}}^\dagger b_{{\bm k}} + \frac{gn}{2} (b_{-{\bm k}} b_{{\bm k}} + b_{{\bm k}}^\dagger b_{-{\bm k}}^\dagger) \right]
		\ee	
where $\epsilon_{\bm k} = 4J-2J \left( \cos k_x + \cos k_y \right )$ is the lattice dispersion relation. 
Hamiltonian (\ref{H-bogo}) is of the form (\ref{H_quadra}) with $A_{{\bm k}}  = \epsilon_{{\bm k}}  + g n$ and $B_{{\bm k}}  = gn$.
				
	The Bogoliubov Hamiltonian of Eq.~\eqref{H-bogo} only describes the quantum dynamics of particles out of the condensate, while the condensate itself is treated as a classical field, and its contribution to entanglement and quantum fluctuations is neglected. Nonetheless, as discussed in Sec.~\ref{tour-d'horizon}  the entanglement entropy of a pure condensate scales as $\ln l$, so that in $d\ge2$ it only contributes a subleading term to the area law expected for the entanglement entropy of the weakly interacting Bose gas; this implies that, within this framework, we can only describe entanglement entropies up to additive logarithmic corrections. On the other hand, the extensive particle-number fluctuations exhibited by a pure condensate (and quoted in Sec.~\ref{tour-d'horizon}) are a ``pathology" of the ideal gas. In the presence of interactions,
the condensate still exhibits extensive fluctuations due to the exchange of particles with the non-condensed part, but the sum of condensed and non-condensed particles exhibits much weaker fluctuations, scaling sub-extensively with the size of $A$. This is predicted by Bogoliubov theory \cite{giorgini, astrakharchik,klawunn} and confirmed via quantum Monte Carlo \cite{song}\footnote{I. Fr\'erot and T. Roscilde, unpublished.}.

		\begin{figure}
	 	{\includegraphics[width = \linewidth]{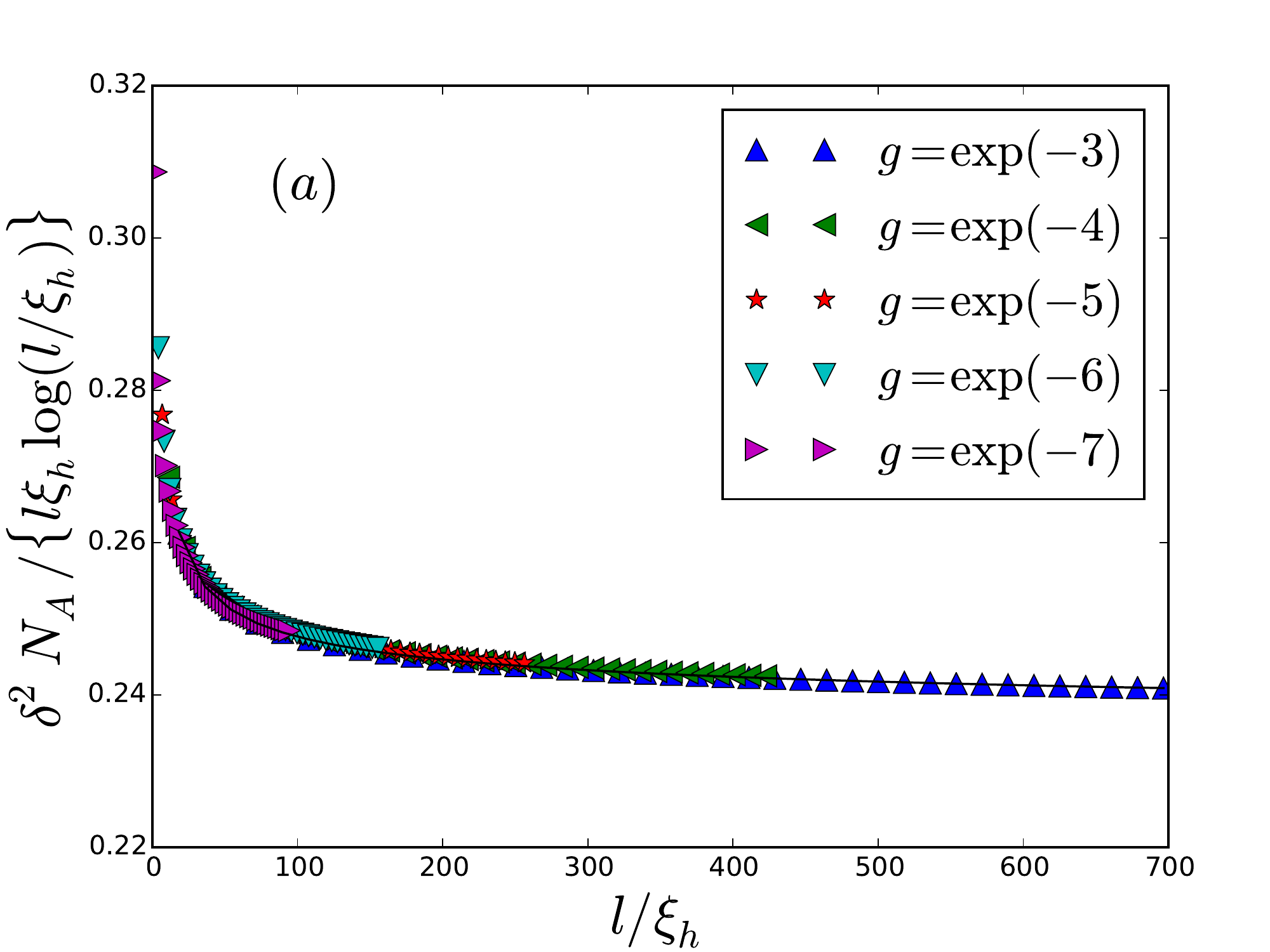}}
		{\includegraphics[width = \linewidth]{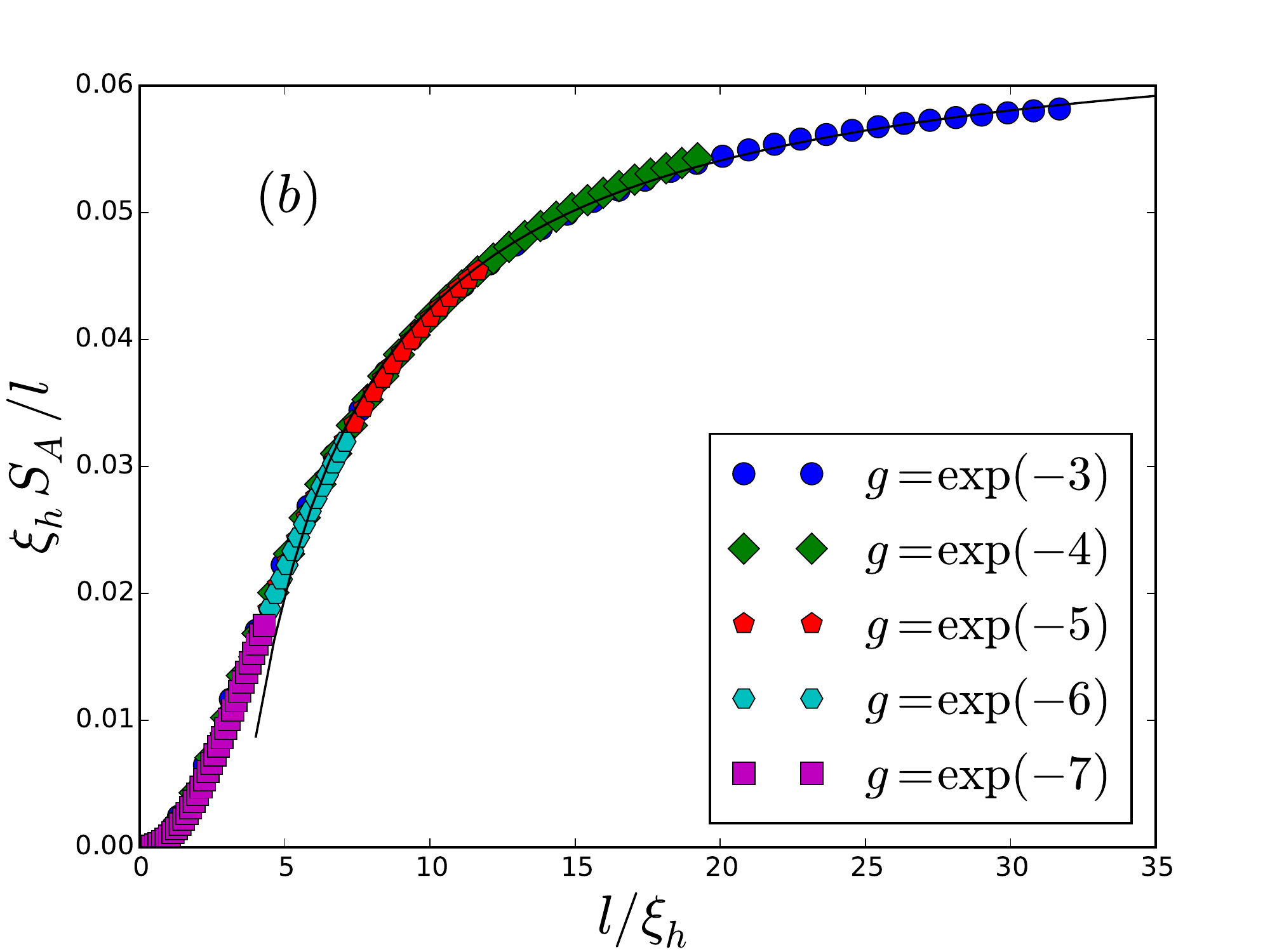}}
		\caption{Particle number fluctuations $(a)$ and entanglement entropy $(b)$ of the two-dimensional weakly interacting Bose gas in the Bogoliubov approximation on a $l/2\times l$ torus, for various values of $g$. For each $l$, $A$ is a cylinder of size $l/2\times l/2$, having a boundary of length $l$. Solid lines show fits of the form $\delta^2N_A/\xi_h^2 = a(l/\xi_h)\ln(l/\xi_h) + b(l/\xi_h) + c$, with $a\approx 0.24$ and $b\approx 0.1$ and $S_A =a'(l/\xi_h)+ b'\ln(l/\xi_h) + c'$, with $a'\approx 0.065$. This corresponds to the scaling forms of Eq.~\eqref{Fn} and Eq.~\eqref{Fs} for $d=2$.}
		\label{varN_S_bogo}
		\end{figure}
		
		\subsubsection{Scaling of fluctuations and entanglement entropy}
		
		From the Hamiltonian in Eq.~\eqref{H-bogo}, it is clear that entanglement entropy and contour depend on the interaction $g$ and density $n$ only through the product $gn$. 
		Moreover, as discussed in  Appendix \ref{dens-dens-correl}, the density-density correlation function is : 
		\be		
		\langle \delta n_i \delta n_j \rangle = \frac{n}{V} \sum_{{\bm k}\neq {\bm 0}} e^{i{\bm k} \cdot ({\bm r}_i-{\bm r}_j)} \frac{\epsilon_{{\bm k}}}{\sqrt{\epsilon_{{\bm k}}(\epsilon_{{\bm k}}+2gn)}}~.
		\label{dens_Bogo}
		\ee
	Hence the density-density correlation function has the form $\langle \delta n_i \delta n_j \rangle  = n f(gn)$, and the fluctuation contour inherits this property.
  	
Fluctuations of particle number in the two dimensional weakly interacting Bose gas were calculated via Bogoliubov theory in Ref.~\cite{klawunn}, where the following asymptotic behavior for a region $A$ with the geometry of a disk of radius $R$ was proved : 
		\be
		\delta^2N_A = nR\xi_h \ln \left(\lambda \frac{R}{\xi_h} \right)~.
		\label{BogodeltaN}
		\ee
Here $\xi_h$ is the healing length -- which on a lattice takes the form $\xi_h=\sqrt{J/(gn)}$ --  and $\lambda$ a numerical constant. Generalizing this result to the case of a square region $A$ of side $l$, and including subleading corrections, one may conjecture the following universal scaling form for the fluctuations
		\be
		\frac{\delta^2N_A}{\xi_h^2} = F_n \left( \frac{l}{\xi_h} \right) 
		\ee
		with the asymptotic behavior 
		\be
		F_n(x\to\infty) \approx a ~n ~x^{d-1}~ \ln x + b ~x^{d-1} + c + ...
		\label{Fn}
		\ee
		A similar guess can be made about the entanglement entropy, namely $S_A = F_S(l/\xi_h)$ with  
		\be 
	      F_S(x\to \infty) = a' ~x^{d-1}  + b' ~\ln x + c'  + ...
	      \label{Fs}
		\ee
		Here $a, a', b, b', c, c'$ are constants independent of (or weakly dependent on) $g$ and $n$.	 
		In the following, unless otherwise specified we fix the density to $n=1$, and we let the interaction $g$ vary over the range $e^{-14} J \dots e^{-1} J$.

The choice of the $F_n$ and $F_S$ functions is motivated by the result of Eq.~\eqref{BogodeltaN} on fluctuations (generalized to account for plausible subleading corrections, including an area law), and by the mounting evidence of entanglement area laws in gapless bosonic systems \cite{Hastingsetal2010, Kallinetal2011, HumeniukR2012, HelmesW2014, Stoudenmireetal2014, Luitzetal2015, Kulchytskyyetal2015}, with additive logarithmic corrections coming from Goldstone modes \cite{metlitski-grover} and corner contributions \cite{CasiniH2007,Stoudenmireetal2014}. 
Fig.~\ref{varN_S_bogo} shows that the above scaling Ans\"atze are well confirmed by our results, with the coefficients of the subleading terms ($b$, $c$, $b'$ and $c'$) depending weakly on $g$. Actually, the scaling forms in Eqs.~\eqref{Fn} and \eqref{Fs} are found to be valid only in the limit $gn/J \ll 1$ (which in any case is the limit of validity of Bogoliubov treatment), and deviations occur outside this regime. 
\footnote{In the case of the entanglement entropy we find $b'\approx -0.1\dots 0.3$ for $g/J \approx 10^{-3}\dots 10^{-1}$ -- but, as anticipated, additive logarithmic terms are not reproduced quantitatively at the level of our treatment, since Bogoliubov theory discards the entanglement coming from the condensate mode, which may also be expected to give an additive logarithmic contribution.}  	
		
	\begin{figure}
		{\includegraphics[width = \linewidth]{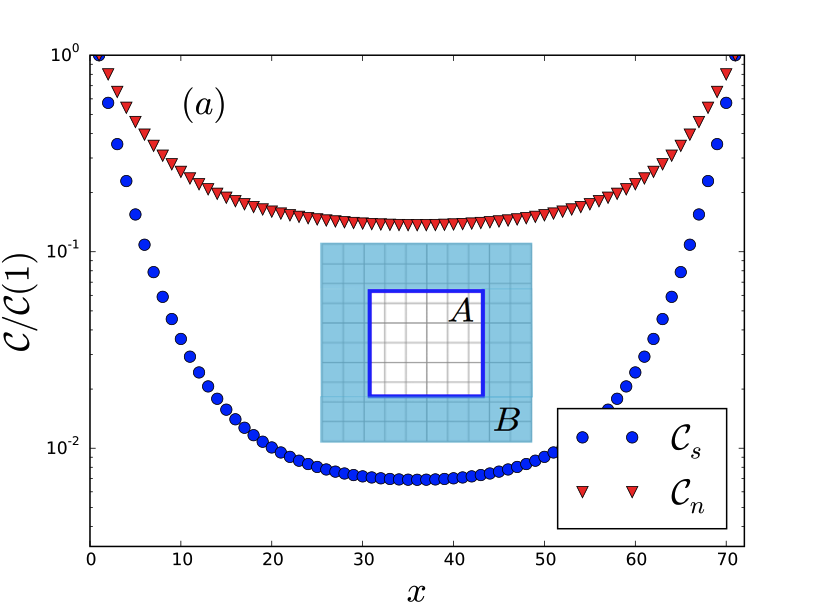}}
		{\includegraphics[width = \linewidth]{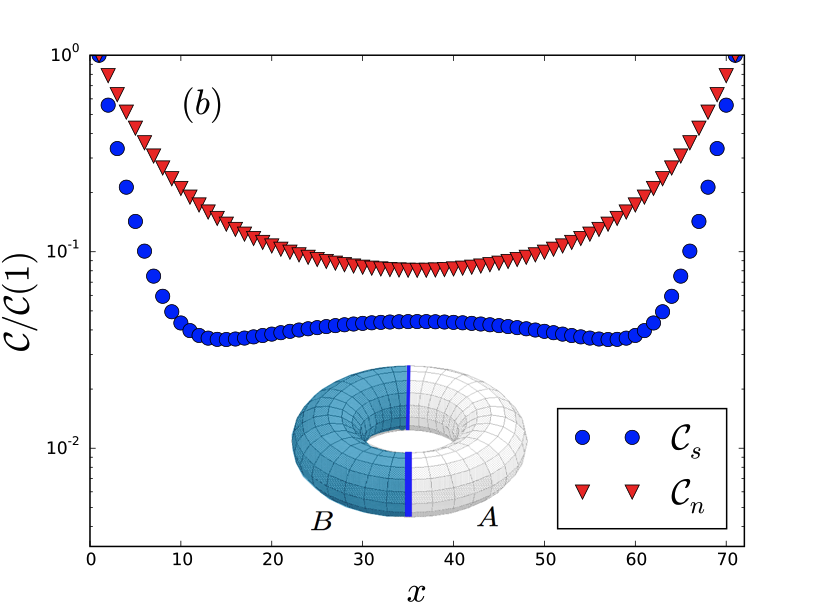}}
		\caption{Entanglement contour $\mathcal{C}_s$ and fluctuation contour $\mathcal{C}_n$ in the weakly interacting Bose gas. $A$ is a $71\times 71$ square embedded in a $1001\times 1001$ torus $(a)$ or a half-torus $(b)$, i.e., a $71\times 71$ cylinder (with PBC in the $y$ direction) in a $143\times 71$ torus. The cut is taken along the $x$ axis, at $y=36$, for $g/J=\exp(-2.5)$ and $n=1$.}
		\label{fig_cont_s_n_bogo_1}
		\end{figure}

\begin{figure}
		{\includegraphics[width = \linewidth]{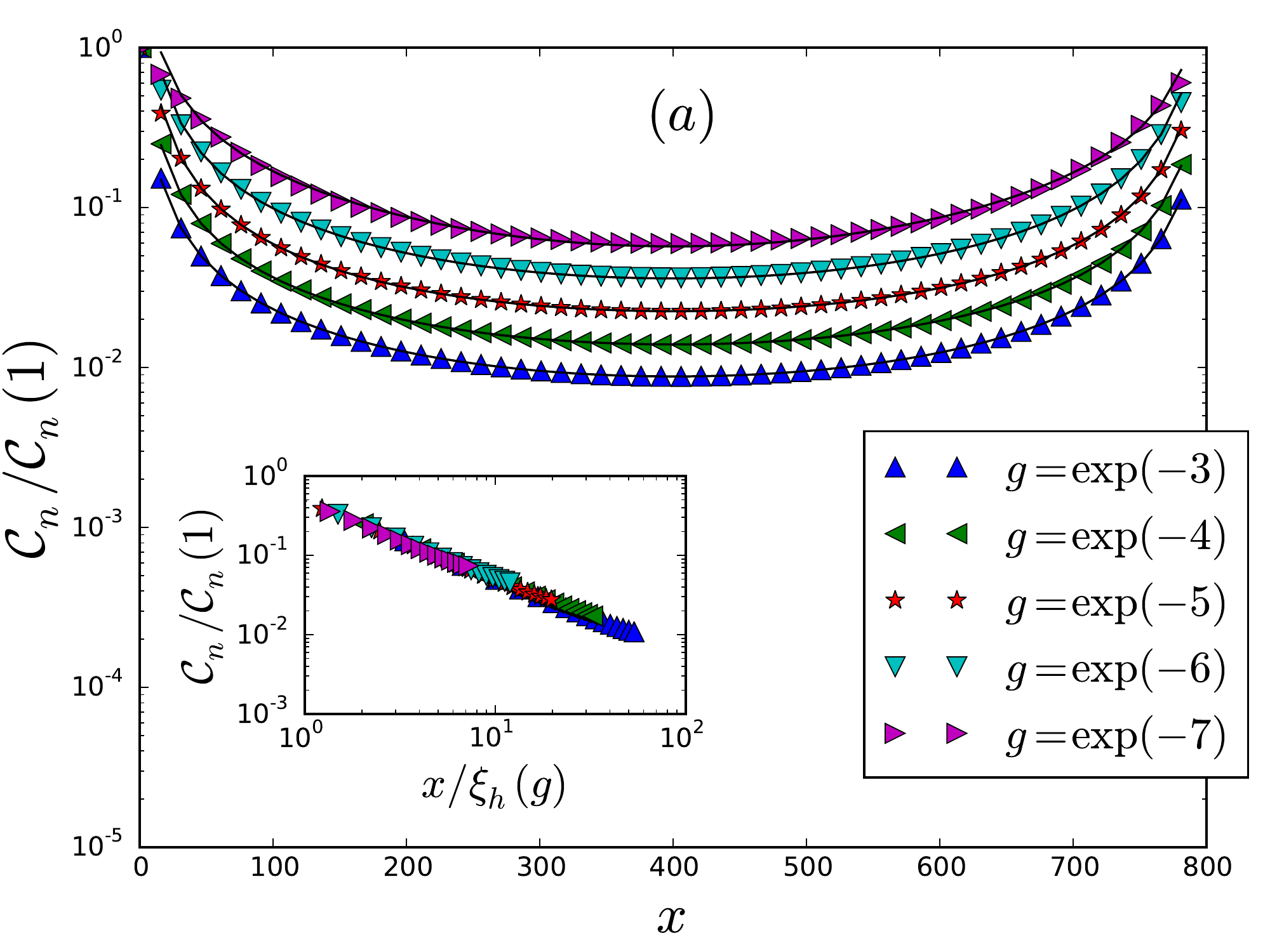}}
		{\includegraphics[width = \linewidth]{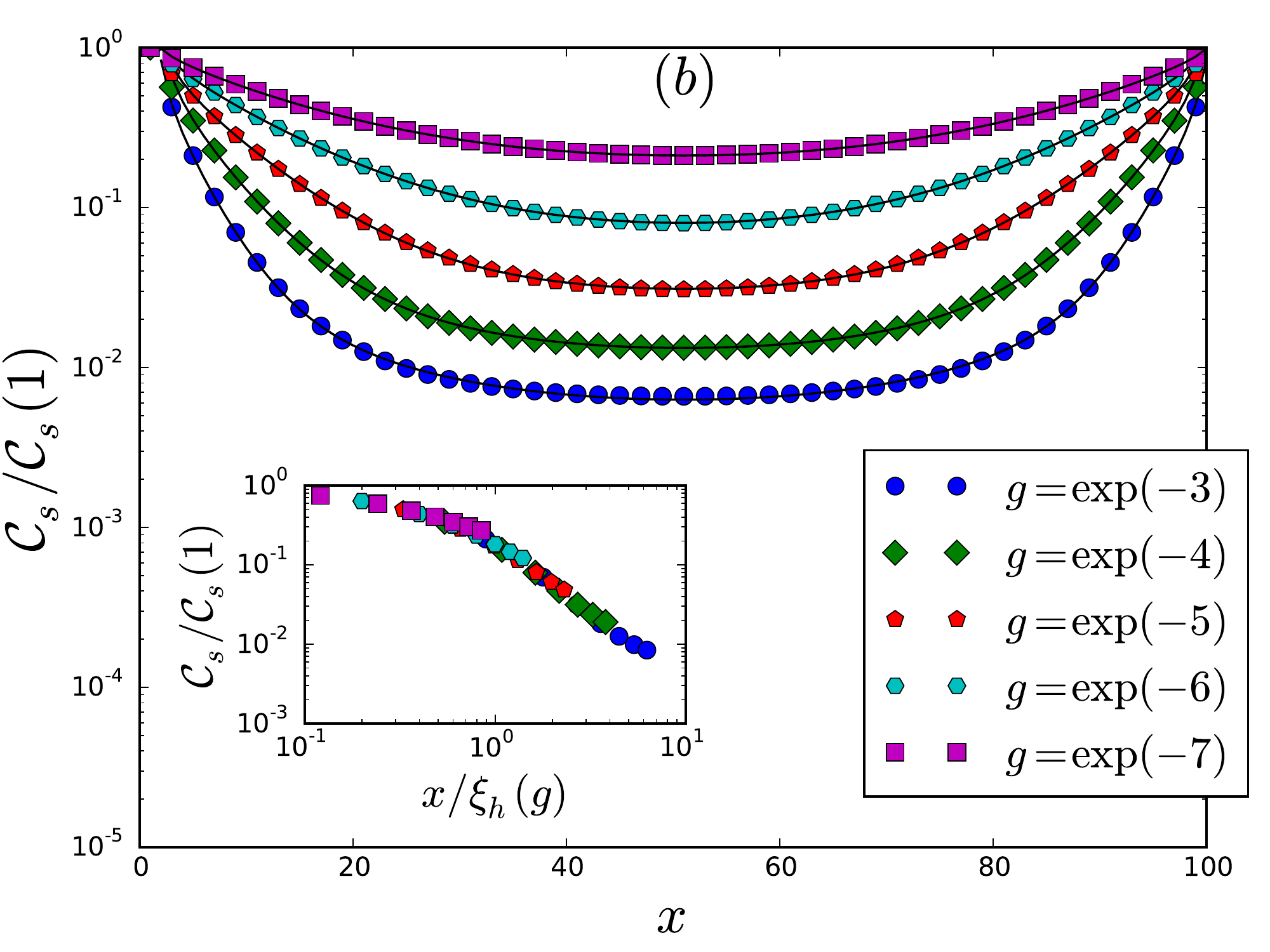}}
		\caption{Fluctuation contour on a half-torus $(a)$ and entanglement contour on a square $(b)$ for the weakly interacting Bose gas. Solid lines show fits of the form $\mathcal{C}_n = A[1/x^\alpha + 1/(l+1-x)^\alpha]$, with $\alpha \approx 1$ and $A\propto \xi_h$ and $\mathcal{C}_s = A'\left[e^{-x/\xi} + e^{-(l+1-x)/\xi}+ A''[1/x^{\alpha'} + 1/(l+1-x)^{\alpha'}]\right]$, with $\xi \approx 0.6 \xi_h$ and $\alpha' \approx 1.3$ for $g=\exp(-4)$ and increases with $g$. Insets: same data plotted as a function of $x/\xi_h(g)$.}
		\label{fit_cont_s_n_bogo}
		\end{figure}

\subsubsection{Fluctuation and entanglement contours}

Eqs.~\eqref{Fn} and \eqref{Fs} establish a fundamental difference between the scaling laws of fluctuations and entanglement: in turn, contours provide an invaluable insight into the origin of such difference. 
Fig.~\ref{fig_cont_s_n_bogo_1} shows the contours $\mathcal{C}_s(x)$ and $\mathcal{C}_n(x)$ as a function of the distance from the boundary of $A$, for $A$ possessing the two geometries described in Fig.~\ref{cont_geom}. While both geometries should give the same result in the limit $l\to\infty$, we observe strong finite-size effects for the size explored in Fig.~\ref{fig_cont_s_n_bogo_1}, and, in the case of the square-in-a-torus geometry, there exists also a significant dependence on the size of the complement $B$ when $A$ and $B$ are comparable. In particular a finite-size, non-monotonic behavior of the entanglement contour complicates the analysis in the case of the half-torus geometry; on the other hand, the latter geometry is best suited to attain the asymptotic regime on smaller sizes. Hence in the following we choose to analyze $\mathcal{C}_s$ on the square-in-a-torus geometry, and $\mathcal{C}_n$ on the half-torus one -- see Fig.~\ref{fit_cont_s_n_bogo}.
		 		 
Figs.~\ref{fig_cont_s_n_bogo_1} and \ref{fit_cont_s_n_bogo} clearly show a vast qualitative difference in the decay of entanglement and fluctuation contours. For points in the bulk of $A$ (namely $1 \ll x \ll l$) the data in Fig.~\ref{fit_cont_s_n_bogo} are very well fitted by the following symmetrized forms
		\be
		\mathcal{C}_n(x)/\mathcal{C}_n(1)  \approx   A_l \left[ \left(\frac{\xi_h}{x}\right)^{\alpha_l} + \left(\frac{\xi_h}{l+1-x}\right)^{\alpha_l} \right]
		\ee
		
		\bearr
		\mathcal{C}_s(x)/\mathcal{C}_s(1)  &\approx   A_{l}'& \left[  ~e^{-x/\xi_l} + e^{-(l+1-x)/\xi_l} \right.\\
&&\left. + A_{l}'' \left(\frac{1}{x^{\alpha'_l}} + \frac{1}{(l+1-x)^{\alpha'_l}} \right) \right]~.
		\label{scaling_cont_s_cont_n_bogo}
		\eearr
The fitting coefficients $A_l, A_{l}',A_{l}'', \alpha_l, \alpha'_l,\xi_l$ should be in principle extrapolated to the limit $l\to\infty$ in order to extract the asymptotic decay of the contours. The fluctuation contour can be investigated on very big sizes $l$ of the $A$ region, given that the density-density correlation function in Eq.~\eqref{dens_Bogo} is readily obtained via a sum over the Brillouin zone. On the other hand, the calculation of the entanglement contour requires the diagonalization of the $(2l)^d\times (2l)^d$ correlation matrix , which is a more demanding numerical task, substantially limiting the system sizes we can access. \footnote{In the case of the half-torus geometry, the diagonalization of the correlation matrix can nonetheless be reduced to $l^{d-1}$ times that of a $2l\times 2l$ matrix by exploiting the translational symmetry along the cut between $A$ and $B$, and even of a $l\times l$ matrix when the correlation matrix is real.}

\begin{figure}
		{\includegraphics[width = \linewidth]{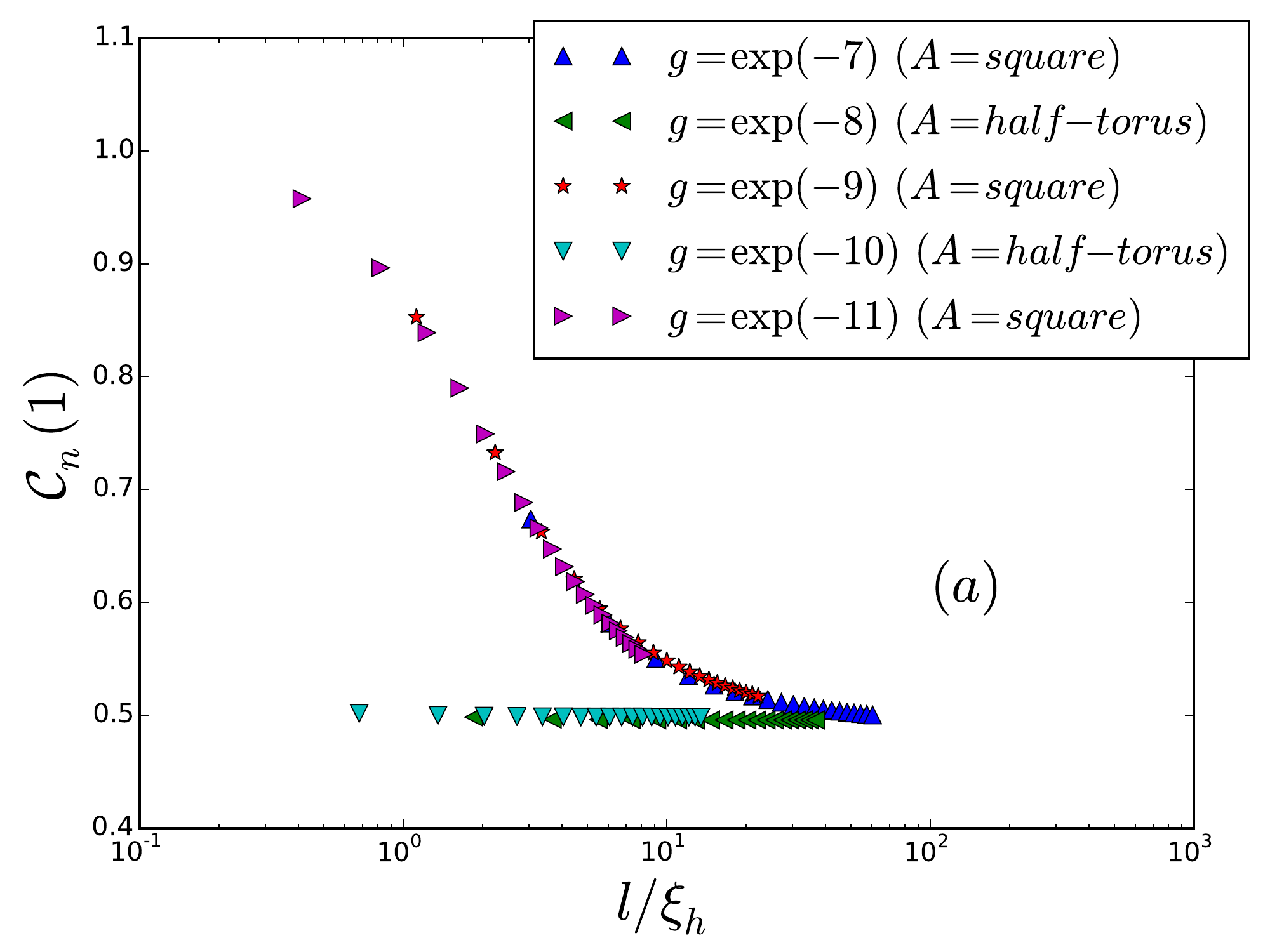}}
		{\includegraphics[width = \linewidth]{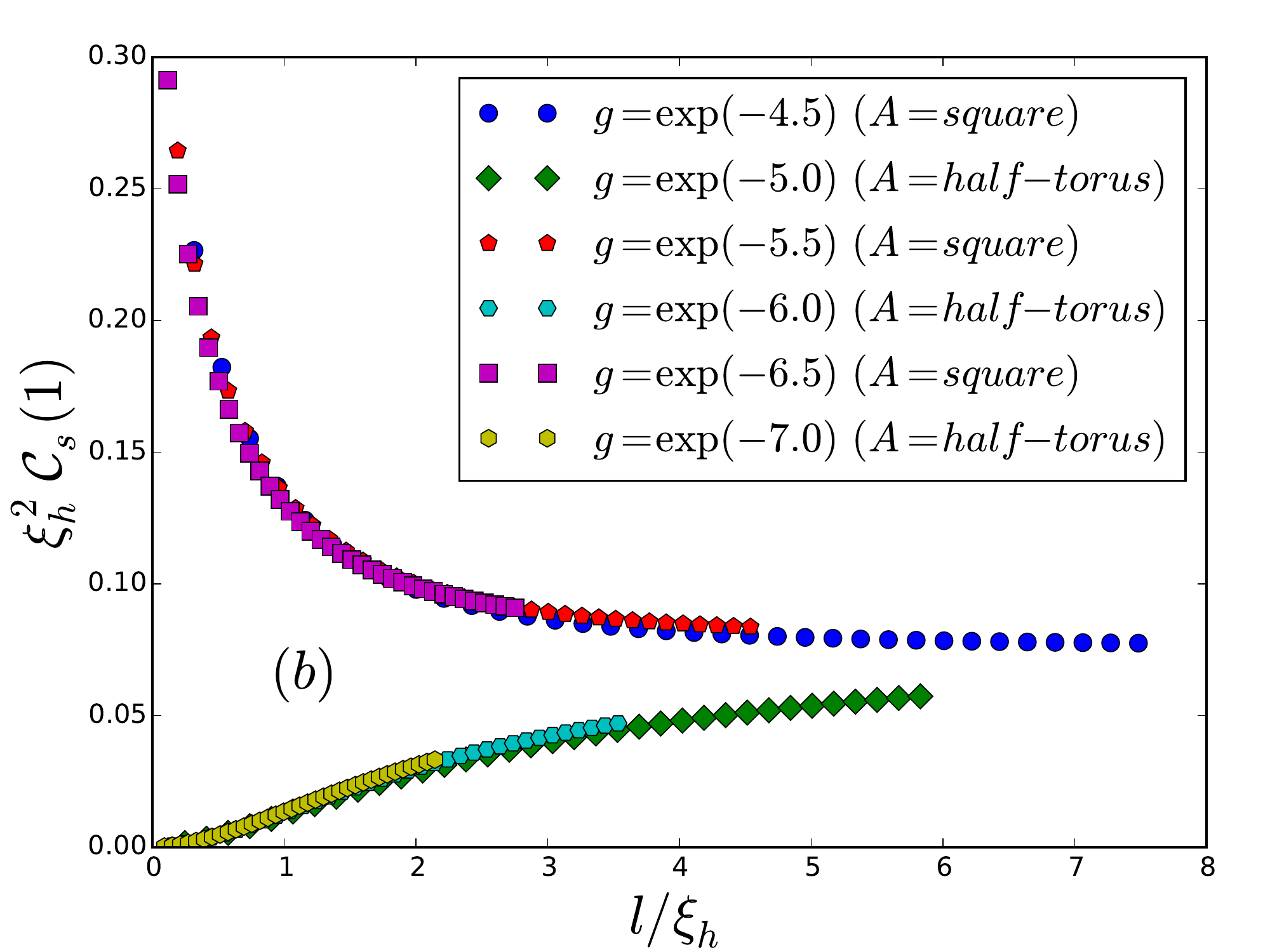}}
		\caption{Fluctuation contour $(a)$ and entanglement contour $(b)$ at the boundary of $A$ for the weakly interacting Bose gas, for various values of of $g$. When $A$ is a square, the chosen site is at the middle of an edge. The dependence on $g$ is almost fully accounted for by rescaling $l$ to $\xi_h$.}
		\label{fig_cont_0_bogo}
		\end{figure}
		
For the fluctuation contour, our data are consistent with $\alpha_l\to 1$ for large $l$ independently of $g$ and $n$, justifying the logarithmic violation of the area law for particle-number fluctuations via Eq.~\eqref{integrals}. The inset of Fig.~\ref{fit_cont_s_n_bogo}(a) shows moreover that plotting $\mathcal{C}_n(x)/\mathcal{C}_n(1)$ as a function of $x/\xi_h$ leads to a collapse of the various curves for different $g$, implying that the $A_l$ constant is nearly independent of $g$ and $n$.
 In addition to that, Fig.~\ref{fig_cont_0_bogo}(a) shows that the boundary value $\mathcal{C}_n(1)$ admits the scaling form $\mathcal{C}_n(1) \approx ~f(l/\xi_h)$ with $f(\infty) \approx 1/2$. Putting everything together this implies that, for $l\to\infty$ and $x \gg 1$ 
 \be
 \mathcal{C}_n(x) \approx \frac{A_{\infty}}{2} \frac{\xi_h}{x} ~. 
 \ee
  The leading term of $\delta^2N_A$ can therefore be obtained by integrating the tail of $\mathcal{C}_n(x)$, namely
           \be
		\delta^2N_A  \approx \frac{2A_\infty ~l}{2}  \int_1^{l/2} \frac{\D x}{x/\xi_h}  =  A_\infty ~l ~\xi_h  \left[ \ln{\left(\frac{l}{\xi_h}\right)} +O(1) \right]    \\
             \ee
which, upon identifying $a$ and $A_{\infty}$, reproduces the dominant term of the scaling form in Eq.~\eqref{Fn} with $n=1$.

For the entanglement contour on the square-in-a-torus geometry, the smaller values of $l$ we can access substantially limit the analysis of the asymptotic behavior. Nonetheless the inset of Fig.~\ref{fit_cont_s_n_bogo}(a) indicates again that plotting $\mathcal{C}_s(x)/\mathcal{C}_s(1)$ as a function of $x/\xi_h$ leads to a collapse of the various curves for different $g$ \footnote{We note that, for both $\mathcal{C}_n(x)/\mathcal{C}_n(1)$ and $\mathcal{C}_s(x)/\mathcal{C}_s(1)$, the collapse occurs only for $gn\ll 1$, which is the range of validity of Bogoliubov approximation. Outside this range, deviations from the scaling behavior are observed.}, indicating that $A'_l$, $A''_l$ and $\alpha_l$ weakly depend upon $g$, and, most importantly, that the length $\xi_l$ must be proportional to $\xi_h$ -- in fact our results are consistent with $\xi_l \approx 0.6~ \xi_h$.  Moreover Fig.~\ref{fig_cont_0_bogo}(b) strongly suggests that the value of the contour at the border is consistent with the scaling law $\mathcal{C}_s(1) \approx \xi_h^{-2} ~h(l/\xi_h)$ with $h(\infty)\approx 0.075$. These elements allow to conclude that for $l\to\infty$ and $x \gg 1$ 
\be
\mathcal{C}_s(x) \approx  \frac{A'_{\infty} h(\infty)}{\xi_h^2}~ \left[e^{-x/\xi_h}+\frac{A''_{\infty}}{x^{\alpha'}} \right] 
\ee 
so that 
\bearr
		S_A  \approx  \frac{2 A'_{\infty} h(\infty)~ l}{\xi_h^2} \int_1^{\infty}  \left[e^{-x/\xi_h}+\frac{A''_{\infty}}{x^{\alpha'}} \right]  \D x \\
		\approx a' \frac{l}{\xi_h} +O\left(\frac{1}{\xi_h^2}\right)
		\label{eq-S_A-cont}
		\eearr
recovering the dominant (area-law) term in the scaling of Eq.~\eqref{Fs} in the limit $\xi_h \to \infty$ (\emph{i.e} $gn\to 0$). The exponent $\alpha'$ is difficult to extract on moderate system sizes, as it requires to study the contour for $l\gg \xi_h$, which in turn requires very large sizes for small values of $gn$. We found that $\alpha' \approx 1.3$ for $gn=\exp(-4)$, and that it increases with $gn$.
	
	\subsubsection{Discussion}
	
	The most striking result of this section is that the healing length $\xi_h$ appears to determine the entanglement properties of the weakly interacting Bose gas in a fundamental way, even though it only enters as a global prefactor in the asymptotic correlation properties of the weakly interacting Bose gas. In particular the healing length controls a crossover in the scaling of the entanglement contour between a short-range exponential decay and a long-range algebraic decay. Given that the algebraic decay is integrable - leaving out an entanglement area law - the entanglement entropy resulting from the integral of the contour is dominated by the short-range part of the latter, namely by the exponentially decaying behavior. Therefore the healing length controls \emph{de facto} the depth of the region of $A$ whose degrees of freedom are significantly entangled with those of $B$. This implies that, for weaker interactions, the size of the portion of $A$ entangled with $B$ increases as so does the healing length; but in fact the entanglement accounted for by our treatment globally decreases, because, according to Eq.~\eqref{eq-S_A-cont}, $S_A \sim \xi_h^{-1} \sim \sqrt{g}$. Indeed as $g\to 0$ particles out of the condensate disappear, but the area law of entanglement is solely due to the modes orthogonal to the condensate. 
	
	 The introduction of a similar notion of contour for density-density correlations indicates that, contrary to entanglement, correlations of particle number fluctuations between $A$ and $B$ involve the whole bulk of $A$: indeed the algebraic decay as $1/x$ of the fluctuation contour implies a logarithmic violation of the area law. Hence the weakly interacting Bose gas is a striking example of a fundamental separation of scales between entanglement and correlations: entanglement is effectively \emph{short-ranged} (in the sense of converging integrals of the contour), while fluctuations are \emph{long-ranged}.

		\subsection{Hardcore bosons and Heisenberg antiferromagnet}
			At this point, it becomes especially interesting to study systems of hardcore bosons, which, as already discussed in Sec.~\ref{tour-d'horizon}, are known to share the scaling properties $S_A \approx a l^{d-1}$ and $\delta^2N_A \propto l^{d-1} \ln l$ of the weakly interacting Bose gas. Nonetheless they do not posses any intrinsic length scale such as the healing length. Yet, the coefficient $a$ of the entanglement area law must still have the dimension of an inverse length, but the only length left at our disposal is the lattice spacing. We know that the predictions of Bogoliubov theory cannot be safely extrapolated outside of its range of validity, namely $gn/J \ll 1$. Nonetheless, taking the limit $gn/J \to \infty$ (or $\xi_h \to 0$) in Eq.~\eqref{scaling_cont_s_cont_n_bogo} could lead to anticipate an algebraic decay of the contours, with a unit exponent for the fluctuation contour, and larger than one for entanglement contour. This scenario is very plausible in light of the critical nature of the hardcore boson gas.


\subsubsection{Hardcore bosons (XX model)}
\label{s.hardcore}
			
We begin our investigation with the square lattice $S=1/2$ XX model, equivalent to hardcore bosons at half filling via the well known Matsubara-Matsuda transformation \cite{MatsubaraM1956}. The Hamiltonian reads : 
		\be 
		\label{H-XY}
		H = - 2J \sum_{\langle ij \rangle} \left( S_i^z S_j^z + S_i^x S_j^x \right) = - J \sum_{\langle ij \rangle} \left ( \tilde{b}_i^{\dagger} \tilde{b}_j + {\rm h. c.} \right ) 
		\ee
		where ${\bm S}$ are $S=1/2$ spin operators and $\tilde{b}_i, \tilde{b}_i^{\dagger}$ are hardcore boson operators, such that $S_i^z = \frac{1}{2} \left( \tilde{b}_i^\dagger + \tilde{b}_i\right )$ and $S_i^x = \frac{1}{2i} \left( \tilde{b}_i^\dagger - \tilde{b}_i \right)$. (Note that we have made the seemingly unconventional choice of coupling the $x$ and $z$ spin components in Eq.~\eqref{H-XY}, in such a way that the $z$ quantization axis lies in the plane where the spin-spin couplings occur). 
		The resulting hardcore boson Hamiltonian can be viewed as the limit $U\to\infty$ of the Bose-Hubbard Hamiltonian (Eq.~\eqref{BH}). 
		The model in question can then be reduced to a quadratic bosonic form via a conventional spin-wave theory \cite{coletta}, built around a ferromagnetic state in the XZ plane, $\otimes_i |s_z = 1/2\rangle_i$, thanks to the Holstein-Primakoff (HP) transformation
\be
S_i^+ = \sqrt{1-b_i^{\dagger}b_i}~~ b_i ~~~~~~~~~ S_i^z = \frac{1}{2}-b_i^{\dagger} b_i
\label{HPtransform}
\ee		
where the HP bosons are constrained to occupations $0 \leq b_i^{\dagger} b_i \leq 1$. Discarding terms beyond quadratic order in the HP bosonic operators leads to the desired quadratic Hamiltonian, whose form in momentum space reads:
		\be 
		{\cal H} = E_0 + 2dJ \sum_{{\bm k}} \left[ (2-\gamma_{{\bm k}}) b_{{\bm k}}^\dagger b_{{\bm k}} - \frac{\gamma_{{\bm k}}}{2}(b_{-{\bm k}} b_{{\bm k}} + b_{{\bm k}}^\dagger b_{-{\bm k}}^\dagger) \right]
		\label{H-hardcore}
		\ee
		Here $E_0 = dL^dJ/4$ is the classical ground state energy, and $\gamma_{{\bm k}} = (1/d) \sum_{p=1}^{d} \cos(k_p)$, where $k_1 = k_x$, $k_2 = k_y$, etc.~. 
		In the following of this section we will specify to the case $d=2$.
		The Hamiltonian in Eq.~\eqref{H-hardcore} is of the form of Eq.~\eqref{H_quadra} with $A_{{\bm k}}= 2- \gamma_{{\bm k}}$ and $B_{{\bm k}}= -\gamma_{{\bm k}}$ (up to a global irrelevant energy scale of $4J$).
		
We would like to stress here that the physical meaning of the quadratic Hamiltonian obtained via spin-wave theory is rather different from that obtained via Bogoliubov theory for the weakly interacting Bose gas. In the latter the reference state is a coherent-state representation of the condensate, and the ``small parameter" is the occupation of modes at finite wave-vector by the physical bosons. In the former the reference state is the classical ferromagnetic state, which is an eigenstate of  $\tilde{b}_i + \tilde{b}_i^{\dagger}$ -- somewhat analog to the coherent state describing the condensate within Bogoliubov theory -- but the quadratic Hamiltonian is obtained by ``re-bosonizing" the physical bosons into HP bosons, and the small parameter is the density of HP bosons, irrespective of their wave vector.
		
	The difference between Bogoliubov theory and spin-wave theory becomes manifest in the treatment of the ${\bm k}=0$ (Goldstone) mode. At variance with the Bogoliubov Hamiltonian, the spin-wave Hamiltonian in Eq.~\eqref{H-hardcore} does contain the term at ${\bm k=0}$, which, on a finite-size system, gives a divergent contribution in the ${\bm k}$-space sums defining the functions $f$ and $g$ in Eq.~\eqref{def_fg}. To cure this divergence in a controlled manner, we adopt the prescription of Ref.~\cite{song}, gapping out the zero mode with a small gap which vanishes in the thermodynamic limit. Such a gap is obtained by applying a term $ -h \sum_i S_i^z$ to the Hamiltonian Eq.~\ref{H-XY}, namely a magnetic field along $z$ which stabilizes the classical ferromagnetic ground state; the magnitude of the field is chosen so as to induce a gap vanishing to zero as $L^{-d}$ in the thermodynamic limit ~\cite{song} (see Section \ref{sec-zero_mode} for a detailed discussion).

		\begin{figure}
		{\includegraphics[width = \linewidth]{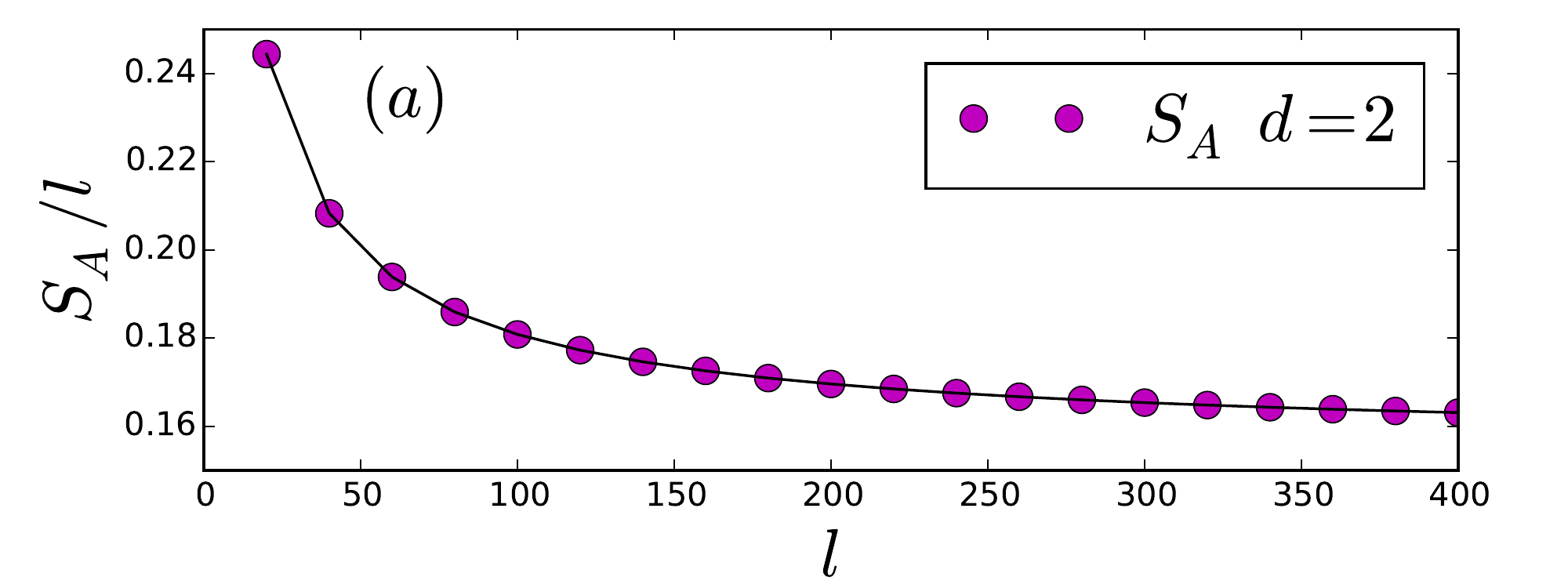}}
		{\includegraphics[width = \linewidth]{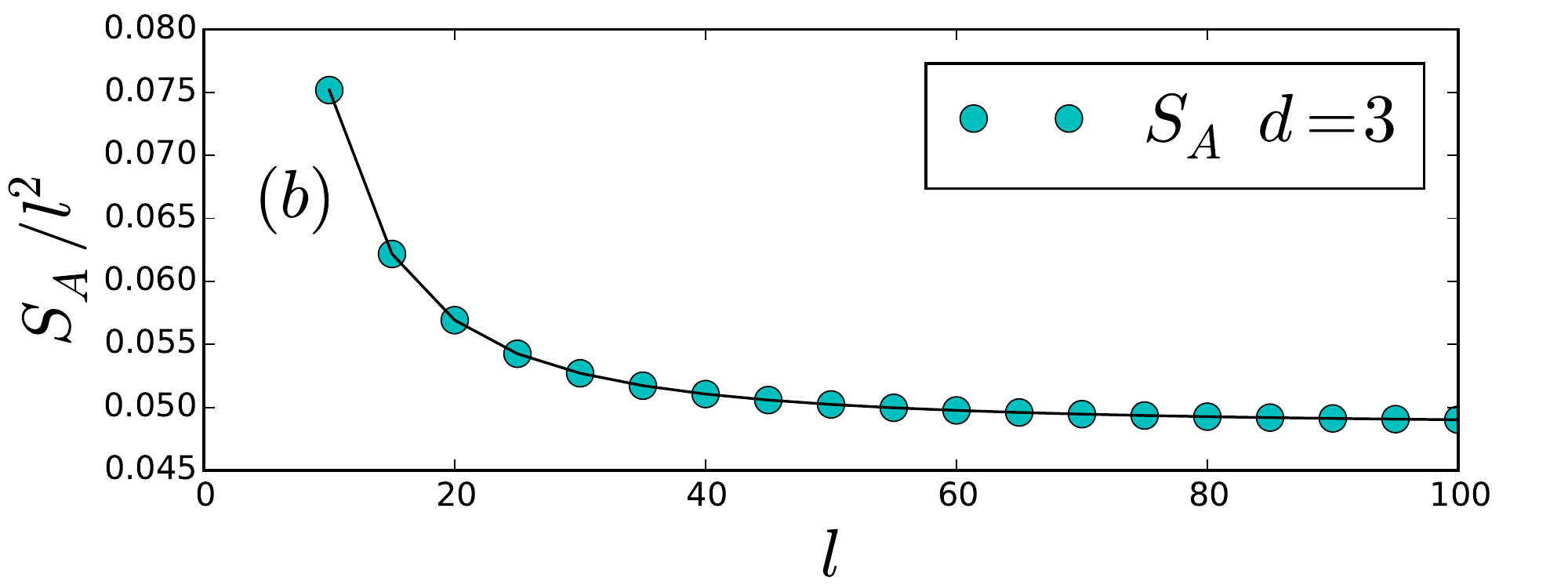}}
		\caption{Scaling of the entanglement entropy for hardcore bosons (XX model) in $d=2$ and 3. For each size $l$, $A$ is taken as half of a $2l\times l^{d-1}$ (hyper)-torus in $d$ dimensions. Black curves show fits of the form $S_A = al^{d-1} +b\ln l +c$. $(a)$  $d=2$, with fit parameters  $a=0.1549$, $b= 0.4995$ and $c=  0.2933$. $(b)$ $d=3$, with fit parameters $a=0.0485$, $b=1.0006$ and $c= 0.3623$.
}
		\label{entropy_scaling_hardcore}
		\end{figure}			
			
			Once the divergence in the $f$ and $g$ functions is cured, one can study the entanglement properties via the diagonalization of the generalized density matrix Eq.~\eqref{relation_C_H}. On the other hand, the density-density correlation function for hardcore bosons reads  
		\be
		\langle \delta n_i \delta n_i \rangle = \frac{n}{V} \sum_{{\bm k}} e^{i{\bm k} \cdot ({\bm r}_i-{\bm r}_j)} \frac{\sqrt{1-\gamma_{{\bm k}}}}{2}~.
		\label{nn-corr}
		\ee
The derivation is detailed in Appendix~\ref{dens-dens-correl}. We notice that Eq.~\eqref{nn-corr} does not require the regularization of the ${\bm k}=0$ mode.

Numerically, we studied the half-torus geometry and found the scaling properties $\delta^2 N_A \sim l \ln l$, and $S_A = a l + b \ln l +c$, with coefficients $a=0.1549$, $b=0.4995$ and $c = 0.2933$, obtained from a fit on sizes of $A$ up to $400^2$, as shown in Fig.~\ref{entropy_scaling_hardcore}(a). The above results confirm the presence of an entanglement area law with additive logarithmic corrections, already reported in Refs.~\cite{HumeniukR2012,HelmesW2014,Kulchytskyyetal2015}, to be contrasted with the multiplicative logarithmic corrections of fluctuations. Fig.~\ref{entropy_scaling_hardcore}(b) shows the scaling of entanglement entropy for $d=3$ and is further discussed in Section \ref{sec-zero_mode}.

			\begin{figure}
			{\includegraphics[width = \linewidth]{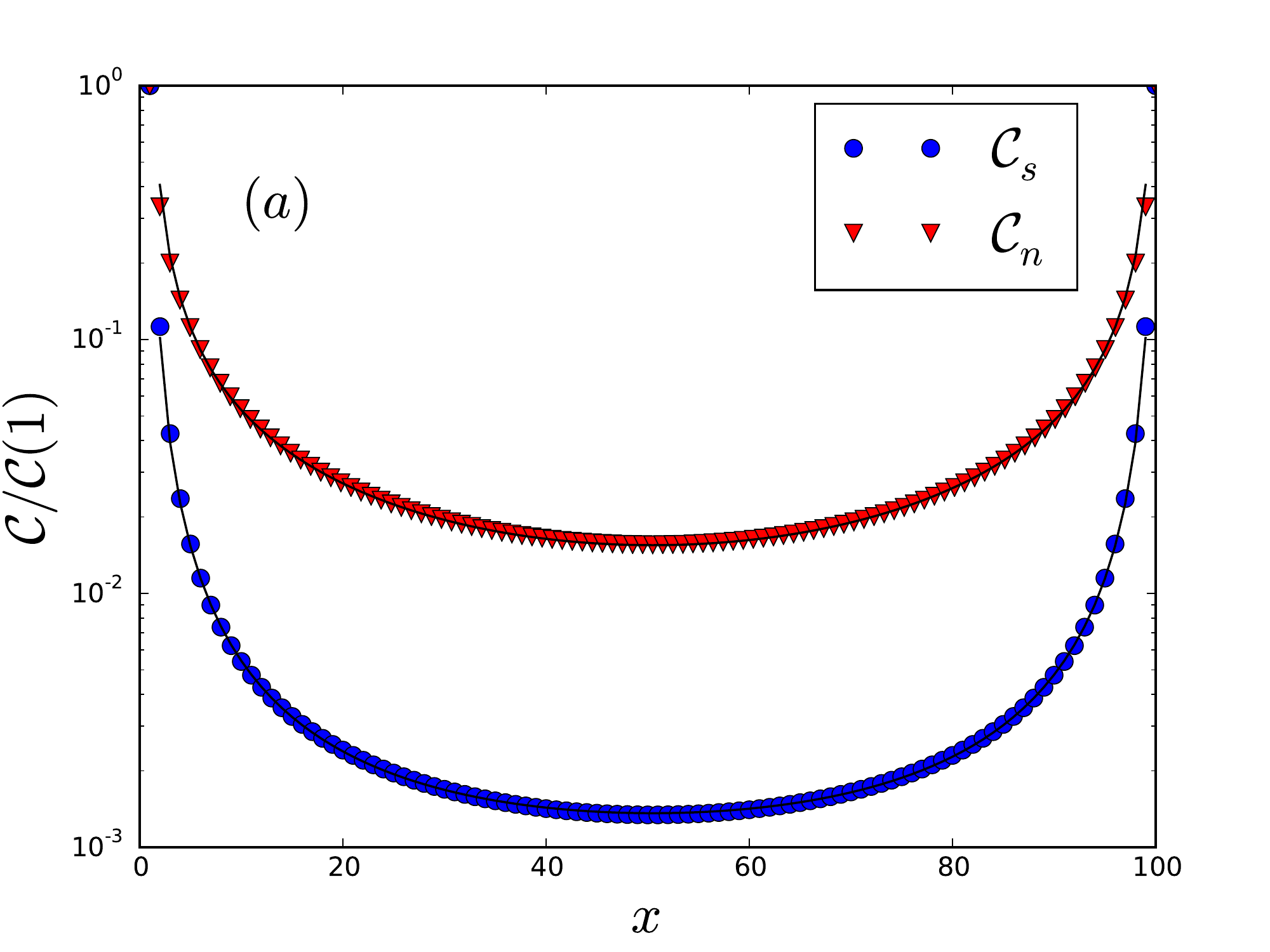}}
			{\includegraphics[width = \linewidth]{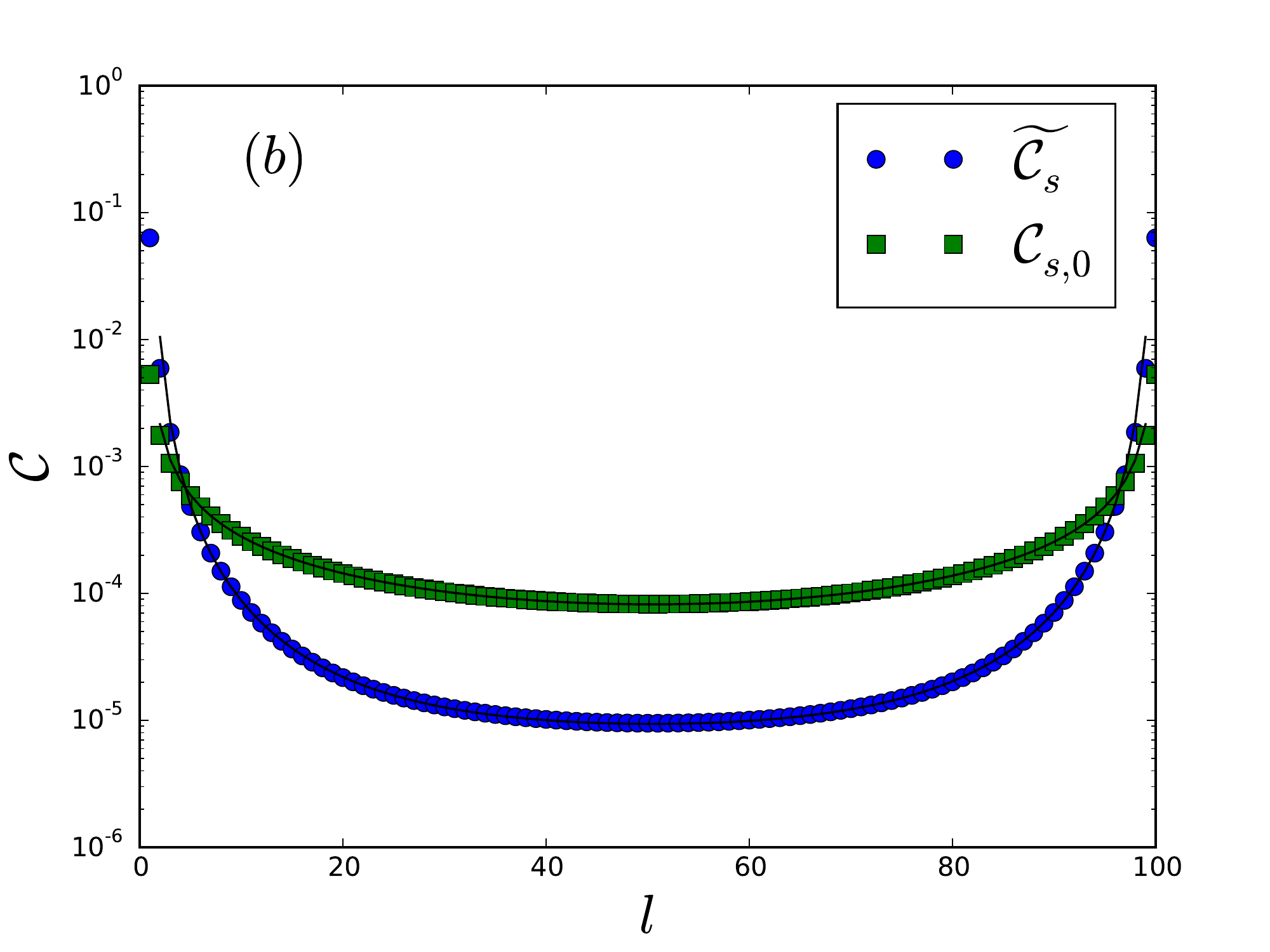}}
			\caption{$(a)$ Contour of density correlations and entanglement contour on a half-torus for hardcore bosons. Solid lines show fits of the form $\mathcal{C} = a[1/x^\beta + 1/(l+1-x)^\beta] + b$, with $\beta \approx 1$ for $\mathcal{C}_n$. $(b)$ Entanglement contour. $\mathcal{C}_{s,0}$ is the contour associated with the entanglement lowest mode, and $\widetilde{\mathcal{C}}_s=\mathcal{C}_s-\mathcal{C}_{s,0}$ the contour without the lowest mode. Solid lines show fits of the form $\mathcal{C}_{s,0} = A[1/x^{\alpha_0} + 1/(l+1-x)^{\alpha_0}] + b$, with $\alpha_0 \approx 1$ and $\widetilde{\mathcal{C}}_s = A'[1/x^\alpha + 1/(l+1-x)^\alpha] + b'$, with $\alpha \approx 2$.}
			\label{cont_s_n_hardcore}
			\end{figure}

Fig.~\ref{cont_s_n_hardcore} shows the entanglement and fluctuation contour for hardcore bosons at half filling. The fluctuation contour is found to decay as $1/x$ with the distance to the boundary, whereas entanglement contour shows a power law with an exponent larger than one. 		
The extraction of the power-law decay exponent of the entanglement contour requires a careful examination. Indeed, on moderate system sizes, the behavior of the contour appears to be dominated by a single contribution coming from the ``entanglement lowest mode" (\emph{i.e.} the eigenmode of the entanglement Hamilonian of lowest ``entanglement energy" $\omega_{\alpha}$, and therefore of highest weight). 
As further discussed in Sec.~\ref{sec-zero_mode}, this mode gives a logarithmically divergent contribution of $\approx 0.57 \ln l$ to the entanglement entropy, and therefore it does not contribute to the area law but to the subdominant terms in the entanglement scaling. This anomalous contribution to the entanglement entropy can be either understood via the scaling of the associated entanglement energy $\omega_0$ (as further discussed in Sec.~\ref{sec-zero_mode}), or via the spatial structure of the mode, reflected in its contribution to the contour, ${\cal C}_{s,\alpha}(i) = w_{\alpha}(i) ~S_{\alpha}$. The latter is found to exhibit the following properties
		\be
		\frac{{\mathcal{C}_{s,0}}(x)}{\mathcal{C}_{s,0}(1)} \approx \frac{A}{x} ~~~~~~~~~~~~
		{\mathcal{C}}_{s,0}(1) \approx \frac{A'}{l}~.
		\ee
The $1/x$ decay of the contour and the $1/l$ scaling of the contour on the boundary produces the $\ln l$ contribution of the mode to the entanglement entropy. In particular the slow decay of the lowest-mode contour masks the actual asymptotic behavior of the total contour on moderate system sizes $l$: it is therefore convenient to eliminate the contribution to the contour from the lowest mode, and to analyze $\widetilde{{\cal C}_s}(x) = {\cal C}_s - {\cal C}_{s,0}$, as done in Fig.~\ref{cont_s_n_hardcore}(b).  
The contour without the lowest mode is found to scale as :
		\be
		\frac{\widetilde{{\cal C}_s}(x)}{\widetilde{\mathcal{C}_s}(1)} \approx \frac{B}{x^\alpha} ~~~~~~~~~~~~
		\widetilde{\mathcal{C}_s}(1) \to B'_\infty 
		\ee
		with $\alpha \approx 2$. The integral of $1/x^2$ being finite, the area law for entanglement entropy follows immediately.
		
		\begin{figure}
		{\includegraphics[width = \linewidth]{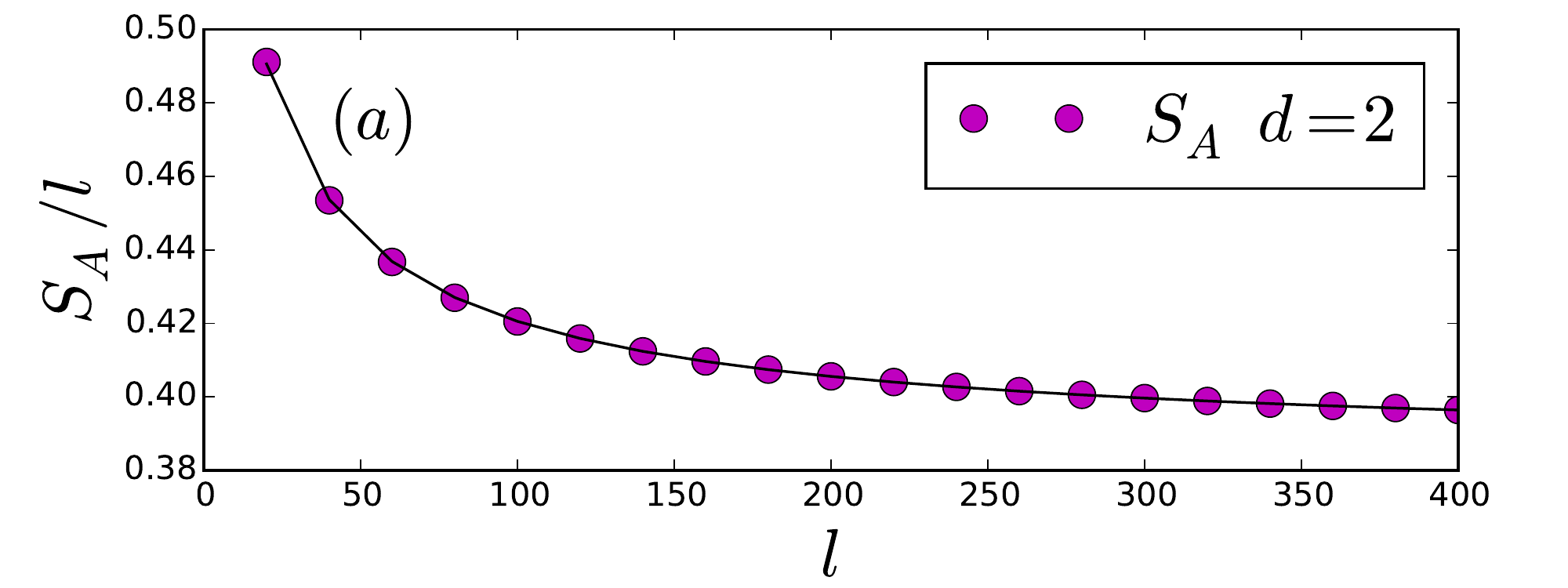}}
		{\includegraphics[width = \linewidth]{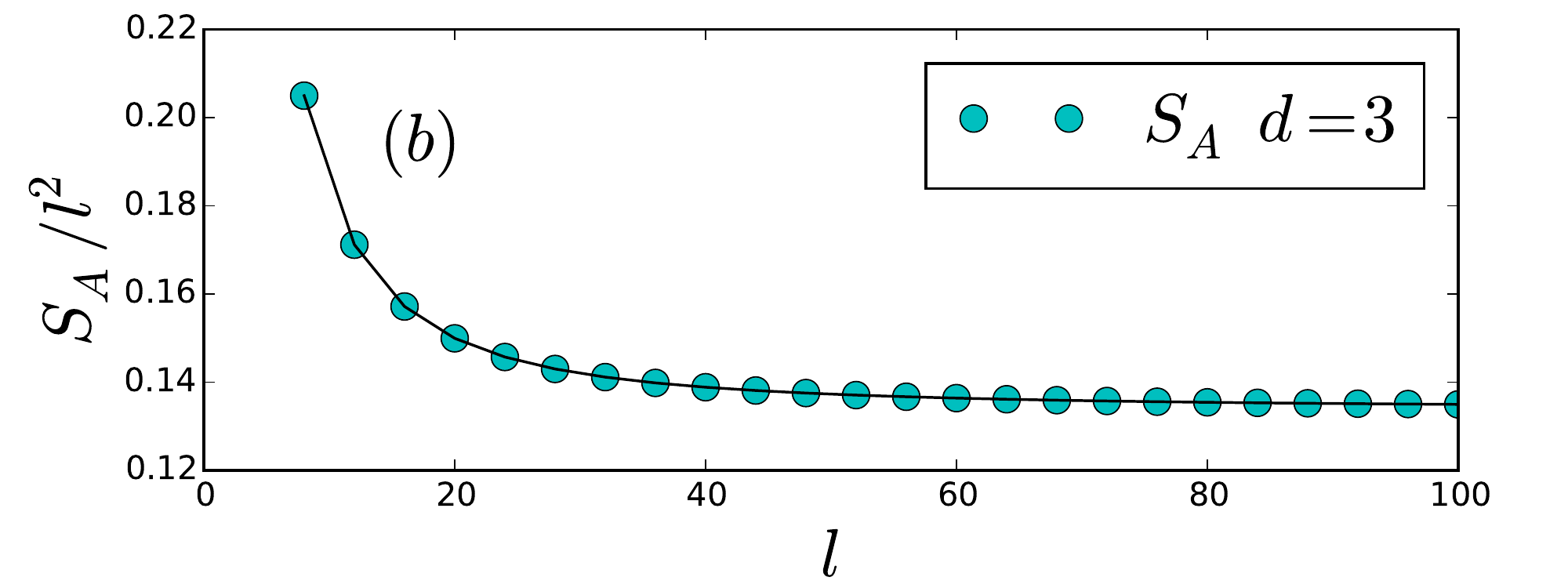}}
		\caption{Scaling of the entanglement entropy for the Heisenberg antiferromagnet in $d=2$ and 3. The geometry of $A$ and the fitting function are the same as for the data in Fig.~\ref{entropy_scaling_hardcore}. $(a)$  $d=2$, with fit parametere $a=0.3840$, $b=  0.9446$ and $c=  -0.6988$. $(b)$ $d=3$, with fit parameters $a=0.1340$, $b=  1.9996$ and $c=0.3834$.}
		\label{entropy_scaling_antiferro}
		\end{figure}	
 \subsubsection{Heisenberg antiferromagnet}
			
		We end our survey with the square lattice $S=1/2$ Heisenberg antiferromagnet, whose fluctuation and entanglement properties have been extensively investigated in the recent past \cite{Hastingsetal2010,Kallinetal2011,song,HumeniukR2012,Luitzetal2015}. The Hamiltonian reads : 
		\be 
		\label{H-Heisenberg}
		H = J \sum_{\langle ij \rangle} {\bm S}_i \cdot {\bm S}_j~. 
		\ee
	Applying spin-wave theory to this model leads to the quadratic Hamiltonian
		\be
		H_{\rm SW} = E_{\textnormal{N\'eel}} + dJ \sum_{{\bm k}} \left[ b_{{\bm k}}^\dagger b_{{\bm k}} - \frac{\gamma_{{\bm k}}}{2}(b_{-{\bm k}} b_{{\bm k}} + b_{{\bm k}}^\dagger b_{-{\bm k}}^\dagger) \right]
		\label{H-SW}
		\ee
		where $E_{\textnormal{N\'eel}} = -dL^dJ/4$ is the classical ground state energy. The Hamiltonian in Eq.~\eqref{H-SW} is hence of the form of Eq.~\eqref{H_quadra}, with $A_{{\bm k}}  = 1$ and $B_{{\bm k}} = -\gamma_{{\bm k}}$ (up to a global energy scale of $dJ$). The divergence in the $f$ and $g$ functions is eliminated similarly to the case of the XX model, by applying a staggered field which vanishes in the thermodynamic limit. This allows to calculate the spin-spin correlations in a controlled way according to the formula \cite{song} 
		\be
		\langle \delta S^z_i \delta S^z_j \rangle= \frac{1}{3}\left[ -\frac{1}{4} \delta_{ij} + f({\bm r}_i-{\bm r}_j)^2 - g({\bm r}_i-{\bm r}_j)^2 \right]~. 
		\label{e.spinspin}
		\ee
		The contour of density correlations (of spin correlations in this case) is then $\mathcal{C}_n(i) = \sum_{j\in B} \langle \delta S^z_{i} \delta S^z_{j} \rangle$.
		
			\begin{figure}
			{\includegraphics[width = \linewidth]{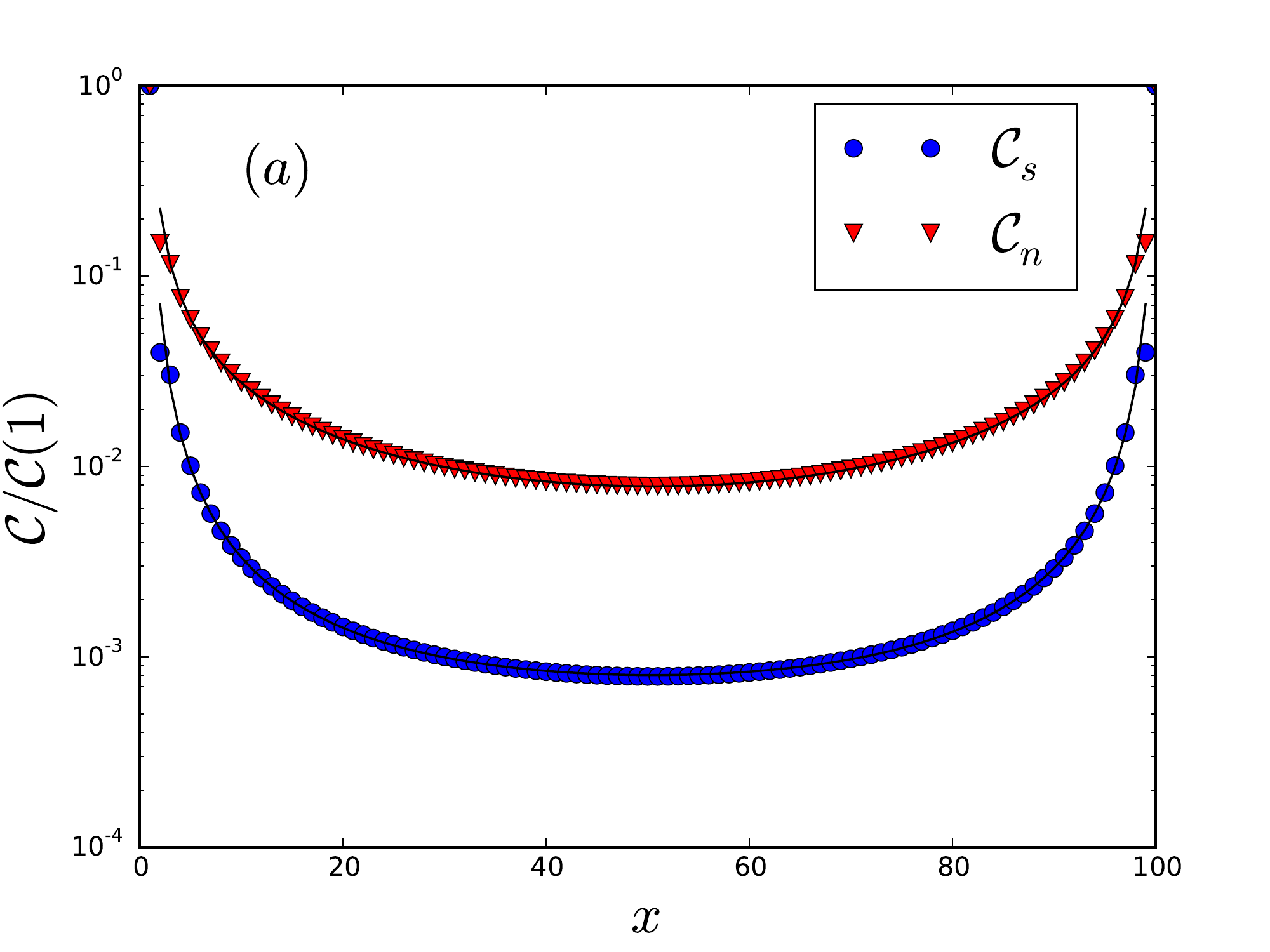}}
			{\includegraphics[width = \linewidth]{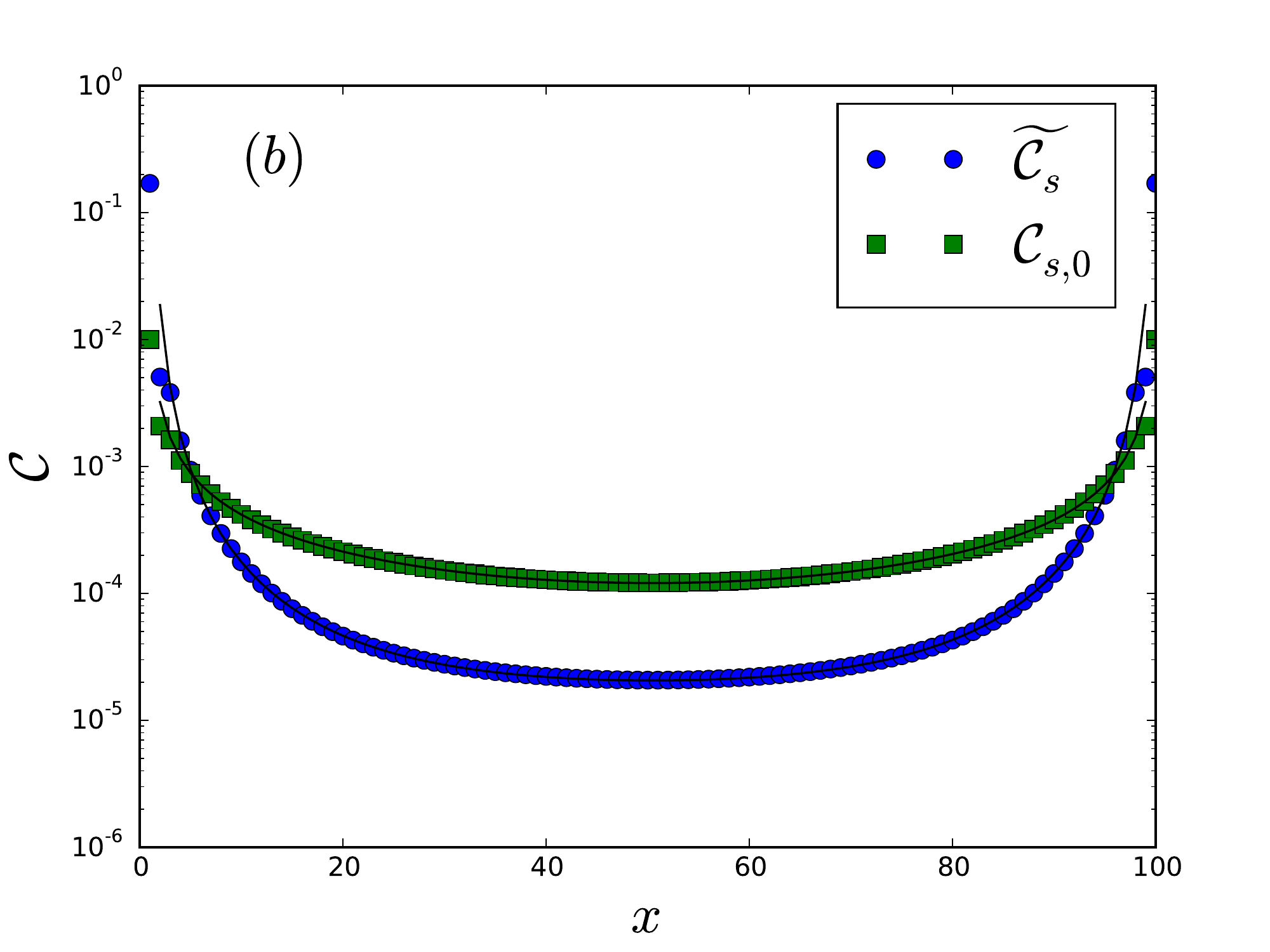}}
			\caption{(a) Fluctuation contour and entanglement contour on a half-torus for the Heisenberg antiferromagnet. Solid lines show fits of the form $\mathcal{C} = a[1/x^\beta + 1/(l+1-x)^\beta] + b$, with $\beta \approx 1$ for $\mathcal{C}_n$. (b) Entanglement contour. $\mathcal{C}_{s,0}$ is the contour associated with the two degenerate entanglement lowest modes, and $\widetilde{\mathcal{C}_s}=\mathcal{C}_s-\mathcal{C}_{s,0}$ the contour without these lowest modes. Solid lines show fits of the form $\mathcal{C}_{s,0} = A[1/x^{\alpha_0} + 1/(l+1-x)^{\alpha_0}] + b$, with $\alpha_0 \approx 1$ and $\widetilde{\mathcal{C}}_s = A'[1/x^\alpha + 1/(l+1-x)^\alpha] + b'$, with $\alpha \approx 2$.}
			\label{cont_s_n_antiferro}
			\end{figure}
		
		The scaling of fluctuations and entanglement entropy where carefully studied in Refs.~\cite{Hastingsetal2010,Kallinetal2011,song,HumeniukR2012,Luitzetal2015}: fluctuations were found to scale as $\delta^2 N_A \sim l\ln l$, and entanglement entropy as $S_A=al +b\ln l +c$. Our results fully confirm these scaling behaviors and extend them to $d=3$, as shown on Fig.~\ref{entropy_scaling_antiferro}, and the study of contours, shown in Fig. \ref{cont_s_n_antiferro}, provides a microscopic insight into these scaling laws. The behavior of contours for Heisenberg antiferromagnet is found to be very close to that of hardcore bosons studied in the previous section, the fluctuation contour decaying as $1/x$, and the entanglement contour decaying as $1/x^\alpha$ with $\alpha>1$. We note however one important difference : the entanglement lowest mode, providing an anomalous contribution to the contour and to the entanglement entropy (as discussed for hardcore bosons) is \emph{twofold} degenerate. The two degenerate lowest entanglement eigenmodes feature a decay as $1/x$ in their entanglement contour, while their contour at the boundary decays as $1/l$: altogether this gives an asymptotic contribution of $\approx 1.1 \ln l$ to the entanglement entropy, and a vanishingly small contribution to the area law coefficient in the thermodynamic limit. In light of their anomalously slow decay, and of their vanishing contribution to the dominant area law of entanglement, it is convenient to remove the lowest modes from the analysis of the contour, extracting in this way the correct asymptotic decay on finite size systems. This is done in Fig. \ref{cont_s_n_antiferro}(b), where the contour $\widetilde{{\cal C}_s}(x)$ is found to decay approximately as $1/x^2$, as found for hardcore bosons in the previous section. Hence a seemingly universal behavior of the contour appears to emerge for gapless hardcore bosons (with or without nearest neighbor repulsion). Future work will be dedicated to assessing the universality of such a behavior.
	
\subsubsection{Discussion}
 We have seen that the fundamental difference between the scaling of entanglement entropy and that of particle-number fluctuations in the case of hardcore bosons / quantum spin models arises from a different power-law decay of the corresponding contours -- with a power law which is integrable for the entanglement contour, and whose integral diverges logarithmically for the fluctuation contour. A careful analysis of the entanglement contour requires a modal decomposition, which in turn highlights an anomalous contribution coming from the lowest mode of the entanglement Hamiltonian (whose role will be further discussed in the next section). Similarly to what is done for fermions in Eq.~\eqref{formule_cont_s_n_fermions}, one could imagine performing a similar modal decomposition for the fluctuation contours, and investigating the contribution of the zero mode. Yet this task cannot be simply accomplished for bosons; in the case of hardcore bosons/ XX model, Eq.~\eqref{e.Crhcbosons} in the Appendix allows apparently for such a modal decomposition, but the resulting expression is non-local in space, and, unlike the single-mode entanglement contour, the single-mode fluctuation contour oscillates in sign, preventing a simple analysis of its spatial decay. In the case of the Heisenberg antiferromagnet, Eq.~\eqref{e.spinspin} for spin-spin correlations does not admit a linear modal decomposition, given the quadratic dependence on the $f$ and $g$ functions. Hence, although the fluctuation contour has a seemingly universal decay in the case of hardcore (and even softcore) bosons, its modal decomposition does not have universal features, unlike the case of the entanglement which we shall now discuss.

\section{Entanglement lowest mode, Goldstone modes and tower of states}
\label{sec-zero_mode}	

	So far, we have mainly focused our attention on the leading term in the scaling of the entanglement entropy and fluctuations, namely the area law or its violation. 
Nonetheless in the case of bosonic models a significant attention been recently devoted to the subleading term in the scaling of the entanglement entropy \cite{metlitski-grover,Kulchytskyyetal2015,Luitzetal2015}, as Ref.~\cite{metlitski-grover} argued that such term is connected to the number of components of the order parameter / number of Goldstone modes in a universal manner. 
 The modal analysis of the entanglement entropy performed in the previous section allows to clarify the origin of the subdominant contributions to the entanglement scaling, as well as their relationship to the number of Goldstone modes, in a rather transparent way. Here we shall focus on hardcore bosons and Heisenberg antiferromagnet, since, as already mentioned, logarithmic corrections to the entanglement entropy are not reliably accounted for by the standard Bogoliubov approximation that we use to describe the weakly interacting Bose gas.
  
  \subsection{Scaling of the lowest spin-wave mode}
   To start our analysis, we come back to the issue of the special treatment of zero energy modes of the spin-wave Hamiltonians, Eq.~\eqref{H-hardcore} and \eqref{H-SW}. 
  For hardcore bosons, there is one mode of zero energy at $\bm k =0$, and for Heisenberg antiferromagnet, there are two modes of zero energy at $\bm k=0$ and $\bm k=(\pi,\pi)$ -- in general, one expects $N_G$ modes of zero energy in a model where the order parameter has $N_G$ independent components, each mode being the zero-energy termination of a branch of gapless excitations (Goldstone modes). As already discussed in Sec.~\ref{s.hardcore} these modes give a divergent contribution to the functions $f$ and $g$, but such a divergence is an artefact of the finite-size system. Indeed in the thermodynamic limit one goes from discrete sums to integrals, and the divergent contribution of the zero modes, going like $1/E_{\bm k} \sim 1/k$  for ${\bm k}\to 0$, is integrable in $d>1$ dimensions (as the integration element is $k^{d-1} \D k$). A similarly integrable divergence can be mimicked with discrete sums on a finite-size system by demanding that the finite-size integration element in ${\bm k}$-space, $\left (\frac{2\pi}{L} \right)^d$, compensates the $1/E_{\bm k}$
term for ${\bm k}\to 0$, namely 
\be
\left(\frac{2\pi}{L} \right)^d \frac{1}{E_{\bm k}} \to {\rm const.} ~~~~~~ ({\bm k}\to 0)
\ee
 which amounts to requiring $E_{\bm k=0} \sim L^{-d}$. As discussed in Ref.~\cite{song}, this condition is achieved by applying a field which gaps out the zero mode, and which scales as $h\sim L^{-2d}$. Nonetheless the actual prefactor relating $h$ and $L^{-2d}$ is rather arbitrary -- in principle, a natural choice would be the one which guarantees the same (finite) value of the order parameter for any finite system size as in the thermodynamic limit. Refs.~\cite{song,metlitski-grover, Luitzetal2015} choose the prefactor so that the order parameter vanishes, arguing that this allows to restore the (inversion) symmetry along the quantization axis by hand -- this statement is questionable, in that the symmetry is rather broken ``twice" (once when the quantization axis is chosen in spin-wave theory, and a second time when the field is applied).  
In fact, we shall argue that the results one obtains for the dominant, subdominant and sub-subdominant terms of the scaling of the entanglement entropy are completely independent of the choice of the prefactor. Our choice is simply $h =L^{-2d}$.
  
  Beside removing divergences, the introduction of a gap scaling as $L^{-d}$ has another, fundamental virtue. Indeed a finite-size realization of a system breaking a continuous symmetry is expected to exhibit low-energy excitations in the form of a tower of states (ToS) \cite{Anderson1952, ZimanN1989}, corresponding to collective, global rotations of the spins, and exhibiting level spacings decreasing as $L^{-d}$. In the case of the XX and Heisenberg model the ToS spectrum is the one of a planar rotor and spherical rotor respectively, with a moment of inertia scaling as $L^d$. This non-linear ToS spectrum is energetically separated from the spin-wave spectrum, whose level spacings scale as $L^{-1}$. Moreover all states of the ToS have zero momentum (as they correspond to global rotations of the spins): populating spin-wave states of opposite momenta does not produce states of the ToS, since the scaling of the energy eigenvalues with $L$ is wrong. The only spin-wave state which can be potentially related to the ToS is the zero mode, as it possesses the right wavevector ${\bm k}=0$; hence the $L^{-d}$ scaling imposed to the energy of the zero mode via the application of a suitable magnetic field is an effective way to mimic the scaling of the \emph{first excited state} in the ToS \cite{WeihongH1993} -- the higher states are not correctly reproduced, because populating the $\bm k=0$ mode with several quasi-particles gives a linear $N-$body spectrum ($nE_0$ with $n = 1, 2,3 \dots$), while the rotor spectrum is quadratic.   

		\begin{figure}
			{\includegraphics[width = \linewidth]{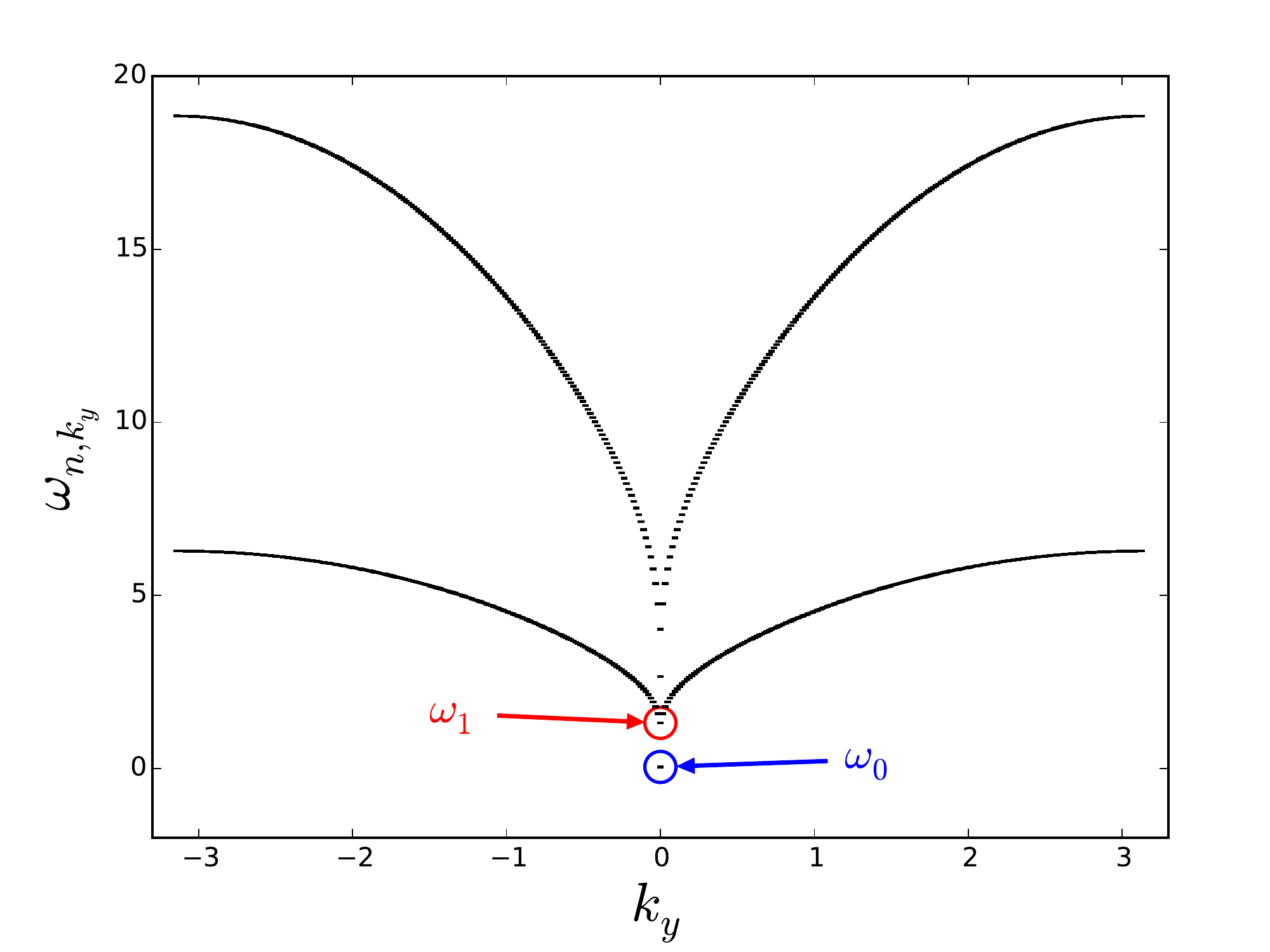}}
			{\includegraphics[width = \linewidth]{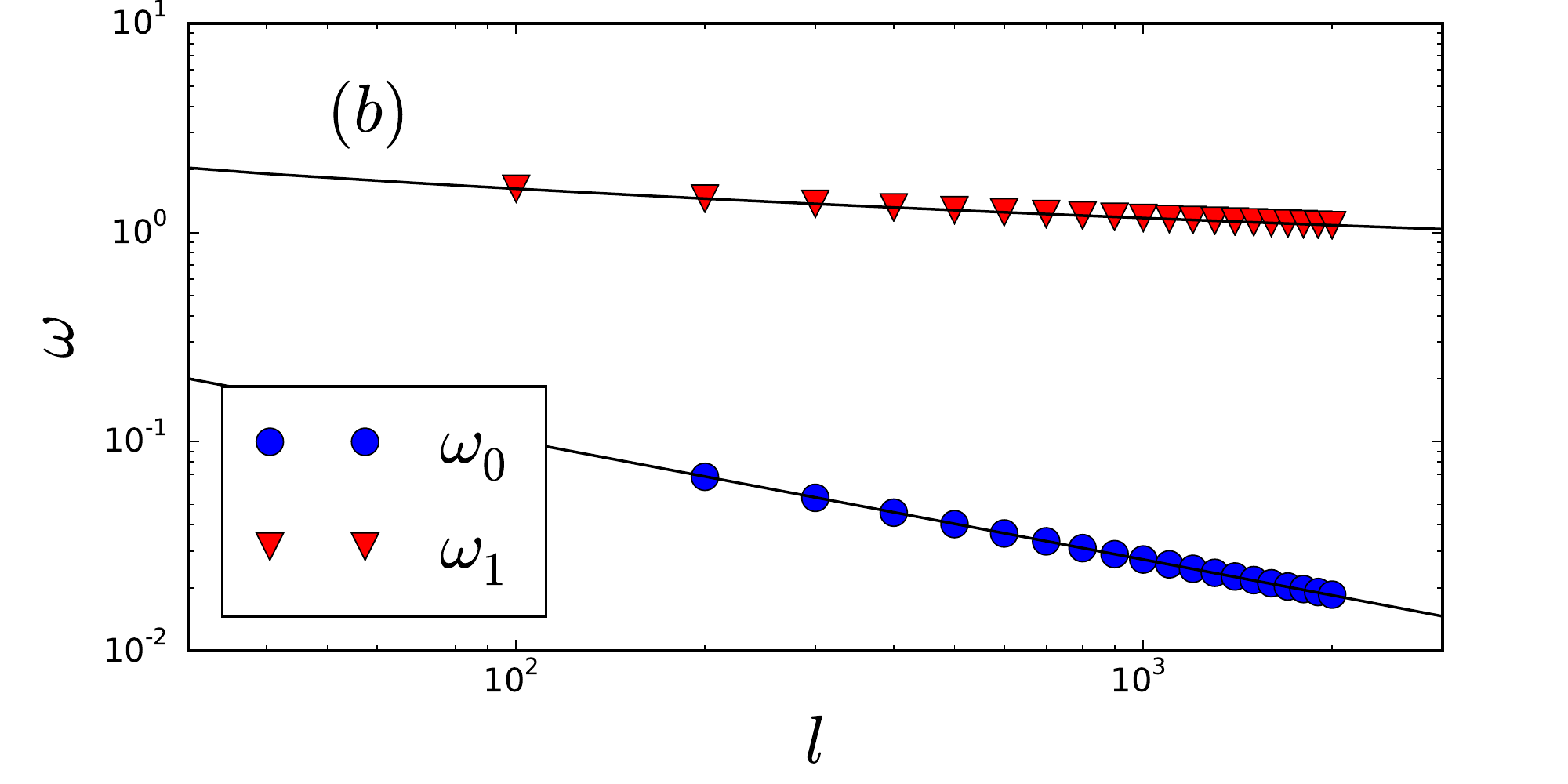}}
			{\includegraphics[width = \linewidth]{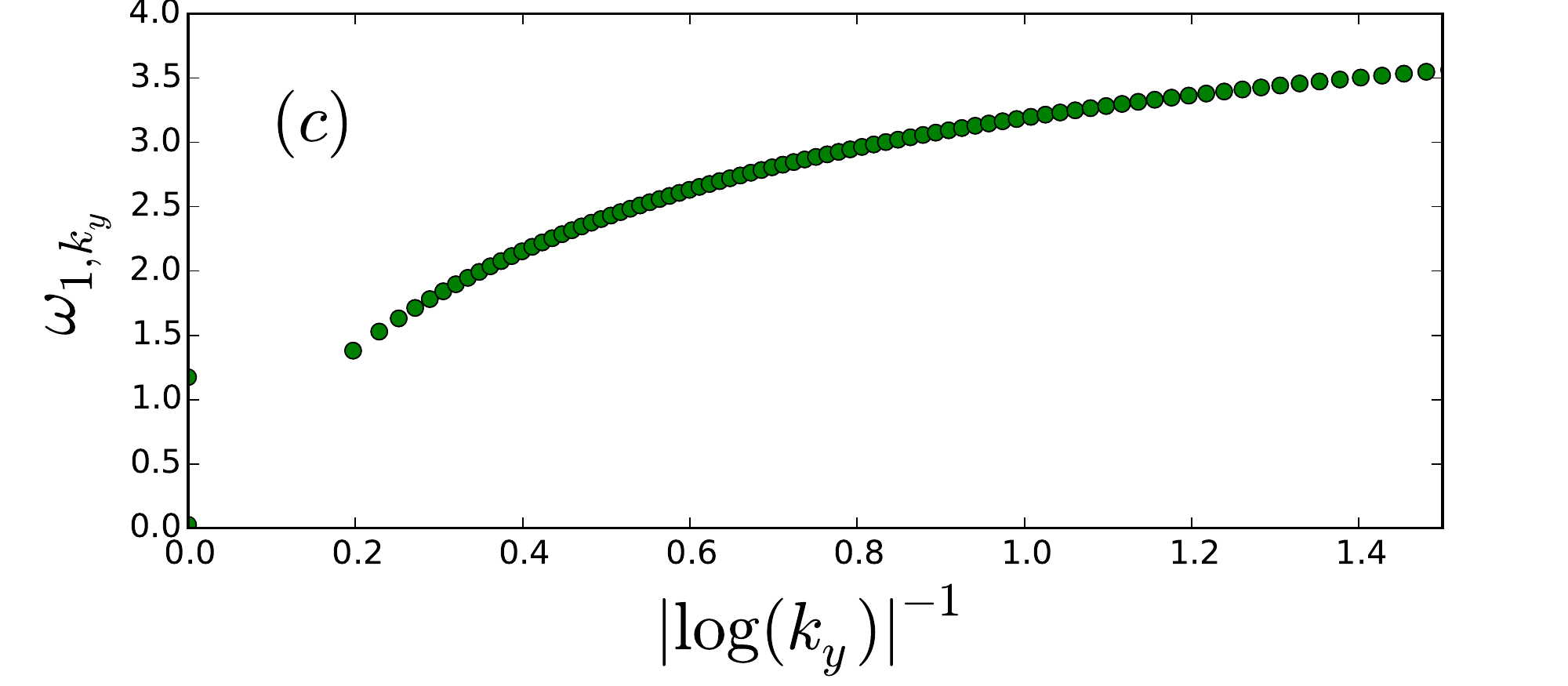}}
			\caption{$(a)$ Entanglement spectrum on a half-torus for hardcore bosons ($l=400$). Only the two lowest bands are shown, each band being two-fold degenerate (apart from the two lowest modes). $(b)$ Scaling of the two lowest entanglement energies $\omega_0$ and $\omega_1$. Solid lines show fits of the form $\omega_0 = a/l^{\gamma}$, with $\gamma \approx 0.57$ and $\omega_1 = a'/\ln(b'l)$, with $a' \approx 10$, $b'\approx 4.3$. (c) Zoom on the lowest branch near $\omega_1$, showing that the dispersion relation is compatible with $\omega_{1,k_y} \sim |\ln(k_y)|^{-1}$ at small $k_y$.}
			\label{sp_int_hardcore}
		\end{figure}
		\begin{figure}
			{\includegraphics[width = \linewidth]{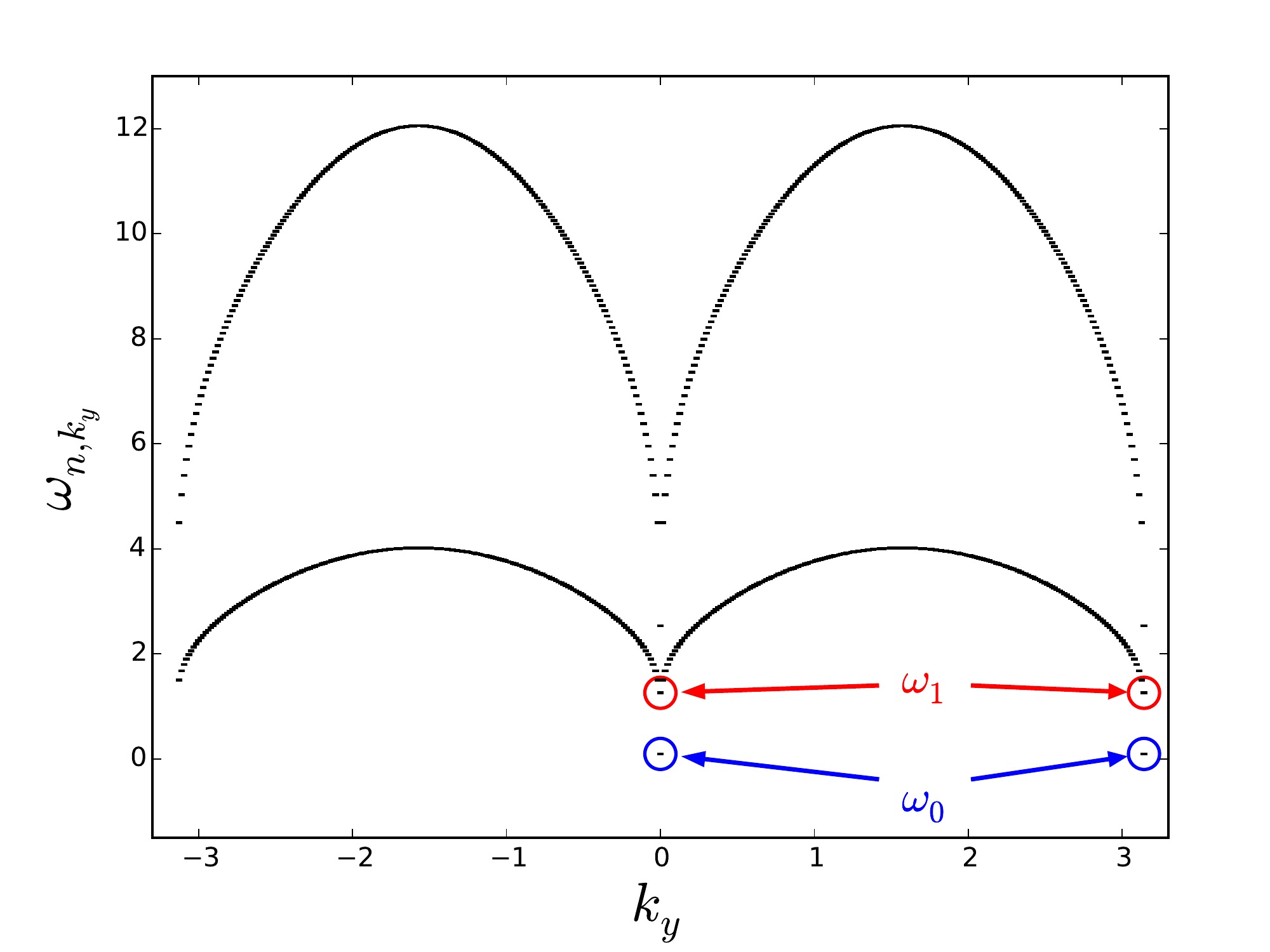}}
			{\includegraphics[width = \linewidth]{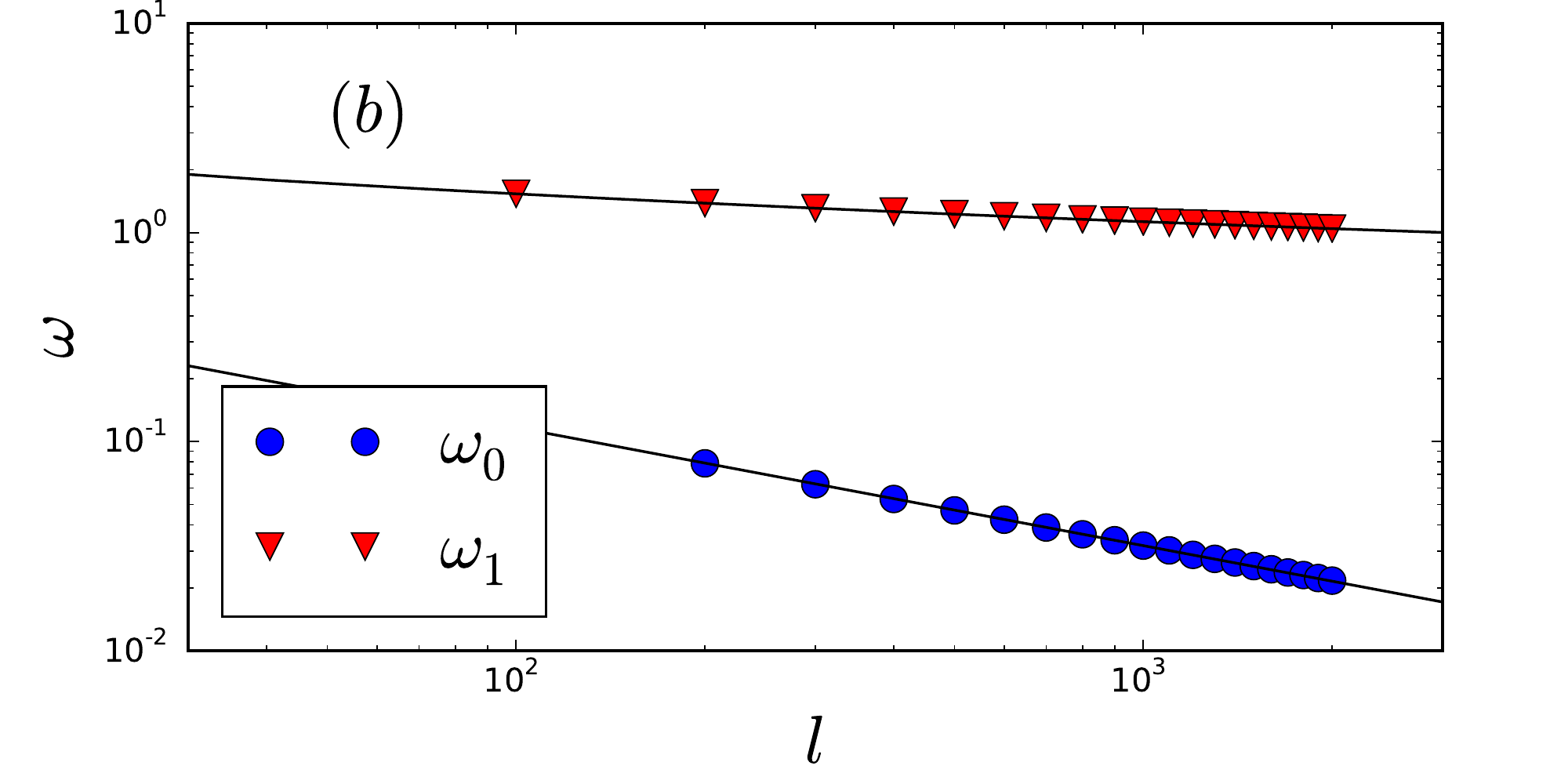}}
			\caption{$(a)$ Entanglement spectrum on a half-torus for Heisenberg antiferromagnet ($l=400)$. $(b)$ Scaling of the two lowest entanglement energies $\omega_0$ and $\omega_1$. Solid lines show fits of the form $\omega_0 = a/l^{\gamma}$, with $\gamma \approx 0.56$ and $\omega_1 = a'/\ln(b'l)$, with $a' \approx 10$, $b'\approx 6$}.
			\label{sp_int_antiferro}
		\end{figure}		
		
\subsection{From the lowest spin-wave mode to the lowest entanglement mode}		
		
	Having identified the special role played by the zero mode in the spin-wave spectrum, we remark that it has a fundamental relationship to a corresponding ``zero mode" in the entanglement spectrum. 
	
	We focus on the half-torus geometry for the $A$ subsystem, which leaves a full translational symmetry along the $A$-$B$ cut and endows therefore the entanglement eigenmodes with $d-1$ quantum numbers corresponding to the wavevector components for this displacement. In $d=2$, the quantum number is $k_y = 2\pi p/l$ with $p = -l/2+1,l/2$, $y$ being the direction along the cut;  hence the (one-body) entanglement spectrum is organized into ``bands" $\omega_{n,k_y}$ ($n=1, ..., l$). Fig.~\ref{sp_int_hardcore}(a) shows the one body entanglement spectrum for hardcore bosons for the two lowest bands; each band is found to consist of doubly degenerate eigenmodes, seemingly related to the existence of two boundaries between $A$ and $B$ on the half-torus geometry. Inspecting the spatial structure of the corresponding eigenmodes, we observe that the lowest band has eigenmodes algebraically localized around the $A$-$B$ boundary, while for higher bands the corresponding eigemodes have a weaker and weaker weight close to the boundary. The high ``energy modes'' play a small role in the entanglement properties, as they are weighted by the bosonic occupation $(e^{\omega_{n,k_y}}-1)^{-1}$ in the modal decomposition of entanglement, Eq.~\eqref{mode_entanglement}. 
	
	In Fig.~\ref{sp_int_hardcore}(a) one can clearly observe the existence of an isolated, lowest energy mode $\omega_{0}$ at $k_y=0$ (hereafter dubbed ``lowest entanglement zero mode"), scaling to zero energy when $l$ grows (as we shall detail shortly), and being non-degenerate. We observe that this entanglement mode disappears if the zero-energy mode of the spin-wave spectrum of the full system is artificially removed from the calculation of the correlation matrix; and that its entanglement energy ceases to scale to zero for growing $l$ if the lowest spin-wave mode is gapped out with a constant gap (namely if a magnetic field $h$ is applied which does not scale with $L$). Hence this establishes a fundamental relationship between the spin-wave zero mode and the lowest entanglement zero mode. A similar relationship between low-energy (ToS) excitations and the low-energy entanglement spectrum has been established numerically (beyond Bogoliubov or spin-wave theory) for bosonic and spin models in Ref.~\cite{alba2013,Kolley2013}.
	 
Coming back to the lowest entanglement zero mode within spin theory, how does it scale to zero? Fig.~\ref{sp_int_hardcore}(b) shows that $\omega_{0}$ scales as $\omega_{0} \sim l^{-\gamma}$ with $\gamma \approx 0.57$. Moreover we observe that there exists an upper $k_y=0$ zero mode in the lowest band (and non-degenerate, as the lowest mode), with an energy $\omega_1$ scaling to zero as $1/\ln l$ (as shown in Fig. \ref{sp_int_hardcore} (b)). The rest of the lowest branch of excitations is shown in Fig.~\ref{sp_int_hardcore}(c), and at small momentum $k_y$, the dispersion relation is found to be compatible with $\omega_{1,k_y} \sim |\ln(k_y)|^{-1}$. Hence the low-$k_y$ modes of the dispersion relation appear to go gapless as $|\ln l|^{-1}$. Ref.~\cite{Swingle2013} predicts an explicit form for the entanglement spectrum of the half space of Lorentz-invariant and massless quantum field theories, namely $\omega_{k} \sim |\ln(k)|^{-1}$ where $k$ is the wave vector parallel to the cut of space. Given that the continuum-space limit of the spin-wave dispersion relation reproduces such a quantum field theory, one expects that the prediction of  Ref.~\cite{Swingle2013} correctly captures the $k_y \to 0$ limit of $\omega_{1,k_y}$. Our findings are clearly not in contradiction with this idea, but there are serious finite-size limitations imposed by the logarithmic dependence of the dispersion relation, requiring prohibitively large systems to observe properly the asymptotic $\omega_{1,k_y}\to 0$ behavior of the dispersion relation.

\subsection{Logarithmic corrections to the entanglement entropy}

The modal decomposition of the entanglement entropy in Eq.~\eqref{mode_entanglement} is dominated by the lowest energy modes $\omega_{n,k_y} \to 0$, giving a contribution $S_{n,k_y} \approx - \ln \omega_{n,k_y}$. Hence from the modal analysis we conclude that, for $l\to\infty$
\be
S_A(l) \approx \frac{l}{2\pi} \sum_{n} \int_{-\pi}^\pi \D k_y~ S_{n,k_y} + \gamma \ln l + \gamma' \ln \ln l~+...
\ee
The entanglement zero mode gives a $\ln l$ term in the entropy scaling, while the other modes going gapless logarithmically give a $\ln \ln l$ contribution. The rest of the band gives instead the dominant area law - as well as further logarithmic contributions. The latter assumption (together with possible finite-size effects) is apparently necessary to reconcile the value of $\gamma =0.57$ obtained from the scaling of the entanglement zero mode with the prefactor $b\approx 0.5$ found for the logarithmic term in the scaling of the entanglement entropy shown in Fig.~\ref{entropy_scaling_hardcore}. Nonetheless the similarity between these two logarithmic terms is not coincidental.  
We stress further that the existence of the additive logarithmic terms in the entanglement entropy is fundamentally related to the $L^{-d}$ scaling of the spin-wave zero mode -- a fit of the entanglement entropy gives a vanishing logarithmic term if we gap out the spin-wave zero mode with a constant gap or artificially eliminate it. Yet we observe that the prefactor of such scaling, while it alters the prefactor of the scaling of $\omega_0$, leaves the $\gamma$ exponent unchanged, and therefore it does not affect the logarithmic contribution of the lowest entanglement zero mode to the modal decomposition of entanglement entropy (this applies as well to the $\gamma'$ coefficient).    

 The relationship between the spin-wave zero mode, the lowest entanglement zero mode and the logarithmic term in the entanglement entropy is further manifest when investigating the Heisenberg antiferromagnet, whose entanglement spectrum is shown in Fig.~\ref{sp_int_antiferro}. There we observe that \emph{all} entanglement eigenmodes -- including the zero modes - acquire a further double degeneracy, which reflects the double ($N_G=2$) degeneracy of the spin-wave spectrum due to the halving of the Brillouin zone. In particular, we observe that the lowest entanglement zero mode exhibits an energy with very similar scaling to the case of hardcore bosons, namely $\omega_0 \sim l^{-\gamma}$  with $\gamma\approx 0.56$. Therefore the existence of $N_G$ lowest entanglement zero modes leads to a $N_G \gamma \ln l$ contribution to the entanglement entropy. 
 
 Ref.~\cite{metlitski-grover} argues that the prefactor of the logarithmic correction to the entanglement entropy is a universal term $b=N_G (d-1)/2$, which has been extensively verified for $d=2$ hardcore bosons and Heisenberg antiferromagnets in Refs.~\cite{Kulchytskyyetal2015,Luitzetal2015} via quantum Monte Carlo and spin-wave theory. Our data for the scaling of entanglement entropy further corroborate this conclusion, showing that for hardcore bosons ($N_G=1$) the coefficient goes from $b\approx 1/2$ to $b\approx 1$ when moving from $d=2$  to $d=3$ (Fig.~\ref{entropy_scaling_hardcore}), and that for Heisenberg antiferromagnet ($N_G=2$), the coefficient goes from $b\approx 1$ to $b\approx 2$ (Fig.~\ref{entropy_scaling_antiferro}). To prove that $b=N_G (d-1)/2$, Ref.~\cite{metlitski-grover} relies on the contribution to the entanglement spectrum coming from the ToS. Here we argue that, within spin-wave theory, the logarithmic correction comes (mostly) from the lowest entanglement zero mode descending from the zero spin-wave mode, which approximates the first ToS excitation when introducing a spin-wave gap which scales like $L^{-d}$. To further corroborate this statement, we verified that going from $d=2$ to $d=3$, the coefficient $\gamma$ in the scaling of the entanglement energy of the zero modes goes from $\approx 0.56$ to $\approx 1.1$, for both the XX and Heisenberg model. The value of $\gamma$ is thus found to be quite close to $(d-1)/2$. It remains highly surprising to us that such a rough account of the ToS spectrum allows to reproduce so precisely the universal logarithmic term in the entanglement scaling.

\section{Conclusions}
\label{s.conclusion}

 In this paper we have addressed the fundamental question of the relationship between entanglement and local quantum fluctuations in extended quantum many-body systems, and we have attacked this problem from the point of view of the spatial structure of both properties. We have presented a survey of widely different quadratic models, encompassing free fermion Hamiltonians (with a gapless metallic, gapless semimetallic, gapped trivial or topological band insulating ground state) and quadratic bosonic ones (describing weakly interacting Bose-condensed bosons, planar quantum magnets and Heisenberg antiferromagnets).
 Entanglement and particle-number fluctuations in a subsystem $A$ have been decomposed into local contributions via the use of contours, which allows to compare/contrast the spatial decay of entanglement (when moving away from the boundary between the subsystem $A$ and its complement $B$) with the decay of correlations -- and in particular of density-density correlations, whose integral produces the fluctuation contours. The study of fermions reveals that the decay of entanglement contour mimics closely that of fluctuations, and that the spatial structure of entanglement is fundamentally governed by that of density-density correlations. In the case of bosons, on the other hand, the decay of the entanglement contour is significantly faster than that of the fluctuation contour. For weakly interacting bosons, if the fluctuation contour decays as the inverse of the distance, the short-range behavior of the entanglement contour appears to be exponentially decaying with a decay length proportional to the healing length, while the long-range behavior is algebraic with a decay exponent larger than one, and seemingly dependent on the interaction. For hardcore bosons and Heisenberg antiferromagnets, the entanglement contour appears to decay approximately as the square of the inverse distance; the lowest mode(s) of the entanglement spectrum contribute(s) an anomalously slow term to the contour, decaying like the inverse of the distance, and producing an additive logarithmic correction to the entanglement entropy. This contribution from the lowest entanglement eigenmode is traced back to the existence of gapless Goldstone modes in the spectrum of the total system, and its prefactor is found to be fundamentally controlled by the number of Goldstone modes and the dimensionality of the system, as recently predicted by Ref.~\cite{metlitski-grover}. 
  
  Our study shows that the entanglement contour provides fundamental new insight into the entanglement properties of quantum many-body systems and their relationship to correlation properties. In the case of fermions, the relationship between entanglement and fluctuations, established rigorously in a number of recent works \cite{calabrese-mintchev-vicari,bipartite-fluct}, is made physically transparent by the use of contours, which exhibit a universal algebraic decay in gapless systems, linked to the equally universal decay of density correlations. In the case of Bose-condensed bosons, on the other hand, universal features seemingly emerge in the hardcore limit, where the contour is found to exhibit the same asymptotic decay in the case of the quantum XX and the Heisenberg model. Yet the case of weakly interacting bosons shows non-universal features, which need therefore to be analyzed as the interaction is changed continuously. Here we adopt substantially different treatments for weakly interacting bosons on the one hand, and hardcore bosons on the other -- which prevents us from bridging the two limits within a single theory. Nonetheless this theory may exist, as no transition is expected in between them; we postpone the discussion of such a theory and of its entanglement properties to an upcoming publication \footnote{I. Fr\'erot and T. Roscilde, in preparation (2015).}.            
 
 Several fundamental questions are left open by our study. First and foremost, the concept of contours should be extended to interacting many-body systems. The formulation of contours stemming from the one-body local density of states, Eqs.~\eqref{e.Wi} and \eqref{e.rhoi} allows to generalize them to quantum many-body systems admitting a description in terms of free quasi-particles at low energy, but a definition for most general many-body systems is still lacking. Searching for such a definition one finds the same difficulty as that encountered for the concept of ``local entropy" in an extended, non-homogeneous system, such as the one described by the entanglement Hamiltonian of an $A$ subsystem. A local entropy could be in principle achieved via a ``local-density approximation" on the entanglement Hamiltonian (provided that such an Hamiltonian is explicitly known), relating the local behavior in $A$ to that of a translationally invariant Hamiltonian which has the same structure (coupling strengths, external potentials, etc.) as the local structure of the entanglement Hamiltonian. The local entropy (or entanglement contour) would then be defined as the entropy per site of the translationally invariant Hamiltonian; a step in this direction has been very recently taken by Ref.~\cite{SwingleM2015}. In particular, the local-density approximation underlying the above construction can be rigorously justified in the case of Lorentz-invariant and conformally invariant quantum field theories: there the reduced density matrix can be written in terms of a uniform local Hamiltonian (corresponding to the microscopic energy density) immersed in a thermal bath with spatially varying entanglement temperature \cite{Wong2013, Swingle2013}. As already mentioned, the relationship between the entanglement temperature and the contours remains a very interesting open question for future studies.
 
 Our study highlights an especially complex relationship between entanglement and correlation properties for bosons. The question concerning which correlation function/ fluctuation property dominates the behavior of entanglement remains wide open, and the answer is expected to be highly non-trivial in light of the complex structure of entanglement contours. Yet this question is of central importance in the endeavor of establishing the relationship between measurable physical properties and entanglement. Finally, even if our work is limited to ground state properties, it opens the path to the investigation of contours in the broader context of excited states, out-of-equilibrium systems (as already undertaken in Ref.~\cite{entanglement-contour}) and thermal states.

 \section{Acknowledgements}
 This work is supported by the Agence Nationale de la Recherche ("ArtiQ" project).

\newpage
\appendix
\section{Entanglement Hamiltonian and correlation functions}
\label{section-wick}
		In this Appendix, we expose in details the general procedure announced in Sec.~\ref{H_ent}. A detailed discussion of the diagonalization of quadratic Hamiltonians is found in \cite{blaizot}. Let ${\cal H}$ be a a quadratic Hamiltonian on a lattice with $N =L^d$ sites, expressed in terms of two $N \times N$ matrices $\cal A$ and $\cal B$ :
		\be
			{\cal H} = \sum_{i,j} a_i^\dagger {\cal A}_{ij} a_j + \frac{1}{2} a_i^\dagger {\cal B}_{ij} a_j^\dagger + \frac{1}{2} a_j {\cal B}_{ij}^* a_i
		\ee
		As ${\cal H}$ is hermitian, we take ${\cal A}^\dagger={\cal A}$ and ${\cal B}^T = -\epsilon {\cal B}$ ($\epsilon = -1$ for bosons, $\epsilon = 1$ for fermions). Introducing the notations : 
		\bearr
		 {\bm a} = \begin{pmatrix} a_1\\ \vdots \\ a_{N}  \end{pmatrix} &~& {\bm \alpha} = \begin{pmatrix}  {\bm a} \\ {\bm a}^\dagger \end{pmatrix} \nonumber \\
		\eta = \begin{pmatrix} \mathbbm{1}_{N} &  0  \\\ 0 &\epsilon \mathbbm{1}_{N}\end{pmatrix}  &~&
		~~ \mathcal{L} = \begin{pmatrix} {\cal A} &  {\cal B}\\ - {\cal B}^*& - {\cal A}^* \end{pmatrix}																			
		\eearr 
		${\cal H}$ is conveniently expressed as : 
		\be
			{\cal H} = \frac{1}{2} {\bm \alpha}^\dagger
						\begin{pmatrix} \eta \mathcal{L} \end{pmatrix} 
						{\bm \alpha}
						+ \frac{\epsilon}{2} {\rm Tr}{\cal A}
		\ee
		The $2N\times 2N$-matrix $\mathcal{L}$ is Hermitian only for fermions, while for bosons $\mathcal{L}^\dagger = \eta \mathcal{L} \eta$. Thanks to this $\eta$-hermiticity, $\mathcal{L}$ can in general be diagonalized according to a Bogoliubov transformation:
		\be
		{\bm \alpha} =  U \begin{pmatrix} {\bm b} \\ {\bm b}^\dagger \end{pmatrix} \equiv U {\bm \beta}
		\ee
		where $U$ is such that $\mathcal{L} = UDU^{-1}$ with $D$ diagonal. Since $\mathcal{L}^\dagger = \eta \mathcal{L} \eta$, $U^{\dagger} = \eta U^{-1} \eta$. Diagonalization is always possible for fermions, and a sufficient condition for bosons (always satisfied in the model studied in this paper) is that $\eta \mathcal{L} >0$. Furthermore, thanks to the property $\sigma \mathcal{L} \sigma = - \mathcal{L}^*$ where $\sigma =\begin{pmatrix} 0 &  \mathbbm{1}\\ \mathbbm{1}&0 \end{pmatrix}$, the eigenvalues $\omega_{\alpha}$ of $\mathcal{L}$ come into pairs $(\omega_{\alpha}, -\omega^*_{\alpha})$, namely if
		\be
		\mathcal{L} \begin{pmatrix} {\bm u}_{\alpha} \\ {\bm v}_{\alpha} \end{pmatrix} = \omega_{\alpha}  \begin{pmatrix} {\bm u}_{\alpha} \\ {\bm v}_{\alpha} \end{pmatrix} 
		\label{diag}
		\ee
		then
		\be
		 -\sigma \mathcal{L} \sigma \begin{pmatrix} {\bm u}_{\alpha}^* \\ {\bm v}_{\alpha}^* \end{pmatrix} = \omega_{\alpha}^* \begin{pmatrix} {\bm u}_{\alpha}^* \\ {\bm v}_{\alpha}^* \end{pmatrix} ~\Longrightarrow \mathcal{L} \begin{pmatrix} {\bm v}_{\alpha}^* \\ {\bm u}_{\alpha}^* \end{pmatrix} = - \omega_{\alpha}^*  \begin{pmatrix} {\bm v}_{\alpha}^* \\ {\bm u}_{\alpha}^* \end{pmatrix} 
		\ee
		
		In addition, if $\eta \mathcal{L}>0$, one can show that the eigenvalues $\omega_\alpha$ of $\mathcal{L}$ are real, namely that Eq.~\eqref{diag} implies
		\be
		 \begin{pmatrix} {\bm u}_{\alpha}^* & {\bm v}_{\alpha}^* \end{pmatrix} \eta \mathcal{L}  \begin{pmatrix} {\bm u}_{\alpha} \\ {\bm v}_{\alpha} \end{pmatrix} = \omega ~\left(|{\bm u}_{\alpha}|^2 +\epsilon |{\bm v}_{\alpha}|^2\right) > 0	\ee
		The last inequality implies that $\omega$ is real, and, incidentally, that $\omega_{\alpha}$ and the $\eta$-norm $|{\bm u}_{\alpha}|^2 +\epsilon |{\bm v}_{\alpha}|^2$ of the associated eigenvector have the same sign.  $\eta \mathcal{L}>0$ is a condition of thermodynamical stability, implying that no perturbation around the ground state can lower the energy. On the other hand, the fact that the frequencies $\omega_{\alpha}$ are real is fundamental requirement for the dynamical stability, as it implies that (initially) small perturbations around the ground state do not grow exponentially in time.
		Under this condition $\mathcal{L}$ takes the form :
		\be
		\mathcal{L} = U \begin{pmatrix} \omega & 0 \\ 0 & -\omega \end{pmatrix} U^{-1}
		\ee
		where $\omega = {\rm diag}(\omega_1,\cdots,\omega_N)$.
		Consequently, with the same notations, the matrix $U$ takes the form 
		\be 
		U = \begin{pmatrix} {\bm u}_1 &\ldots &{\bm u}_N&{\bm v}_1^*&\ldots&{\bm v}_N^*\\
						 {\bm v}_1 &\ldots &{\bm v}_N&{\bm u}_1^*&\ldots&{\bm u}_N^* \end{pmatrix}
		\ee
		The property $\eta U^\dagger \eta U=\mathbbm{1}$ means that the eigenbasis is $\eta-$orthonormal : 
		\bearr
		{\bm u}_\alpha^*\cdot {\bm u}_{\beta} + \epsilon~ {\bm v}_{\alpha}^*\cdot {\bm v}_{\beta} &=& \delta_{\alpha\beta} \\
		{\bm u}_\alpha\cdot {\bm v}_{\beta} + \epsilon ~{\bm v}_{\alpha}\cdot {\bm u}_{\beta} &=& 0 
		\eearr
		In the $b$ basis, ${\cal H}$ takes the diagonal form : 
		\be
		{\cal H} = {\bm b}^\dagger \omega {\bm b} - \frac{\epsilon}{2} + \frac{\epsilon}{2} {\rm Tr} {\cal A}
		\ee
		
		We introduce the generalized one-body correlation matrix : 
		\be \widetilde{C} \equiv \langle {\bm \alpha} {\bm \alpha}^\dagger \rangle = U \langle {\bm \beta} {\bm \beta}^\dagger \rangle U^\dagger
		\ee
		which can be expressed in terms of the one body correlation matrices $C_{ij} = \langle a_i^\dagger a_j \rangle$ and $F_{ij} = \langle a_i a_j \rangle$, which satisfy $C^\dagger = C$ and $F^\dagger = -\epsilon F^*$: 
		\be
		\widetilde{C} = \begin{pmatrix} 1-\epsilon C^* & F\\ -\epsilon F^*&C \end{pmatrix}~.
		\ee
		If the total density matrix has a thermal form $\rho = \exp(-{\cal H})$, the matrix of $ \langle {\bm \beta} {\bm \beta}^\dagger \rangle $ is simply diagonal in the $b$ basis, each mode $b_\alpha$ being occupied by $n_\alpha = 1/(\exp{(\omega_\alpha)}+\epsilon)$ quasi-particles on average : 
		\be
		\widetilde{C} = \begin{pmatrix} 1-\epsilon C^* & F\\ -\epsilon F^*&C \end{pmatrix}
		= U \begin{pmatrix} {\rm diag}(1-\epsilon n_\alpha) & 0 & \\ 
								0 &{\rm diag}(n_\alpha)& \end{pmatrix} U^\dagger
		\ee
		we then multiply on the right by $\eta$ :
		\be
		\widetilde{C} \eta = U \begin{pmatrix} {\rm diag}(1-\epsilon n_\alpha) & 0 & \\ 
								0 &{\rm diag}(n_\alpha)& \end{pmatrix} \eta~ \eta U^\dagger \eta~.
		\ee
		Since $\eta U^\dagger \eta = U^{-1}$ and multiplying by $\epsilon$, we finally obtain : 
		\be
		 \begin{pmatrix} \epsilon - C^* &  F\\ - F^*& C \end{pmatrix} = U \begin{pmatrix} {\rm diag}(\epsilon- n_\alpha) & 0 & \\ 
								0 &~ {\rm diag}(n_\alpha)& \end{pmatrix} U^{-1}~.
		\ee
		Note that when $C$ and $F$ are real matrices, a simplified treatment is possible (detailed for instance in \cite{botero-reznik} for bosons) which enables to diagonalize a $N\times N$ matrix, instead of our $2N\times 2 N$ generalized correlation matrix. However, our treatment is fully general, and applies as well for bosons and fermions.

\section{Density-density correlations for free fermions}
\label{dens-dens-fermions}		
		
The density-density correlation function for free fermions can be obtained from the one-body correlation function $G({\bm r}) = \langle c^{\dagger}_i c_{i+{\bm r}} \rangle$ via the Wick's theorem
\begin{equation}
\langle \delta n_i \delta n_{i+{\bm r}} \rangle =  \bar{n} ~ \delta_{\bm r, 0} - G^2({\bm r})~.
\end{equation}		
For fermions with a finite Fermi-surface, the one-body correlation function is generally estimated in the continuum limit (and assuming a spherical Fermi surface for $d>1$) as
\begin{equation}
G({\bm r}) \approx \int_{|{\bm k}| < k_F}  \frac{d^dk}{(2\pi)^d} ~ e^{i{\bm k}\cdot{\bm r}}~. 
\end{equation}
For $d=1$ this produces the well-known results for large $x$
\begin{equation}
G({\bm x}) \approx \frac{\sin(k_F x)}{\pi x} ~~~~~~~~~~~ \langle \delta n_i \delta n_{i+{\bm x}} \rangle \approx - \frac{1-\cos(2k_F  x)}{2\pi^2 x^2}~.
\end{equation}
Applying the formula Eq.~\eqref{intuitive_decay_cont_n_2} for the calculation of the contour, one readily obtains that 
\begin{equation}
{\cal C}_n(i) \approx \int_{-\infty}^0 dx' ~  \frac{1-\cos[2k_F  (x_i-x')]}{2\pi^2 (x_i-x')^2} \approx \frac{1}{2\pi^2 x_i}  + ...		
\end{equation}		
where the integral of the oscillating term is not calculated explicitly. 
		
\section{Density-density correlations for bosons}
\label{dens-dens-correl}
In this Appendix we explain how to obtain the expression for density-density correlations in terms of the operators appearing in the quadratic bosonic Hamiltonians. The quantity to be evaluated is:
	\be
	C({\bm r}_i-{\bm r}_j) = \langle \delta n_i  \delta n_j \rangle 
	\ee
	\subsection{Weakly interacting Bose gas}
	
	 The lattice Bose operator can be decomposed onto the momentum basis as 
	\be
	b_i = \frac{1}{\sqrt{L^d}} \sum_{{\bm k}\neq 0} e^{i{\bm k}\cdot {\bm r}_i} b_{\bm k} + \frac{b_0}{\sqrt{L^d}}~.
	\ee
	Within the Bogoliubov approximation, one makes the substitution $b_0 \to \sqrt{N}$ and introducing the operators $\delta b_i = b_i - \sqrt{n}$ ($n = N/L^d$), one obtains 
	\be
	C({\bm r}_i-{\bm r}_j) \approx n \left ( \langle \delta b_i^\dagger \delta b_{j}^\dagger  + {\rm h.c.} \rangle + \langle \delta b_i^\dagger \delta b_{j} + {\rm h.c.}\rangle + \delta_{i,j} \right )
	\ee
	where we neglected terms of order 3 and 4 in the $\delta b$ operators. 
	Using the fact that  
	\be
	\langle \delta b_i^\dagger \delta b_j \rangle = \frac{1}{L^d}\sum_{{\bm k}\neq 0} e^{i{\bm k}\cdot ({\bm r}_i-{\bm r}_j)} v_{\bm k}^2
	\ee
	and
	 \be
	 \langle \delta b_i \delta b_j \rangle = \frac{1}{L^d}\sum_{{\bm k}\neq 0} e^{i{\bm k}\cdot ({\bm r}_i-{\bm r}_j)} (-u_{\bm k} v_{\bm k}) 
	\ee 
	we obtain
	\be
	C({\bm r}) = \frac{2n}{V} \sum_{{\bm k}\neq 0} e^{i{\bm k}\cdot{\bm r}} (v_{\bm k}^2 - u_{\bm k} v_{\bm k} + 1/2) 
	\ee
	Finally, using Eq.~\eqref{expressions_u_v} for $u_{\bm k}$ and $v_{\bm k}$, and after a little algebra, we obtain
	\be
	C({\bm r}) = \frac{n}{V} \sum_{{\bm k}\neq 0} e^{i{\bm k}\cdot {\bm r}} \frac{\epsilon_{\bm k}}{E_{\bm k}}
	\ee
	with $E_{\bm k} = \sqrt{\epsilon_{\bm k}(\epsilon_{\bm k}+2gn)}$. This expression was used in \cite{astrakharchik} and \cite{klawunn} to calculate atom-number fluctuations.
	 
 	\subsection{Hardcore bosons}
	In the case of hardcore bosons, the Matsubara-Matsuda transformation introduced in Sec.~\ref{s.hardcore} leads to the identification ${\tilde b}_i^{\dagger} {\tilde b}_i - 1/2 = S^y_i$. Hence the density-density correlation function for hardcore bosons becomes:
	\be
	C({\bm r}_i-{\bm r}_j) = \langle S^y_i S^y_j \rangle~.
	\ee
	Rebosonizing the spins via the Holstein-Primakoff (HP) transformation, Eq.~\eqref{HPtransform}, with quantization axis along $z$, one obtains after linearization of the HP transformation 
     \be
	S^y_i \approx \frac{1}{2i} (b_i - b_i^{\dagger})
	\ee
	which leads to the following expression for the correlation function :
	\bearr
	C({\bm r}_i-{\bm r}_j) &=& \frac{1}{4}\left( \langle b_i b^\dagger_j  + b_i^\dagger b_j \rangle - \langle b_i b_j + b_i^\dagger b_j^\dagger \rangle  \right)\nonumber \\
	 & =& \frac{1}{2} \left[ f({\bm r}_i-{\bm r}_j)  - g({\bm r}_i-{\bm r}_j) \right ]\nonumber \\
	 &=& \frac{1}{4L^d}  \sum_{\bm k} e^{i{\bm k}\cdot ({\bm r}_i-{\bm r}_j)} \frac{A_{\bm k} + B_{\bm k}}{\sqrt{A_{\bm k}^2-B_{\bm k}^2}}~.
	 \label{e.Crhcbosons}
	 \eearr
	 Since $A_{\bm k} = 2-\gamma_{\bm k}$ and $B_{\bm k} = -\gamma_{\bm k}$, one obtains : 
	 \be
	 C({\bm r}) = \frac{n}{L^d} \sum_{\bm k} e^{i{\bm k}\cdot {\bm r}} \frac{\sqrt{1-\gamma_{\bm k}}}{2}~.
	 \ee

\bibliographystyle{unsrt}
\bibliography{biblio}

\begin{thebibliography}{10}

\bibitem{entanglement-contour}
Yangang Chen and Guifre Vidal.
\newblock Entanglement contour.
\newblock {\em Journal of Statistical Mechanics: Theory and Experiment},
  2014(10):P10011, 2014.

\bibitem{many-body}
Luigi Amico, Rosario Fazio, Andreas Osterloh, and Vlatko Vedral.
\newblock Entanglement in many-body systems.
\newblock {\em Rev. Mod. Phys.}, 80:517--576, May 2008.

\bibitem{RMP-area-laws}
J.~Eisert, M.~Cramer, and M.~B. Plenio.
\newblock \textit{Colloquium} : Area laws for the entanglement entropy.
\newblock {\em Rev. Mod. Phys.}, 82:277--306, Feb 2010.

\bibitem{Wen2012}
X.-G. Wen.
\newblock Topological order: from long-range entangled quantum matter to an
  unification of light and electrons.
\newblock {\em arXiv}, 1210.1281, 2012.

\bibitem{bipartite-fluct}
H.~Francis Song, Stephan Rachel, Christian Flindt, Israel Klich, Nicolas
  Laflorencie, and Karyn Le~Hur.
\newblock Bipartite fluctuations as a probe of many-body entanglement.
\newblock {\em Phys. Rev. B}, 85:035409, Jan 2012.

\bibitem{song}
H.~Francis Song, Nicolas Laflorencie, Stephan Rachel, and Karyn Le~Hur.
\newblock Entanglement entropy of the two-dimensional heisenberg
  antiferromagnet.
\newblock {\em Phys. Rev. B}, 83:224410, Jun 2011.

\bibitem{Songetal2010}
H.~Francis Song, Stephan Rachel, and Karyn Le~Hur.
\newblock General relation between entanglement and fluctuations in one
  dimension.
\newblock {\em Phys. Rev. B}, 82:012405, Jul 2010.

\bibitem{jordan-buttiker}
Andrew~N. Jordan and Markus B\"uttiker.
\newblock Entanglement energetics at zero temperature.
\newblock {\em Phys. Rev. Lett.}, 92:247901, Jun 2004.

\bibitem{calabrese-mintchev-vicari}
Pasquale Calabrese, Mihail Mintchev, and Ettore Vicari.
\newblock Exact relations between particle fluctuations and entanglement in
  fermi gases.
\newblock {\em EPL (Europhysics Letters)}, 98(2):20003, 2012.

\bibitem{Hastings2007}
M~B Hastings.
\newblock An area law for one-dimensional quantum systems.
\newblock {\em Journal of Statistical Mechanics: Theory and Experiment},
  2007(08):P08024, 2007.

\bibitem{BrandaoH2015}
Fernando~G.S.L. Brand\~ao and Michal Horodecki.
\newblock Exponential decay of correlations implies area law.
\newblock {\em Communications in Mathematical Physics}, 333(2):761--798, 2015.

\bibitem{Hastingsetal2010}
Matthew~B. Hastings, Iv\'an Gonz\'alez, Ann~B. Kallin, and Roger~G. Melko.
\newblock Measuring renyi entanglement entropy in quantum monte carlo
  simulations.
\newblock {\em Phys. Rev. Lett.}, 104:157201, Apr 2010.

\bibitem{Kallinetal2011}
Ann~B. Kallin, Matthew~B. Hastings, Roger~G. Melko, and Rajiv R.~P. Singh.
\newblock Anomalies in the entanglement properties of the square-lattice
  heisenberg model.
\newblock {\em Phys. Rev. B}, 84:165134, Oct 2011.

\bibitem{HumeniukR2012}
Stephan Humeniuk and Tommaso Roscilde.
\newblock Quantum monte carlo calculation of entanglement r\'enyi entropies for
  generic quantum systems.
\newblock {\em Phys. Rev. B}, 86:235116, Dec 2012.

\bibitem{Luitzetal2015}
David~J. Luitz, Xavier Plat, Fabien Alet, and Nicolas Laflorencie.
\newblock Universal logarithmic corrections to entanglement entropies in two
  dimensions with spontaneously broken continuous symmetries.
\newblock {\em Phys. Rev. B}, 91:155145, Apr 2015.

\bibitem{Kulchytskyyetal2015}
B.~Kulchytskyy, C.~M. Herdman, S.~Inglis, and R.~G. Melko.
\newblock Detecting goldstone modes with entanglement entropy.
\newblock {\em arXiv}, 1502.01722, 2015.

\bibitem{gioev-klich}
Dimitri Gioev and Israel Klich.
\newblock Entanglement entropy of fermions in any dimension and the widom
  conjecture.
\newblock {\em Phys. Rev. Lett.}, 96:100503, Mar 2006.

\bibitem{wolf}
Michael~M. Wolf.
\newblock Violation of the entropic area law for fermions.
\newblock {\em Phys. Rev. Lett.}, 96:010404, Jan 2006.

\bibitem{astrakharchik}
G.~E. Astrakharchik, R.~Combescot, and L.~P. Pitaevskii.
\newblock Fluctuations of the number of particles within a given volume in cold
  quantum gases.
\newblock {\em Phys. Rev. A}, 76:063616, Dec 2007.

\bibitem{klawunn}
M.~Klawunn, A.~Recati, L.~P. Pitaevskii, and S.~Stringari.
\newblock Local atom-number fluctuations in quantum gases at finite
  temperature.
\newblock {\em Phys. Rev. A}, 84:033612, Sep 2011.

\bibitem{calabrese-cardy}
Pasquale Calabrese and John Cardy.
\newblock Entanglement entropy and quantum field theory.
\newblock {\em Journal of Statistical Mechanics: Theory and Experiment},
  2004(06):P06002, 2004.

\bibitem{storms-singh}
Michelle Storms and Rajiv R.~P. Singh.
\newblock Entanglement in ground and excited states of gapped free-fermion
  systems and their relationship with fermi surface and thermodynamic
  equilibrium properties.
\newblock {\em Phys. Rev. E}, 89:012125, Jan 2014.

\bibitem{S-HE-entier}
Iv\'an~D. Rodr\'iguez and Germ\'an Sierra.
\newblock Entanglement entropy of integer quantum hall states.
\newblock {\em Phys. Rev. B}, 80:153303, Oct 2009.

\bibitem{ding-yang}
Wenxin Ding and Kun Yang.
\newblock Entanglement entropy and mutual information in bose-einstein
  condensates.
\newblock {\em Phys. Rev. A}, 80:012329, Jul 2009.

\bibitem{blaizot}
J.P. Blaizot and G.~Ripka.
\newblock {\em Quantum Theory of Finite Systems}.
\newblock The MIT Press, Cambridge, MA, 1986.

\bibitem{Peschel-Eisler}
Ingo Peschel and Viktor Eisler.
\newblock Reduced density matrices and entanglement entropy in free lattice
  models.
\newblock {\em Journal of Physics A: Mathematical and Theoretical},
  42(50):504003, 2009.

\bibitem{casini-huerta}
H~Casini and M~Huerta.
\newblock Entanglement entropy in free quantum field theory.
\newblock {\em Journal of Physics A: Mathematical and Theoretical},
  42(50):504007, 2009.

\bibitem{botero-reznik}
Alonso Botero and Benni Reznik.
\newblock Spatial structures and localization of vacuum entanglement in the
  linear harmonic chain.
\newblock {\em Phys. Rev. A}, 70:052329, Nov 2004.

\bibitem{Fetter-Walecka}
A.~L. Fetter and J.~D. Walecka.
\newblock {\em Quantum Theory of Many-Particle Systems}.
\newblock Dover publications, Dover, NY, 2012.

\bibitem{TersoffH1985}
J.~Tersoff and D.~R. Hamann.
\newblock Theory of the scanning tunneling microscope.
\newblock {\em Phys. Rev. B}, 31:805--813, Jan 1985.

\bibitem{Kreiseletal2014}
A.~Kreisel, Peayush Choubey, T.~Berlijn, B.~M. Andersen, and P.~J. Hirschfeld.
\newblock Interpretation of scanning tunneling quasiparticle interference and
  impurity states in cuprates.
\newblock {\em arXiv}, 1407.1846, 2014.

\bibitem{Wenbook}
X.-G. Wen.
\newblock {\em Quantum Field Theory of Many-Body Systems}.
\newblock Oxford, 2004.

\bibitem{Wong2013}
Gabriel Wong, Israel Klich, LeopoldoA.Pando Zayas, and Diana Vaman.
\newblock Entanglement temperature and entanglement entropy of excited states.
\newblock {\em Journal of High Energy Physics}, 2013(12), 2013.

\bibitem{Swingle2013}
B.~{Swingle}.
\newblock {Structure of entanglement in regulated Lorentz invariant field
  theories}.
\newblock {\em ArXiv e-prints}, April 2013.

\bibitem{You2014}
Wen-Long You.
\newblock The scaling of entanglement entropy in a honeycomb lattice on a
  torus.
\newblock {\em Journal of Physics A: Mathematical and Theoretical},
  47(25):255301, 2014.

\bibitem{levin-wen}
Michael Levin and Xiao-Gang Wen.
\newblock Detecting topological order in a ground state wave function.
\newblock {\em Phys. Rev. Lett.}, 96:110405, Mar 2006.

\bibitem{kitaev-preskill}
Alexei Kitaev and John Preskill.
\newblock Topological entanglement entropy.
\newblock {\em Phys. Rev. Lett.}, 96:110404, Mar 2006.

\bibitem{PitaevskiiStringari}
L.~P. Pitaevskii and S.~Stringari.
\newblock {\em Bose-Einstein Condensation}.
\newblock Oxford, 2003.

\bibitem{giorgini}
S.~Giorgini, L.~P. Pitaevskii, and S.~Stringari.
\newblock Anomalous fluctuations of the condensate in interacting bose gases.
\newblock {\em Phys. Rev. Lett.}, 80:5040--5043, Jun 1998.

\bibitem{HelmesW2014}
Johannes Helmes and Stefan Wessel.
\newblock Entanglement entropy scaling in the bilayer heisenberg spin system.
\newblock {\em Phys. Rev. B}, 89:245120, Jun 2014.

\bibitem{Stoudenmireetal2014}
E.~M. Stoudenmire, Peter Gustainis, Ravi Johal, Stefan Wessel, and Roger~G.
  Melko.
\newblock Corner contribution to the entanglement entropy of strongly
  interacting o(2) quantum critical systems in 2+1 dimensions.
\newblock {\em Phys. Rev. B}, 90:235106, Dec 2014.

\bibitem{metlitski-grover}
M.~A. {Metlitski} and T.~{Grover}.
\newblock {Entanglement Entropy of Systems with Spontaneously Broken Continuous
  Symmetry}.
\newblock {\em ArXiv e-prints}, December 2011.

\bibitem{CasiniH2007}
H.~Casini and M.~Huerta.
\newblock Universal terms for the entanglement entropy in dimensions.
\newblock {\em Nuclear Physics B}, 764(3):183 -- 201, 2007.

\bibitem{MatsubaraM1956}
Takeo Matsubara and Hirotsugu Matsuda.
\newblock A lattice model of liquid helium, i.
\newblock {\em Progress of Theoretical Physics}, 16(6):569--582, 1956.

\bibitem{coletta}
Tommaso Coletta, Nicolas Laflorencie, and Fr\'ed\'eric Mila.
\newblock Semiclassical approach to ground-state properties of hard-core bosons
  in two dimensions.
\newblock {\em Phys. Rev. B}, 85:104421, Mar 2012.

\bibitem{Anderson1952}
P.~W. Anderson.
\newblock An approximate quantum theory of the antiferromagnetic ground state.
\newblock {\em Phys. Rev.}, 86:694--701, Jun 1952.

\bibitem{ZimanN1989}
Herbert Neuberger and Timothy Ziman.
\newblock Finite-size effects in heisenberg antiferromagnets.
\newblock {\em Phys. Rev. B}, 39:2608--2618, Feb 1989.

\bibitem{WeihongH1993}
Zheng Weihong and C.~J. Hamer.
\newblock Spin-wave theory and finite-size scaling for the heisenberg
  antiferromagnet.
\newblock {\em Phys. Rev. B}, 47:7961--7970, Apr 1993.

\bibitem{alba2013}
Vincenzo Alba, Masudul Haque, and Andreas~M. L\"auchli.
\newblock Entanglement spectrum of the two-dimensional bose-hubbard model.
\newblock {\em Phys. Rev. Lett.}, 110:260403, Jun 2013.

\bibitem{Kolley2013}
F.~Kolley, S.~Depenbrock, I.~P. McCulloch, U.~Schollw\"ock, and V.~Alba.
\newblock Entanglement spectroscopy of su(2)-broken phases in two dimensions.
\newblock {\em Phys. Rev. B}, 88:144426, Oct 2013.

\bibitem{SwingleM2015}
B.~Swingle and J~McGreevy.
\newblock Area law for gapless states from local entanglement thermodynamics.
\newblock {\em arXiv}, 1505.07106, 2015.

\bibitem{FagottiC2011}
Maurizio Fagotti and Pasquale Calabrese.
\newblock Universal parity effects in the entanglement entropy of xx chains
  with open boundary conditions.
\newblock {\em Journal of Statistical Mechanics: Theory and Experiment},
  2011(01):P01017, 2011.

\end{thebibliography}

\end{document}